\documentclass[
    amsfonts,
    amsmath,
    amssymb,
    aps,
    floatfix,
    longbibliography,
    nofootinbib,
    prx,
    reprint,
    superscriptaddress,
    twoside,
]{revtex4-2}
\usepackage[english]{babel}
\usepackage{amsthm}
\usepackage{array}
\usepackage{ragged2e}
\usepackage{enumitem}
\usepackage[T1]{fontenc}
\usepackage{graphicx}
\usepackage[colorlinks=true,linkcolor=blue,citecolor=red]{hyperref}
\usepackage{multirow}
\usepackage{orcidlink}
\usepackage{physics}
\usepackage{placeins}

\theoremstyle{plain}
\newtheorem{theorem}{Theorem}[section]
\newtheorem{proposition}[theorem]{Proposition}

\theoremstyle{definition}
\newtheorem{definition}[theorem]{Definition}
\theoremstyle{plain}

\hypersetup{
    pdftitle={Interference-caged quantum many-body scars: Local ETH-violation witnesses and deformation stability},
    pdfauthor={Tao-Lin Tan; Yi-Ping Huang},
    pdflang={en-US}
}

\begin{document}

\title{Interference-caged quantum many-body scars: Local ETH-violation witnesses and deformation stability}

\author{Tao-Lin Tan\,\orcidlink{0009-0005-5398-1206}}
    \email[Contact author: ]{tao-lin.tan@gapp.nthu.edu.tw} %
    \affiliation{Department of Physics, National Tsing Hua University, Hsinchu 30013, Taiwan}

\author{Yi-Ping Huang\,\orcidlink{0000-0001-6453-1191}}
    \email[Contact author: ]{yphuang@phys.nthu.edu.tw} %
    \affiliation{Department of Physics, National Tsing Hua University, Hsinchu 30013, Taiwan}
    \affiliation{Physics Division, National Center for Theoretical Sciences, Taipei 10617, Taiwan}
    \affiliation{Institute of Physics, Academia Sinica, Taipei 115, Taiwan}

\date{\today} %

\begin{abstract}
Many-body interference can render an exact eigenstate locally nonthermal by canceling a microscopic process active in thermal states. We formulate the Fock-space analogue of compact flat-band localization as \emph{interference caging}: escape amplitudes cancel exactly across the active boundary of a set of product configurations. We give sufficient conditions to reduce this configuration-dependent cancellation to a fixed real-space region, yielding a bounded local caging operator. Its positive square vanishes on the caged eigenstate and witnesses violation of the eigenstate thermalization hypothesis (ETH) if its microcanonical expectation for the same Hamiltonian and energy density remains bounded away from zero in the thermodynamic limit. We call these states \emph{interference-caged quantum many-body scars} (ICQMBS) when they form a vanishing fraction of a symmetry-resolved bulk energy window whose typical eigenstates are locally thermal for every bounded local observable. The mechanism is independent of quasiparticle, algebraic, or symmetry-based constructions and compatible with Hilbert-space fragmentation. Our chiral-shell search constructs exact caged eigenstates from allowed transitions without a preselected wavefunction, target subspace, or ladder operator. We identify local ETH-violation witnesses, test their deformation stability, and demonstrate thermal contrast in two systems. In the spin-1 XY chain, we recover the known scar tower, also describable through quasiparticles and projector embedding, and prove ICQMBS throughout an open interval of compatible exchange couplings in the thermodynamic limit. In the square-lattice quantum dimer model, we construct exact caged eigenstates and local operators from plaquette flips on fixed-width strips under a checkerboard phase deformation; finite-size microcanonical comparisons support thermal contrast.
\end{abstract}

\maketitle

\section{Introduction} \label{sec: introduction}

Why can an isolated quantum many-body system look thermal under local measurements even though its evolution is unitary?
The eigenstate thermalization hypothesis (ETH) answers this at the level of individual energy eigenstates: in a generic interacting system, a bounded local observable essentially takes its equilibrium value in a typical eigenstate at the same finite energy density \cite{deutsch1991,srednicki1994,rigol2008thermalization,dalessio_quantum_2016,mori_thermalization_2018}.
Thus, ETH says more than agreement after averaging many eigenstates: asymptotically, a fixed local measurement cannot distinguish a typical eigenstate from equilibrium.

Quantum many-body scars (QMBS) are striking exceptions: a sparse set of atypical eigenstates can remain locally nonthermal inside an otherwise thermal spectrum.
Their modern study was motivated by persistent coherent oscillations in Rydberg-atom experiments and the subsequent identification of special eigenstates in constrained models \cite{bernien2017probing,turner2018,turner2018weak}; broader perspectives are reviewed in Refs.~\cite{serbyn_quantum_2021,papic2022weak,moudgalya2022quantum,chandran_quantum_2023}.
Exact constructions based on projector embeddings, spectrum-generating algebras, quasiparticle towers, commutant algebras, and Hilbert-space fragmentation now explain how special states or subspaces can remain embedded in an exponentially large many-body spectrum \cite{shiraishi_systematic_2017,mark_unified_2020,moudgalya2020eta,Moudgalya2022,moudgalya_exhaustive_2023,pai2019,khemani2020,sala2020}.
Building on these exact constructions, we ask how the microscopic mechanism
protecting a state can supply a physical measurement of its local
nonthermal behavior.
This is the question we address: \emph{when does the local structure that protects an exact eigenstate also make it locally distinguishable from equilibrium, and when can both properties survive a Hamiltonian deformation?}

To make this local distinction precise, consider a system of size $L$ on a physical lattice $\Omega_L$ and a bounded observable $\widehat O_R$ supported on a fixed region $R$, with support and norm independent of $L$.
After resolving the relevant exact global symmetries, diagonal ETH requires, for a typical bulk eigenstate $|E_n\rangle$,
\begin{equation}
    \langle E_n|\widehat O_R|E_n\rangle
    =\operatorname{Tr}\!\left[
        \rho_{\mathrm{mc}}(E_n)\widehat O_R
      \right]+o(1),
    \qquad L\to\infty,
    \label{eq: intro_diagonal_eth}
\end{equation}
where $\rho_{\mathrm{mc}}(E_n)$ is the corresponding microcanonical ensemble and $o(1)$ vanishes with system size.
Throughout, by diagonal ETH we mean this typical-eigenstate local statement within the symmetry-resolved energy window.
A persistent thermodynamic difference between QMBS and typical eigenstates for one fixed local observable is precisely the kind of local diagnostic we seek.

Our starting point is a familiar interference picture.
In a single-particle flat-band lattice, a compact localized state can remain trapped on a finite cluster because hopping amplitudes into neighboring zero-amplitude sites cancel exactly \cite{Sutherland1986,Bergman2008,leykam2018artificial}.
For a many-body Hamiltonian, the analogous interference occurs not directly in real space but on the weighted configuration graph of the microscopic product basis: vertices are many-body configurations and edges are local physical rearrangements.
\emph{Interference caging} on such many-body configuration graphs has recently emerged as an active direction, with related many-body and Fock-space constructions exploring configuration-space confinement and its consequences \cite{ben2025many,nicolau2025fragmentation,jonay2025localized,mohapatra_unraveling_2025}.
We call such a state an \emph{interference-caged eigenstate}, or \emph{caged eigenstate} for short, when its occupied configuration support has an active boundary---individual transitions out of the support are allowed, but their coherent amplitudes cancel.
The caged eigenstates studied below are not accidental finite-size null vectors: their boundary cancellation reduces to repeated local interference rules.
Given a caged eigenstate and its Hamiltonian, we ask when the repeated
cancellation yields a bounded local operator that detects a violation of
diagonal ETH, and whether this mechanism survives Hamiltonian deformation.
Interference caging can be identified from the Hamiltonian and its Hilbert space alone, independently of quasiparticle, algebraic, or symmetry-based constructions of QMBS, which rely on additional structural input. It can also occur within a Hilbert-space fragment (see Appendix~\ref{appendix: relations_other_formalisms} for connections to other QMBS constructions).

The first step is to ask whether the same destructive-interference rule can be expressed by a \emph{bounded local caging operator} $\widehat L_R$ acting on a fixed physical region $R$.
One route follows the canceled kinetic process itself; another uses a locally sharp potential value when the interference-caged eigenstate carries one.
Once such an operator exists, its positive square
$\widehat Q_R=\widehat L_R^\dagger\widehat L_R$
defines the \emph{local ETH-violation witness} for the microcanonical comparison.
For a size-compatible caged sequence $|\psi_L\rangle$, with
$\rho_{\mathrm{mc},L}$ the corresponding resolved microcanonical reference, the simple conditions
\begin{subequations}
\label{eq: intro_thermal_activity}
\begin{gather}
    \widehat L_R|\psi_L\rangle=0,
    \label{eq: intro_local_annihilation}
    \\
    \liminf_{L\to\infty}
    \operatorname{Tr}\!\left[
        \rho_{\mathrm{mc},L}\widehat Q_R
    \right]>0,
    \label{eq: intro_positive_thermal_activity}
\end{gather}
\end{subequations}
are sufficient for persistent local diagonal-ETH violation. The witness is \emph{dark} as it
vanishes on the caged eigenstate, and its microcanonical expectation at the
same energy density stays bounded away from zero.
If the same resolved window forms a locally thermal background in this diagonal-ETH sense and the family of caged eigenstates occupies a vanishing spectral fraction, we call the sequence an \emph{interference-caged quantum many-body scar} (ICQMBS).
Section~\ref{sec: interference_caging} gives the precise thermodynamic definition.

The second step is to ask whether this structure admits a \emph{deformation continuation} as the Hamiltonian couplings change.
A deformation can create leakage out of the configuration support; readjusting
the state's amplitudes and energy may repair it.  Section~\ref{sec: deformation_stability}
formulates the corresponding first-order obstruction and asks whether the
bounded local caging operator can continue with the state.
Thus, deformation continuation of an exact eigenstate, of its local physical mechanism, and of the ICQMBS diagnosis remain three distinct questions.

The framework does not depend on a particular discovery algorithm.
For practical discovery, Sec.~\ref{sec: practical_search} narrows the candidate
support to configurations sharing a potential value and lacking internal
kinetic transitions before solving the exact boundary equation.  This helps
identify the repeated interference rules used by the framework.

We apply the framework in two complementary settings.
In the spin-1 $XY$ chain, we recover the
known staggered $\pi$-bimagnon tower, built from paired spin raises on
individual sites with alternating sign \cite{schecter_weak_2019}, expose its
bounded local interference mechanism, preserve that mechanism under a genuine
Hamiltonian deformation, and prove that the resulting caged sequence is a
thermodynamic ICQMBS throughout an open compatible exchange family.
The square-lattice quantum dimer model (QDM) provides a constrained-Hilbert-space
realization with an exact repeated interference-caged construction and a
fixed-width local thermal test
\cite{rokhsar_superconductivity_1988,moessner_short-ranged_2001,moessner2010quantum,banerjee_quantum_2021,biswas_scars_2022,wildeboer2021,sau2024}.
The exact state and bounded local rule continue analytically under a
checkerboard phase deformation, while its thermodynamic ICQMBS
classification remains open despite the strong finite-size evidence.

\begin{figure*}[!htbp]
    \centering
    \includegraphics[width=\textwidth]{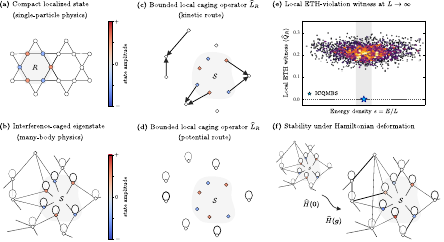}
    \caption{From destructive-interference caging to a local ETH test and its
    deformation stability.  Warm- and cool-colored vertices carry amplitudes of
    opposite sign; open vertices have zero amplitude.
    \textbf{(a)} A single-particle compact localized state is trapped on a
    finite real-space motif $R$ because hopping to neighboring sites cancels.
    \textbf{(b)} In the many-body product basis, the analogous cancellation
    occurs at the active boundary of a configuration support $\mathcal S$.
    \textbf{(c),(d)} When that cancellation admits a bounded real-space
    reduction, a local caging operator $\widehat L_R$ can arise from a kinetic
    process \textbf{(c)} or a locally sharp potential value \textbf{(d)}.
    \textbf{(e)} The witness $\widehat Q_R=\widehat L_R^\dagger\widehat L_R$
    vanishes on the caged eigenstate (star) and has a positive microcanonical
    value in the energy-matched window. The color scale shows schematic
    scatter-point density around the microcanonical center, rather than the
    many-body density of states.
    \textbf{(f)} Along a preserving deformation parameterized by $\boldsymbol g$, the amplitudes and local
    matrix elements may change while the caged eigenstate and its bounded local diagnostic
    continue together.}
    \label{fig: overview_idea}
\end{figure*}

Figure~\ref{fig: overview_idea} summarizes the full logic.
Sections~\ref{sec: product_basis_graph}--\ref{sec: deformation_stability} develop the configuration-space criterion, its bounded local reduction, the ETH test, and deformation stability; Sec.~\ref{sec: practical_search} gives a practical search construction; and Secs.~\ref{sec: spin1_xy} and~\ref{sec: square_qdm} present the two model applications.

\section{From compact localized states to interference caging}
\label{sec: product_basis_graph}

A compact localized state remains confined even though hopping across its
boundary is allowed: the outgoing amplitudes cancel.  The same mechanism can
occur in a many-body Hilbert space, where the relevant ``sites'' are microscopic
many-body configurations.  This section turns that picture into an exact
support--boundary condition.  Section~\ref{sec: interference_caging} then asks
the harder question: when can the same cancellation be detected by a bounded
real-space observable?

\subsection{Compact localization as the guiding picture}

Consider a single particle hopping on a lattice.  A compact localized state
(CLS) has nonzero amplitude on a finite cluster and vanishes outside it even
when individual hoppings across the cluster boundary are nonzero.  At every
neighboring zero-amplitude site, the outgoing paths cancel coherently.  In the
simplest two-path example,
\begin{equation}
    t_{hu_1}\phi_{u_1}+t_{hu_2}\phi_{u_2}=0,
    \label{eq: cls_two_path_cancellation}
\end{equation}
where $\phi_{u_1}$ and $\phi_{u_2}$ are amplitudes on two occupied sites, $h$ an adjacent site with zero amplitude, and $t_{hu}$ the single-particle hopping strength.  Together with the eigenvalue
equation inside the cluster, this boundary cancellation makes the compact wave
function an exact eigenstate.  The same mechanism underlies many flat-band
CLSs and Aharonov--Bohm cages
\cite{Sutherland1986,vidal1998,Bergman2008,leykam2018artificial}.

For a many-body problem, we replace the single-particle sites
$(u_1,u_2,h)$ by microscopic product configurations.  An edge between two
configurations then represents an allowed local rearrangement generated by the
Hamiltonian, and Eq.~\eqref{eq: cls_two_path_cancellation} becomes a
cancellation among many-body transition amplitudes.  What we carry over from
flat-band physics is this \emph{active-boundary interference mechanism}.  We do
not require the many-body configuration support itself to remain finite, nor do
we require dynamical caging of an entire band or invariant subspace.

\subsection{Hamiltonian connectivity in the microscopic product basis}

Let $\Omega$ denote the physical lattice and let
$\mathcal B=\{\ket v\}$ be the microscopic product basis fixed by its local
degrees of freedom.  In a constrained model, $\mathcal B$ means the allowed
subset of those product configurations.  If an exact superselection label is
diagonal in this basis---for example, total magnetization, particle number, or a
dimer winding number---we may restrict $\mathcal B$ to that block.  A symmetry
such as translation momentum is different: because its eigenstates are
coherent superpositions of product configurations, it is not imposed by
deleting vertices from $\mathcal B$.  Such symmetries are instead resolved in
the thermal reference of Sec.~\ref{ssec: thermodynamic_eth_criterion}.

All configuration-space statements below are therefore relative to this
\emph{declared microscopic product basis}.  A different basis would generally
produce a different connectivity graph and a different configuration support;
we make no basis-independent caging claim.  This basis dependence is physically meaningful because each graph edge
retains the interpretation of a microscopic local process.

We assume a real-space-local Hamiltonian
\begin{equation}
    \widehat H=\sum_X \widehat h_X,
\end{equation}
where the physical diameter of every $X\subset\Omega$ and the norm
$\|\widehat h_X\|$ are bounded independently of system size, and each site
belongs to only a bounded number of such terms.  When useful, we write
\begin{equation}
    \widehat H=\widehat K+\widehat V,
    \label{eq: general_kinetic_potential_split}
\end{equation}
with $\widehat K$ containing off-diagonal transitions and $\widehat V$ the
diagonal matrix elements in $\mathcal B$.  This split is basis dependent and
is not a decomposition into conserved pieces; generally
$[\widehat K,\widehat V]\neq0$ because a local transition can change the
diagonal energy.

The corresponding \emph{weighted configuration graph} has one vertex for each
$\ket v\in\mathcal B$ and an edge between distinct configurations whenever
$\langle v|\widehat H|u\rangle\neq0$.  The matrix elements are the edge weights,
while diagonal terms may be retained as weighted self-loops.  Hermiticity only
requires
$\langle v|\widehat H|u\rangle=\langle u|\widehat H|v\rangle^{*}$; both the edge
weights and the state amplitudes may otherwise be complex.  The real weights
used in Fig.~\ref{fig: overview_idea} are only a visual simplification, and
Sec.~\ref{sec: deformation_stability} likewise allows complex Hermitian
couplings.  A non-Hermitian algebraic extension is recorded in the appendix
and does not enter the ETH claims.

The graph records \emph{which configurations are connected by physical local
processes}; its drawn geometry has no physical meaning.\footnote{For
visualization, we often use force-directed graph-layout algorithms.  Their
coordinates are chosen from graph connectivity for readability and do not
represent physical distances or an embedding geometry.}  Related configuration
graphs are standard tools in quantum dynamics and many-body localization
\cite{logan1990quantum,altshuler1997quasiparticle,roy2024fock}.  This is also
where graph locality must be distinguished from real-space locality.  Two
configurations can be adjacent because they differ by one local move, while a
projector selecting one complete outside configuration can still act on the
entire system.  We therefore reserve ``local'' for physical real-space
locality and say ``local on the configuration graph'' when that weaker notion
is intended.  Likewise, $\mathcal S$ below denotes a state's
\emph{configuration support}, whereas $R$ in Sec.~\ref{sec: interference_caging}
denotes the physical support of an operator.

\subsection{Configuration support, active boundary, and interference caging}

For a state $\ket\psi$, its actual configuration support is
\begin{equation}
    \mathcal S_\psi
    =\operatorname{supp}_{\mathcal B}\ket\psi
    =\{\ket u\in\mathcal B:\langle u|\psi\rangle\neq0\}.
\end{equation}
We henceforth write $\mathcal S\equiv\mathcal S_\psi$, with complement
$\overline{\mathcal S}=\mathcal B\setminus\mathcal S$, and denote the span of
$\mathcal S$ by $\mathcal H_{\mathcal S}$.  The associated projectors are
\begin{equation}
    \widehat P_{\mathcal S}
    =\sum_{\ket u\in\mathcal S}\ket u\!\bra u,
    \qquad
    \widehat P_{\overline{\mathcal S}}
    =\widehat{\mathbf 1}-\widehat P_{\mathcal S}.
\end{equation}

Only some outside configurations can receive amplitude directly from
$\mathcal S$.  They form the \emph{active configuration boundary}
\begin{equation}
    \partial\mathcal S
    =\left\{
        \ket h\in\overline{\mathcal S}:
        \bra h\widehat H\widehat P_{\mathcal S}\neq0
      \right\}.
\end{equation}
The corresponding boundary map
\begin{equation}
    B_{\mathcal S}
    \equiv
    \widehat P_{\partial\mathcal S}
    \widehat H
    \widehat P_{\mathcal S}:
    \mathcal H_{\mathcal S}
    \longrightarrow
    \mathcal H_{\partial\mathcal S}
    \label{eq: boundary_map_definition}
\end{equation}
collects every possible leakage from the occupied configuration support.

Write
$\ket\psi=\sum_{u\in\mathcal S}x_u\ket u$ and collect its nonzero amplitudes
in $\boldsymbol x$.  For $X,Y\in\{\mathcal S,\overline{\mathcal S}\}$, let
$H_{XY}=\widehat P_X\widehat H\widehat P_Y$.  Relative to
$\mathcal H_{\mathcal S}\oplus\mathcal H_{\overline{\mathcal S}}$,
\begin{subequations}
\label{eq: support_boundary_characterization}
\begin{align}
    \widehat H
    &=
    \begin{pmatrix}
        H_{\mathcal S\mathcal S}
        &H_{\mathcal S\overline{\mathcal S}}
        \\
        H_{\overline{\mathcal S}\mathcal S}
        &H_{\overline{\mathcal S}\overline{\mathcal S}}
    \end{pmatrix},
    \label{eq: subgraph-matrix-model}
    \\
    \widehat H
    \begin{pmatrix}\boldsymbol x\\0\end{pmatrix}
    =E\begin{pmatrix}\boldsymbol x\\0\end{pmatrix}
    &\quad\Longleftrightarrow\quad
    \begin{cases}
        H_{\mathcal S\mathcal S}\boldsymbol x=E\boldsymbol x,\\
        B_{\mathcal S}\boldsymbol x=0.
    \end{cases}
    \label{eq: boundary_kernel_condition}
\end{align}
\end{subequations}
The first equation fixes the state inside $\mathcal S$.  The second says that
although boundary transitions are present, their total outgoing amplitude into
every $\ket h\in\partial\mathcal S$ vanishes exactly,
\begin{equation}
    \sum_{\ket u\in\mathcal S}H_{hu}x_u=0,
    \label{eq: boundary_row_cancellation}
\end{equation}
as in the single-particle cancellation of
Eq.~\eqref{eq: cls_two_path_cancellation}.  We call such an active boundary
configuration an \emph{interference zero}.

The support--boundary decomposition itself is an identity for any eigenstate
once its complete nonzero support is known.  Its physical content therefore
comes from requiring that leakage processes actually exist and cancel rather
than being absent altogether.

\begin{definition}[Interference-caged eigenstate]
\label{def: interference_caged_eigenstate}
Fix the microscopic product basis and let $\mathcal S$ be the complete nonzero
configuration support of an eigenstate
$\ket\psi=\sum_{u\in\mathcal S}x_u\ket u$.  We call $\ket\psi$ an
\emph{interference-caged eigenstate}, or simply a \emph{caged eigenstate}, when
\begin{equation}
    H_{\mathcal S\mathcal S}\boldsymbol x=E\boldsymbol x,
    \qquad
    B_{\mathcal S}\boldsymbol x=0,
    \qquad
    B_{\mathcal S}\neq0.
    \label{eq: interference_caged_definition}
\end{equation}
Equivalently, the state has a proper configuration support with a nonempty
active boundary: individual leakage processes are allowed, but their coherent
sum vanishes exactly.
\end{definition}

The condition $B_{\mathcal S}\neq0$ excludes a disconnected invariant
component.  At fixed system size $L$, however, Definition~\ref{def: interference_caged_eigenstate}
is intentionally broad: neither $|\mathcal S|$ nor the diameter of a graphical
motif has to remain bounded.  Across system sizes we speak of a \emph{family of interference-caged eigenstates},
or simply a \emph{family of caged eigenstates}, only when one recurring
construction can be followed along the sequence; its configuration support may
still grow.

There is one further fixed-size subtlety.  Inside an exactly degenerate
energy eigenspace, a basis rotation can change the product-basis support of an
individual eigenvector.  We therefore attribute caging to a vector only when
it is fixed by additional exact labels or by the declared analytic
construction.  For a degenerate collection we first identify the complete
caged subspace and only afterward choose convenient sparse representatives;
Appendix~\ref{ssec: degenerate_cage_localization} gives the basis-independent
subspace convention.

Finally, interference caging is not yet a statement about ETH violation.  
The equations above certify a configuration-space interference mechanism 
at fixed size, but they do not tell us whether the cancellation can be expressed 
by a bounded real-space operator, nor whether such an operator remains distinguishable 
from the energy-matched thermal ensemble.  Those are precisely the two steps of
Sec.~\ref{sec: interference_caging}.

\section{From interference caging to local ETH-violation witnesses}
\label{sec: interference_caging}

Section~\ref{sec: product_basis_graph} explains how destructive interference can
confine an exact eigenstate to a configuration support \(\mathcal S\).  That
configuration-space statement is not yet a local diagnostic of thermalization.
A canceled boundary row of $B_{\mathcal S}$ (an interference zero)
specifies a physical transition only after the rest of
the many-body configuration has been fixed; as an operator, it therefore
carries an outside projector that can extend across the system.  The key
question is whether this outside condition can be removed while preserving the
same darkness.

\begin{figure*}[!t]
    \centering
    \includegraphics[width=\textwidth]{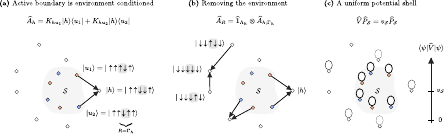}
    \caption{From configuration-space cancellation to a local ETH-violation
    witness.  Filled vertices belong to the configuration support \(\mathcal S\);
    open vertices lie outside it.  The spin-$1/2$ product-basis configurations are schematic and only illustrate the local transitions. \textbf{(a)} A boundary configuration
    \(\ket h\) receives canceling amplitudes from occupied parents.  The active
    transitions lie in a microscopic region \(\Gamma_h\), but the exact row
    operator is still conditioned on the outside configuration in
    \(\Lambda_h=\Omega\setminus\Gamma_h\), where \(\Omega\) denotes the
    physical lattice in this schematic.  \textbf{(b)} A kinetic reduction is
    local only when the same operator on \(R=\Gamma_h\) annihilates every
    occupied outside environment; orthogonal environments cannot cancel one
    another.  \textbf{(c)} The illustrated global uniform-potential shell is a
    precursor to the diagonal route.  A bounded local caging operator additionally
    requires a sharp potential value on the bounded physical region $R$, giving
    \(\widehat Y_R=\widehat V_R-v_R\widehat{\mathbf1}_R\).  Either
    route gives a local caging operator \(\widehat L_R\) and hence the positive
    witness \(\widehat Q_R=\widehat L_R^\dagger\widehat L_R\).}
    \label{fig:local_caging_operator_schematic}
\end{figure*}

When it can, the cancellation is represented by a \emph{bounded local caging
operator} \(\widehat L_R\) acting on a fixed physical region \(R\), with
\(\widehat L_R\ket\psi=0\).  Along a thermodynamic sequence, the diameter of
\(R\) and \(\|\widehat L_R\|\) remain bounded.  Its positive square,
\(\widehat Q_R=\widehat L_R^\dagger\widehat L_R\), is a bounded, positive
witness. It vanishes on the caged eigenstate, and its microcanonical value
tests whether the same local condition holds in equilibrium.
Figure~\ref{fig:local_caging_operator_schematic}
shows the physical reduction.

There are two complementary ways to obtain \(\widehat L_R\).  The
\emph{kinetic route} starts from one or several canceled boundary transitions
and removes their outside conditioning.  The independent \emph{diagonal route}
uses a local potential value that is sharp throughout the caged superposition;
it does not require \([\widehat K,\widehat V]=0\).  Both routes end with the
same thermal question.  For the kinetic route, the logic is
\begin{equation}
    \boxed{\begin{aligned}
    B_{\mathcal S}\boldsymbol x=0
    &\xrightarrow{\text{bounded local reduction}}
    \widehat L_R\ket{\psi_L}=0,\\[-0.2ex]
    &\xrightarrow{\text{thermal comparison}}
    \liminf_{L\to\infty}\langle\widehat Q_R\rangle_{\rm mc}>0 .
    \end{aligned}}
    \label{eq: caging_to_local_eth_chain}
\end{equation}
The first implication is a locality problem; the second is an ETH test.  Neither
is automatic from interference caging alone.

We first present the kinetic reduction in an ordinary tensor-product Hilbert
space, where the outside conditioning is easiest to see.  Constrained models
obey the same logic after resolving the locally allowed conditional blocks;
Appendix~\ref{appendix: local_iz_technical_details} gives that formulation.
The chiral and uniform-shell restrictions introduced later for searching are
not assumptions here.  After establishing the two local routes, we formulate
the common thermodynamic criterion and then ask in
Sec.~\ref{sec: deformation_stability} whether the caged eigenstate, its local operator, and
its local ETH mismatch survive the same Hamiltonian deformation.

\subsection{Kinetic route: localizing a boundary cancellation}
\label{ssec: interference_zeros_local_reduction}

\subsubsection{One canceled boundary row}

We first consider Eq.~\eqref{eq: boundary_row_cancellation} for a single
interference zero \(\ket h\in\partial\mathcal S\).  This is the simplest
starting point for a bounded local reduction because the corresponding boundary
row involves only the kinetic moves from the occupied support into \(h\).
Writing \(K_{hu}=\bra h\widehat K\ket u\), the occupied parents are
\begin{equation}
    N_{\mathcal S}(h)
    =\{u\in\mathcal S:K_{hu}\neq0\},
    \label{eq: occupied_boundary_neighborhood}
\end{equation}
and the canceled row is
\begin{equation}
    \sum_{u\in N_{\mathcal S}(h)}K_{hu}x_u=0.
    \label{eq: Iz}
\end{equation}
The question is whether this configuration-space cancellation can be read as a
process confined to a bounded physical region.

Decompose the local kinetic Hamiltonian as
\(\widehat K=\sum_X\widehat K_X\).  Let \(\Gamma_h\) collect the physical
supports of all local terms that generate the occupied transitions
\(u\to h\) in Eq.~\eqref{eq: Iz}, and let
\(\Lambda_h=\Omega\setminus\Gamma_h\).  Although every \(\widehat K_X\) is
local, the union \(\Gamma_h\) need not be: distinct parents could feed the same
row through processes far apart.  A single boundary row therefore defines a
bounded physical motif only when
\(\operatorname{diam}(\Gamma_h)=O(1)\) along the size sequence.

In an ordinary product Hilbert space,
\begin{equation}
    \mathcal H
    =\mathcal H_{\Lambda_h}\otimes\mathcal H_{\Gamma_h}.
    \label{eq: ordinary_product_decomposition}
\end{equation}
Every parent in the selected row agrees with \(h\) outside \(\Gamma_h\); call
that common outside configuration \(\alpha_h\).  Restricting the row to the
microscopic motif gives the \emph{directed kinetic operator}
\begin{equation}
    \widehat A_{h;\Gamma_h}
    =\ket h_{\Gamma_h}
      \sum_{u\in N_{\mathcal S}(h)}
      K_{hu}\,{}_{\Gamma_h}\!\bra u,
    \label{eq: directed_local_iz}
\end{equation}
which returns the coherent amplitude entering the local configuration \(h\).
Its Hermitian symmetrization,
\begin{equation}
    \widehat Z_{h;\Gamma_h}
    =\widehat A_{h;\Gamma_h}
     +\widehat A_{h;\Gamma_h}^{\dagger},
    \label{eq: local_iz_kernel_operator}
\end{equation}
adds the reverse process and will be useful when a Hermitian local observable is
preferred.

At this stage, however, the row still knows which outside configuration was
selected.  With
\(\widehat P_{\alpha_h;\Lambda_h}
=\ket{\alpha_h}\!\bra{\alpha_h}_{\Lambda_h}\), the exact cancellation gives
\begin{align}
    \bigl(
    \widehat P_{\alpha_h;\Lambda_h}
    \otimes\widehat A_{h;\Gamma_h}
    \bigr)\ket\psi&=0,
    \label{eq: conditioned_directed_operator}\\
    \bigl(
    \widehat P_{\alpha_h;\Lambda_h}
    \otimes\widehat Z_{h;\Gamma_h}
    \bigr)\ket\psi&=0.
    \label{eq: conditioned_iz_operator}
\end{align}
The second relation also uses the zero amplitude of the boundary configuration
in that environment.  These operators reproduce the boundary row exactly, but
the projector on \(\Lambda_h\) can be extensive.  They are therefore not yet
local caging operators.

\subsubsection{Removing the outside condition}
\label{ssec: environment_condition_removal}

A genuine local reduction asks the same microscopic operator to work without
checking the rest of the system.  We therefore extend the local rule by the
identity,
\begin{align}
    \widehat A_R
    &=\widehat{\mathbf 1}_{\Lambda_h}
      \otimes\widehat A_{h;\Gamma_h},
      \label{eq: reduced_directed_operator}\\
    \widehat Z_R
    &=\widehat{\mathbf 1}_{\Lambda_h}
      \otimes\widehat Z_{h;\Gamma_h},
      \label{eq: reduced_iz_operator}
\end{align}
with \(R=\Gamma_h\).  If \(R\) remains bounded, these are bounded local
operators.  The nontrivial question is whether they still annihilate the full
many-body state.

To see the obstruction directly, resolve the normalized state by its outside
environment,
\begin{equation}
    \ket\psi
    =\sum_\alpha c_\alpha
      \ket\alpha_{\Lambda_h}\otimes
      \ket{\psi_{R|\alpha}}_{\Gamma_h},
    \qquad
    \sum_\alpha|c_\alpha|^2=1.
    \label{eq: common_sector_resolution}
\end{equation}
Here \(\ket{\psi_{R|\alpha}}\) is the conditional state on the bounded motif
\(R=\Gamma_h\).  The originally selected environment
\(\alpha=\alpha_h\) is dark by construction;
locality requires the same rule to be dark in every other environment that the
state actually occupies.

\begin{proposition}[Environment-independence criterion for local reduction]
\label{prop: complementary_sector_reduction}
Let \(\widehat L_{h;\Gamma_h}\) denote either
\(\widehat A_{h;\Gamma_h}\) or \(\widehat Z_{h;\Gamma_h}\), and define
\(\widehat L_R
=\widehat{\mathbf 1}_{\Lambda_h}\otimes\widehat L_{h;\Gamma_h}\)
with \(R=\Gamma_h\).  Then
\begin{equation}
    \widehat L_R\ket\psi=0
    \quad\Longleftrightarrow\quad
    c_\alpha\,
    \widehat L_{h;\Gamma_h}\ket{\psi_{R|\alpha}}=0
    \quad\text{for every }\alpha.
    \label{eq: theorem_sectorwise_reduction}
\end{equation}
If \(\operatorname{diam}(R)=O(1)\), a successful reduction yields a bounded
local caging operator.  For \(\widehat L=\widehat A\), the condition tests the
coherent incoming amplitude; for \(\widehat L=\widehat Z\), it also tests the
reverse process and occupation of the local boundary configuration.
\end{proposition}

The criterion has a simple physical reading.  If \(c_\alpha=0\), that outside
environment is absent and imposes no condition.  If \(c_\alpha\neq0\), the same
local operator must annihilate \(\ket{\psi_{R|\alpha}}\).  Because different
outside configurations are orthogonal, a failure in one occupied environment
cannot be repaired by a cancellation in another.  In a constrained Hilbert
space, the locally allowed conditional blocks themselves can depend on
\(\alpha\); Appendix~\ref{ssec: appendix_complementary_sector_proof} gives the
corresponding formulation.

Once this criterion is satisfied, we suppress the row label \(h\) and retain
only the physical support \(R\): the boundary row has served its purpose of
constructing a local rule.  The directed choice gives the positive witness
\(\widehat A_R^\dagger\widehat A_R\), while the Hermitian choice gives
\(\widehat Z_R^2\).  Appendix~\ref{ssec: appendix_directed_positive_constraint}
relates these two witnesses for a single boundary row.
Neither local annihilation by itself nor positivity establishes an ETH
violation; that requires the thermal comparison in
Sec.~\ref{ssec: thermodynamic_eth_criterion}.

\subsubsection{When one boundary row is not enough}
\label{ssec: local_reduction_hierarchy}

A single row can fail the environment-independence test even when the physical
interference motif is local.  The reason is simple: several nearby boundary
configurations may have to be combined before the outside dependence cancels.
The kinetic route should therefore ask whether the \emph{whole} required
combination still fits inside one bounded region.

Let \(\mathcal C\) be a finite set of boundary rows with
\(\Gamma_h\subseteq R\), let \(\widehat A_{h;R}\) denote each local row block
padded to the common region \(R\), and let \(\eta_h\) be their coherent
coefficients.  The corresponding directed operator is
\begin{equation}
    \widehat A_{\mathcal C;R}
    =\widehat{\mathbf1}_{\Omega\setminus R}\otimes
      \sum_{h\in\mathcal C}\eta_h\widehat A_{h;R}.
    \label{eq: bounded_collective_annihilator}
\end{equation}
A \emph{bounded multi-row kinetic reduction} is present when
\(\widehat A_{\mathcal C;R}\ket\psi=0\), the diameter of \(R\) remains
\(O(1)\), and one size-compatible normalization keeps the coefficients
\(\eta_h\) bounded.  Individual row extensions need not be dark; only their
coherent sum inside \(R\) must be.  Because the local Hilbert space and \(R\)
are both bounded, the number of distinct row blocks in such a motif is also
bounded independently of system size.

Hermitian symmetrization is performed only after the rows have been combined,
\begin{equation}
    \widehat Z_{\mathcal C;R}
    =\widehat A_{\mathcal C;R}
     +\widehat A_{\mathcal C;R}^{\dagger}.
    \label{eq: bounded_multirow_hermitian_channel}
\end{equation}
Since the directed combination is already dark, this Hermitian combination
qualifies as a local caging operator only when the reverse combination
\(\widehat A_{\mathcal C;R}^{\dagger}\) also annihilates every occupied
conditional state.  Under the directed condition alone, the positive witness
\(\widehat A_{\mathcal C;R}^{\dagger}\widehat A_{\mathcal C;R}\)
remains valid.

If every annihilating combination requires a region whose diameter grows with
system size, the eigenstate can remain exactly caged in configuration space,
but this kinetic route has not produced a local caging operator.  Conversely,
failure to find one bounded combination is only a negative result for the
searched construction, not a no-go theorem for all local observables.
Appendix~\ref{ssec: appendix_multirow_kinetic_reduction} gives the corresponding
multi-row algebra.

The diagonal route below is independent of this hierarchy.  A caged eigenstate may
possess a kinetic reduction, a shifted-potential operator, both, or neither.

\subsection{Diagonal route: a locally fixed potential value}
\label{ssec: potential_shell_constraints}

Boundary interference is not the only way a caged eigenstate can become locally dark.  In
many constrained models, every product configuration in the caged
superposition carries the same diagonal energy.  The potential then cannot tell
those occupied configurations apart.  If the same sharp value is already
visible on a bounded region, it directly supplies a second local caging
operator.

Using the kinetic--potential split of
Eq.~\eqref{eq: general_kinetic_potential_split}, let
\(\mathcal S=\operatorname{supp}_{\mathcal B}\ket\psi\).  We call
\(\mathcal S\) a \emph{uniform-potential shell} when
\begin{equation}
    \widehat V\widehat P_{\mathcal S}
    =v_{\mathcal S}\widehat P_{\mathcal S}.
    \label{eq: uniform_shell_scalar_action}
\end{equation}
Thus every occupied configuration contributes the same potential energy
\(v_{\mathcal S}\).  If the kinetic support--boundary problem gives
\begin{equation}
    \widehat K\ket\psi=\varepsilon\ket\psi,
    \label{eq: shell_kinetic_eigenstate}
\end{equation}
then the full energy is simply
\begin{equation}
    E=\varepsilon+v_{\mathcal S}.
    \label{eq: shell_energy_shift_general}
\end{equation}
No commutation of \(\widehat K\) and \(\widehat V\) on the full Hilbert space
is assumed.

Such a shell can be structural rather than fine-tuned.  Local constraints or
fixed motif counts may force the occupied configurations to share the same
potential value throughout a coupling family; equality that instead relies on
a special relation among independently variable local contributions is tuned.
Appendix~\ref{ssec: appendix_uniform_shell_structure} gives the linear
conditions, and Sec.~\ref{sec: deformation_stability} asks which shells persist
under deformation.

For a tunable potential
\(\widehat V(\boldsymbol\lambda)=\sum_\mu
\lambda_\mu\widehat N_\mu\), the $\lambda_\mu$ are independent
potential couplings and the $\widehat N_\mu$ are their fixed diagonal
operators.  A structural shell already has a useful finite-size spectral consequence.  The caged eigenstate remains an eigenstate
of \(\widehat V(\boldsymbol\lambda)\) with the common shell value
\(v_{\mathcal S}(\boldsymbol\lambda)\), whereas neighboring eigenstates
that are not confined to the same shell generically shift or hybridize
differently as \(\boldsymbol\lambda\) varies.  When this tuning makes the
caged eigenstate spectrally exceptional relative to those neighbors, it
realizes the mechanism described as \emph{order-by-disorder in the Hilbert
space} in constrained systems hosting QMBS~\cite{banerjee_quantum_2021,biswas_scars_2022}.
Here the phrase denotes only this finite-size spectral exceptionality: it
neither establishes a local ETH mismatch nor its persistence in the
thermodynamic limit.  Those require the bounded local test below and the
thermodynamic criterion of Sec.~\ref{ssec: thermodynamic_eth_criterion}.

A sharp \emph{total} potential energy is still not a local ETH diagnostic.
The additional local condition can be summarized as follows.

\begin{proposition}[Diagonal local-caging criterion]
\label{prop: diagonal_local_caging_criterion}
Fix a declared microscopic decomposition
\(\widehat V=\sum_X\widehat v_X\) into bounded local terms and, on a bounded
region \(R\), define
\begin{equation}
    \widehat V_R=\sum_{X\subseteq R}\widehat v_X.
    \label{eq: local_potential_operator}
\end{equation}
If a caged eigenstate has the sharp local value
\begin{equation}
    \widehat V_R\ket\psi=v_R\ket\psi,
    \label{eq: local_potential_shell_condition}
\end{equation}
then the \emph{shifted-potential operator}
\begin{equation}
    \widehat Y_R
    =\widehat V_R-v_R\widehat{\mathbf 1}_R,
    \qquad
    \widehat Y_R\ket\psi=0,
    \label{eq: local_potential_annihilator}
\end{equation}
is a bounded local caging operator.  The witness
\(\widehat Q_R=\widehat Y_R^2\) vanishes on the caged eigenstate. A
microcanonical value bounded away from zero as the system grows then
establishes a persistent local ETH mismatch.
\end{proposition}

This route is complementary to the kinetic one.  The Hermitian kinetic operator
\(\widehat Z_R\) probes a process suppressed by destructive interference;
\(\widehat Y_R\) probes the additional fact that the caged eigenstate selects one local
potential value.  The latter does not follow from
\(B_{\mathcal S}\boldsymbol x=0\), and a caged eigenstate may have either or both
constraints.  In a constrained Hilbert space, the sharp local value must hold
in every occupied conditional block; Appendix~\ref{ssec: appendix_projected_potential_reduction} gives the precise statement.

A bounded local reduction of the internal kinetic eigenvalue equation can
likewise yield a \emph{shifted kinetic caging operator}, provided the same
shifted rule is dark in every occupied outside environment.
Appendix~\ref{ssec: appendix_shifted_kinetic_construction} gives this extension,
including the local projector shift required for a selected internal row.

In either route, exact local darkness is only the first half of the ETH test.
The same operator must remain active in the surrounding thermal ensemble.  We
now formulate that common comparison.

\subsection{Thermal test: zero on the caged eigenstate, nonzero in the ensemble}
\label{ssec: thermodynamic_eth_criterion}

A local caging operator turns the interference mechanism into a physical
measurement.  The caged eigenstate is exactly dark, whereas a thermal state at the
same energy density will generally activate the same local process.  The
positive operator \(\widehat Q_R=\widehat L_R^\dagger\widehat L_R\) measures
this local process without relying on a sign convention.

Because multiplying \(\widehat L_R\) by a nonzero scalar does not change its
dark state, we first remove any common local energy scale and use one
size-independent normalization.  Relative coupling ratios and phases remain
part of the operator.  With the detailed convention deferred to
Appendix~\ref{ssec: appendix_caging_operator_normalization}, we take
\begin{equation}
    \widehat L_R\ket{\psi_L}=0,
    \qquad
    \widehat Q_R=\widehat L_R^\dagger\widehat L_R,
    \qquad
    0\le\widehat Q_R\le\widehat{\mathbf1},
    \label{eq: thermo_local_annihilation}
\end{equation}
so that \(\langle\widehat Q_R\rangle_{\psi_L}=0\) exactly at every size.

\subsubsection{The local microcanonical criterion}
\label{sssec: thermal_reference_protocol}

The relevant comparison is with the same Hamiltonian at the same energy
density.  At each size we first resolve the exact global symmetries and
superselection labels that define the thermodynamic sector.  If the microscopic
local dynamics independently selects a finer invariant fragment, the reference
may be restricted to the fragment containing the caged eigenstate; arbitrary projectors
chosen from spectral degeneracies are not additional sectors.  For an ICQMBS
interpretation, this resolved reference space must still contain an extensive
many-body bath.

For a caged eigenstate with energy \(E_L\) on a physical lattice \(\Omega_L\), take the raw
energy window
\(\mathcal W_L=\{E:|E-E_L|\le\delta E_L\}\) inside the resolved reference
space, where \(\delta E_L\) is the window half-width.  With projector
\(\widehat P_{\mathcal W_L}\), the microcanonical state is
\begin{equation}
    \rho_L^{\rm mc}
    =\frac{\widehat P_{\mathcal W_L}}{N_{\rm win,L}},
    \qquad
    N_{\rm win,L}=\operatorname{Tr}\widehat P_{\mathcal W_L}.
    \label{eq: thermal_reference_microcanonical_state}
\end{equation}
A thermal bulk sequence should sharpen in energy density while retaining a
positive entropy density,
\begin{equation}
    \frac{\delta E_L}{|\Omega_L|}\longrightarrow0,
    \qquad
    s_{\rm win}\equiv
    \liminf_{L\to\infty}
    \frac{\log N_{\rm win,L}}{|\Omega_L|}>0.
    \label{eq: main_microcanonical_window_scaling}
\end{equation}
The second condition rules out an isolated or merely polynomially large
fragment as the thermal bath of a QMBS.

The microcanonical value of the normalized witness is
\begin{equation}
    \tau_{Q,L}
    =\operatorname{Tr}(\rho_L^{\rm mc}\widehat Q_R).
    \label{eq: undeformed_thermal_activity}
\end{equation}
At one size, \(\tau_{Q,L}>0\) only says that the caged eigenstate differs locally from that
finite-size ensemble.  A thermodynamic local ETH violation requires the same
bounded local rule to remain distinguishable,
\begin{equation}
    \boxed{
    \tau_*\equiv\liminf_{L\to\infty}\tau_{Q,L}>0.}
    \label{eq: thermo_mean_variance_condition}
\end{equation}
Because \(\widehat Q_R\) is exactly zero in \(\ket{\psi_L}\), a positive
\(\tau_*\) is already a persistent bounded-local mismatch.  It says that the
caged sequence violates diagonal ETH; it does \emph{not} yet say that the
surrounding eigenstates are otherwise thermal.

Finite-size cleaning of independently identified joint-dark states can make the
surrounding scatter easier to inspect, but the raw window above is the defining
thermal reference.  Projector conventions, exact-degeneracy handling,
raw--cleaned bounds, and window systematics are collected in
Appendix~\ref{ssec: appendix_energy_matching_and_canonical_role}.

\subsubsection{Witnesses from Hermitian local caging operators}
\label{sssec: hermitian_mean_fluctuation_refinement}

For the Hermitian kinetic and shifted-potential routes,
\(\widehat L_R=\widehat L_R^\dagger\) is respectively \(\widehat Z_R\) or
\(\widehat Y_R\).  The positive witness then has a simple physical
interpretation.  Writing
\begin{align}
    m_{R,L}&=\operatorname{Tr}(\rho_L^{\rm mc}\widehat L_R),\\
    \sigma_{R,L}^2
    &=\operatorname{Tr}(\rho_L^{\rm mc}\widehat L_R^2)-m_{R,L}^2,
\end{align}
we have
\begin{equation}
    \tau_{Q,L}
    =\operatorname{Tr}(\rho_L^{\rm mc}\widehat L_R^2)
    =\sigma_{R,L}^2+m_{R,L}^2.
    \label{eq: mean_variance_activity_identity}
\end{equation}
Thus the thermal ensemble can differ from the dark caged eigenstate in two ways: its mean
local process is displaced from zero, or the process continues to fluctuate
even when its mean vanishes.  For \(\widehat Y_R\), these are simply
\(m_{R,L}=\langle\widehat V_R\rangle_{\rm mc}-v_R\) and
\(\sigma_{R,L}^2=\operatorname{Var}_{\rm mc}(V_R)\).  The same interpretation
applies to \(\widehat Z_R\) as a Hermitian kinetic process.

Here \(\sigma_{R,L}^2\) is an equilibrium measurement fluctuation of a local
operator, not the eigenstate-to-eigenstate spread used below to test whether the
background obeys ETH.  The directed operator \(\widehat A_R\) is generally
non-Hermitian; its canonical positive observable is instead the one-sided
incoming-transition weight \(\widehat A_R^\dagger\widehat A_R\).

\subsubsection{From local ETH violation to ICQMBS}
\label{sssec: dark_channel_to_icqmbs}

A persistent mismatch for one local observable identifies an exceptional caged sequence.
To classify the sequence as a QMBS, we also ask whether typical eigenstates in
the surrounding window are locally thermal.  By diagonal ETH in
Eq.~\eqref{eq: intro_diagonal_eth}, local thermality on a fixed region \(R\)
means that typical eigenstates reproduce the microcanonical expectation of
every bounded Hermitian observable on \(R\) as the system grows.  We use the
weak form of diagonal ETH, which requires a vanishing fraction of outliers
at any fixed tolerance for each observable in the raw window
\cite{mori_thermalization_2018,lydzba_normal_2024}.
It therefore does not assume strong ETH, Wigner--Dyson or other random-matrix level statistics, 
or a dynamical thermalization theorem.  Those stronger notions are logically independent 
diagnostics and are not part of the ICQMBS definition.

Exact degeneracies should not make this test depend on the eigenbasis returned
by a numerical solver.  For each distinct energy \(E\) in \(\mathcal W_L\),
let \(\widehat P_E\) be its spectral projector.  For a bounded Hermitian local
observable \(\widehat O_R\), denote by \(o_{E,j}^{(R)}\) the eigenvalues of the
compressed operator \(\widehat P_E\widehat O_R\widehat P_E\) within that
energy block, and write the corresponding microcanonical center as
\(\langle\widehat O_R\rangle_{{\rm mc},L}
=\operatorname{Tr}(\rho_L^{\rm mc}\widehat O_R)\).  For the particular
choice \(\widehat O_R=\widehat Q_R\), this center is exactly
\(\tau_{Q,L}\).  The fraction of locally atypical directions is
\begin{equation}
    f_{L,\epsilon}[\widehat O_R]
    =\frac{1}{N_{\rm win,L}}
      \sum_{E\in\mathcal W_L}
      \#\left\{j:
      |o_{E,j}^{(R)}-\langle\widehat O_R\rangle_{{\rm mc},L}|>\epsilon
      \right\}.
    \label{eq: main_background_eth_concentration}
\end{equation}
For a nondegenerate spectrum, this reduces to the usual fraction of eigenstates
whose diagonal matrix element lies outside an \(\epsilon\) neighborhood of the
microcanonical value.
Appendix~\ref{ssec: appendix_thermodynamic_witness_details} records the 
corresponding root-mean-square spread for one observable. Maximizing this spread 
over normalized observables on \(R\) gives the raw concentration width
\(w_L^{\rm raw}\) used in the model sections.

\begin{definition}[Interference-caged quantum many-body scar (ICQMBS)]
\label{def: icqmbs}
Consider a thermodynamic sequence of local Hamiltonians \(\widehat H_L\) and a
caged sequence \(\{\ket{\psi_L}\}\) with
\(E_L/|\Omega_L|\to e\), together with raw microcanonical windows satisfying
Eq.~\eqref{eq: main_microcanonical_window_scaling}.  We call it an
\emph{interference-caged quantum many-body scar} (ICQMBS) sequence when:

(i) the target caged subspace specified by the construction, independently
of the thermal diagnostic, is spectrally sparse: if
\(\mathcal C_L\subseteq\operatorname{Ran}\widehat P_{\mathcal W_L}\)
denotes this declared subspace, contains \(\ket{\psi_L}\), and
\(N_{\rm cage,L}=\dim\mathcal C_L\), then
\begin{equation}
    \frac{N_{\rm cage,L}}{N_{\rm win,L}}\longrightarrow0,
    \label{eq: icqmbs_vanishing_cage_fraction}
\end{equation}
and the raw background concentrates locally, so that for every fixed bounded
region \(R\), every bounded Hermitian \(\widehat O_R\), and every
\(\epsilon>0\),
\begin{equation}
    f_{L,\epsilon}[\widehat O_R]\longrightarrow0;
    \label{eq: icqmbs_background_condition}
\end{equation}

(ii) one size-compatible local observable construction, with uniformly
bounded physical support and norm, remains thermodynamically distinguishable
from the caged eigenstate,
\begin{equation}
    \liminf_{L\to\infty}
    \left|
      \langle\psi_L|\widehat O_R|\psi_L\rangle
      -\operatorname{Tr}(\rho_L^{\rm mc}\widehat O_R)
    \right|>0.
    \label{eq: icqmbs_definition}
\end{equation}
\end{definition}

On a fixed finite region, a spanning Hermitian operator basis suffices to
test concentration.  Establishing it on every bounded region requires
analogous larger-region tests or an independent ETH argument.  The positive
witness constructed above supplies the local mismatch in condition (ii)
whenever Eq.~\eqref{eq: thermo_mean_variance_condition} holds.
Appendix~\ref{ssec: appendix_thermodynamic_witness_details} relates the
mismatch to a local-state distance and details the concentration test.

For a spectrally sparse caged sequence in the same raw windows, the
persistent local mismatch in Eq.~\eqref{eq: thermo_mean_variance_condition}
and concentration of the same witness, \(f_{L,\epsilon}[\widehat Q_R]\to0\)
for every \(\epsilon>0\), together establish \emph{scarring with respect to
\(\widehat Q_R\)}.  The witness expectation stays bounded away from zero for
typical eigenstates in the window and vanishes exactly on the caged eigenstate.
Classifying such a sequence as an ICQMBS also requires concentration for all
bounded local observables, as in Eq.~\eqref{eq: icqmbs_background_condition}.
In general, the interference caging mechanism can still occur in spectra whose background fails weak diagonal ETH, 
while retaining a persistent local ETH violation.

The three local constructions now have a common interpretation.  The directed
kinetic operator \(\widehat A_R\), Hermitian kinetic operator \(\widehat Z_R\),
and shifted-potential operator \(\widehat Y_R\) are different microscopic ways
of making the caged eigenstate locally dark; their positive observables are all instances
of \(\widehat Q_R=\widehat L_R^\dagger\widehat L_R\).  Failure of these three
constructions does not exclude an ICQMBS whose local mismatch is established by
another bounded observable.

\subsubsection{Ensemble matching as a computational continuation}
\label{sssec: thermodynamic_implementation}

The microcanonical ensemble defines the ETH test, but it need not be the most
efficient way to evaluate the thermal target at large size.  In a regular
thermodynamic regime, the rest of a large system acts as a bath for a bounded
region, so locally equivalent canonical calculations can be used after that
equivalence has been established for the resolved sequence.

With \(\widehat P_{\rm ref,L}\) the projector onto the same resolved reference
space and $\beta$ the inverse temperature, define
\begin{equation}
    \rho^{\rm can}_{\beta,L}
    =\frac{\widehat P_{\rm ref,L}e^{-\beta\widehat H_L}}
           {\operatorname{Tr}(\widehat P_{\rm ref,L}e^{-\beta\widehat H_L})}.
    \label{eq: main_resolved_canonical_state}
\end{equation}
Choose \(\beta_L\) so that the canonical mean energy matches the energy of the caged eigenstate,
\begin{equation}
    \operatorname{Tr}(\rho^{\rm can}_{\beta_L,L}\widehat H_L)=E_L,
    \label{eq: main_canonical_energy_matching}
\end{equation}
and verify on overlapping sizes that the local target agrees,
\begin{equation}
    \operatorname{Tr}(\rho_L^{\rm mc}\widehat Q_R)
    -\operatorname{Tr}(\rho^{\rm can}_{\beta_L,L}\widehat Q_R)
    \longrightarrow0.
    \label{eq: main_local_ensemble_matching}
\end{equation}
Only then may the canonical calculation be used as a continuation of the
microcanonical target.

At infinite temperature this reduces to the normalized identity on the resolved
space,
\begin{equation}
    \rho^{\rm can}_{0,L}
    =\frac{\widehat P_{\rm ref,L}}
           {\operatorname{Tr}\widehat P_{\rm ref,L}},
    \label{eq: main_beta_zero_resolved_trace}
\end{equation}
whose local expectation values can often be obtained by counting or transfer
methods far beyond exact diagonalization.  The simplification does not remove
the matching requirement: the energy density of the caged eigenstate must approach the resolved
\(\beta=0\) energy density, and the local trace and microcanonical values must
approach the same limit.  The two model applications below deliberately exercise
both regimes rather than tuning both to the spectral center: the spin-1
thermodynamic proof uses a slice with an exact \(\beta=0\) bridge, whereas the
QDM thermal study retains a nonzero uniform potential and therefore requires an
energy-matched finite-temperature bridge.

The model sections therefore have a simple division of labor.  They identify a
bounded local caging operator, test its raw same-Hamiltonian microcanonical
mismatch, and separately test whether the surrounding reference window becomes
locally thermal.  Canonical or \(\beta=0\) traces are computational extensions
only after a local bridge is established.  The physical support of the witness
must remain bounded throughout; in particular, a fixed-width sequence defines
a quasi-one-dimensional rather than a genuine two-dimensional thermodynamic
limit.  Section~\ref{sec: deformation_stability} now asks which parts of this
structure survive a Hamiltonian deformation.

\section{Stability under Hamiltonian deformations}
\label{sec: deformation_stability}

A compact localized state in a flat-band lattice survives because the matrix
entries on its occupied motif and its destructive-interference boundary continue
to satisfy the same cancellation rule.  Matrix elements farther away may change
without acting on that state.\footnote{This motif-based mechanism was
historically called ``local topology'' in early flat-band work
\cite{Sutherland1986}.  Here the phrase refers only to local connectivity, not
to topological order, a bulk invariant, or generic protection against local
perturbations.}  The many-body picture is similar, except that the amplitudes
inside the configuration support may themselves readjust when the Hamiltonian
changes.  Stability therefore asks whether that readjustment can keep every
outgoing boundary amplitude canceled.

The stability of exact QMBS under perturbations has been studied in
several settings \cite{lin2020_stab,surace2021}.  For an ICQMBS, we separate
this question into three layers.  First, can the caged eigenstate itself
continue under the deformation?  Second, can the \emph{same bounded physical
construction} continue to annihilate it, so that the local caging operator
remains meaningful?  Third, does the microcanonical witness value stay
bounded away from zero as the system grows?  Exact continuation of one layer
does not imply the next.

Local-parent and commutant approaches already organize Hamiltonian families
preserving fixed target states \cite{shiraishi_systematic_2017,moudgalya_exhaustive_2023}.
Here the amplitudes may readjust, and a selected boundary-derived operator
must continue with them.  The resulting obstruction is relative to the
declared support and bounded operator class; it is a first-order compatibility
test, not an exhaustive classification of QMBS-preserving Hamiltonians.

\begin{figure*}[!t]
    \centering
    \includegraphics[width=\textwidth]{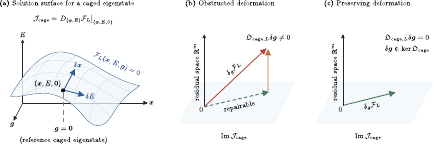}
    \caption{Linearized geometry of continuation of a caged eigenstate.  \textbf{(a)} Near a
    regular reference caged eigenstate, the exact solutions form a local surface in the
    combined state--energy--coupling space.  Varying the state and energy spans
    the repairable residuals, represented by the image of the cage Jacobian
    \(\mathcal J_{\rm cage}\).  \textbf{(b)} An obstructed coupling displacement
    generates a residual with a component outside that image and is detected
    by the cage obstruction map,
    \(\mathfrak O_{{\rm cage},L}\delta\boldsymbol g\neq0\).  \textbf{(c)} For a
    preserving direction, the deformation-induced residual lies entirely in the
    repairable image and can be canceled to first order by a state--energy
    correction.}
    \label{fig:cage_deformation_geometry_schematic}
\end{figure*}

The first layer has a simple linearized geometry.  A coupling change produces a
residual in the internal eigenvalue and boundary-cancellation equations.  By
changing the amplitudes and energy of the caged eigenstate, one can cancel only the part of that
residual lying in the corresponding repairable subspace.  Any remaining
component is a first-order obstruction.  Figure~\ref{fig:cage_deformation_geometry_schematic}
shows this geometry; the equations below make it precise.  Throughout this
section, the microscopic Hilbert space and constraint structure are held fixed.
Changing the constraint itself requires a new configuration-space description,
not a continuation in the coordinates used here.  Appendix~\ref{appendix: deformation_stability_details} collects the component formulas, complex-Hermitian
extension, and constrained-space details.

\subsection{Can the caged eigenstate readjust?}
\label{ssec: deformation_general_criteria}
\label{ssec: exact_cage_continuation}

\subsubsection{Fixed-support continuation}

Consider a smooth family \(\widehat H_L(\boldsymbol g)\) on the same microscopic
Hilbert space and constraint structure, with
\(\widehat H_L(\boldsymbol 0)=\widehat H_L\).  Near the reference point,
\(\boldsymbol g\) denotes independent real coupling coordinates.  We first ask
whether the caged eigenstate can survive while keeping the same configuration support
\(\mathcal S\).  In the block form of Sec.~\ref{sec: product_basis_graph},
\begin{equation}
    \widehat H_L(\boldsymbol g)=
    \begin{pmatrix}
        H_{\mathcal S\mathcal S}(\boldsymbol g)
        & H_{\mathcal S\overline{\mathcal S}}(\boldsymbol g)\\
        H_{\overline{\mathcal S}\mathcal S}(\boldsymbol g)
        & H_{\overline{\mathcal S}\overline{\mathcal S}}(\boldsymbol g)
    \end{pmatrix},
    \qquad
    \ket{\psi_L(\boldsymbol g)}=
    \begin{pmatrix}\boldsymbol x(\boldsymbol g)\\0\end{pmatrix},
    \label{eq: undeformed_cage_block_stability}
\end{equation}
where \(\boldsymbol x\) contains the nonzero amplitudes on \(\mathcal S\).
The boundary map \(B_{\mathcal S}(\boldsymbol g)\) uses one fixed row set that
contains every boundary configuration that can become active in the declared
parameter neighborhood; rows that vanish identically throughout that
neighborhood are omitted.  The exact conditions are then
\begin{equation}
    H_{\mathcal S\mathcal S}(\boldsymbol g)\boldsymbol x(\boldsymbol g)
      =E(\boldsymbol g)\boldsymbol x(\boldsymbol g),
    \qquad
    B_{\mathcal S}(\boldsymbol g)\boldsymbol x(\boldsymbol g)=0.
    \label{eq: cage_equations_stability_entry}
\end{equation}
The first equation determines the motion inside the support; the second keeps
all outgoing leakage canceled.

The outside block
\(H_{\overline{\mathcal S}\overline{\mathcal S}}\) does not act on the caged
vector.  Algebraically it can therefore change without changing the caged
eigenpair, even though it may reorganize most of the surrounding spectrum.  A
bounded real-space perturbation need not be able to vary that block
independently, so physical locality and size compatibility are imposed only
when we formulate deformation stability below.

A smooth solution
\(\boldsymbol g\mapsto[\boldsymbol x(\boldsymbol g),E(\boldsymbol g)]\) of
Eq.~\eqref{eq: cage_equations_stability_entry} on this support will be called a
\emph{continued caged family}, or simply a \emph{continued family}.  Failure in
this fixed-support chart does not rule out a nearby caged eigenstate whose support
reorganizes or enlarges.  If an exact degeneracy leaves the caged eigenstate
basis-dependent, the object to continue is the corresponding projector; the
basis-independent formulation is given in
Appendix~\ref{ssec: appendix_exact_cage_continuation}.

\subsubsection{Repairable leakage and the obstruction}

The boundary equation already contains the essential physics.  To first order,
\begin{equation}
    B_{\mathcal S}\,\delta\boldsymbol x
    =-\delta B_{\mathcal S}\,\boldsymbol x.
    \label{eq: first_order_boundary_continuation}
\end{equation}
A perturbation creates leakage on the right-hand side; redistributing the amplitudes of the caged eigenstate must cancel it.  The same redistribution must simultaneously repair
the internal eigenvalue equation.  We therefore collect both conditions in one
residual map,
\begin{equation}
    \mathcal F_L(\boldsymbol x,E,\boldsymbol g)
    =\begin{pmatrix}
       [H_{\mathcal S\mathcal S}(\boldsymbol g)-E]\boldsymbol x\\
       B_{\mathcal S}(\boldsymbol g)\boldsymbol x
     \end{pmatrix}=0.
    \label{eq: full_cage_equation_map}
\end{equation}
For the main-text tangent analysis we use real coordinates; the general complex
Hermitian formulation is given in
Appendix~\ref{ssec: appendix_exact_cage_continuation}.

At the reference caged eigenstate, changing only the support amplitudes and energy spans the
residuals that the state itself can repair.  After fixing the normalization of
\(\boldsymbol x\), the corresponding linear map is the cage Jacobian
\begin{equation}
    \boxed{
    \mathcal J_{\rm cage}
    \equiv
    \left.
    D_{(\boldsymbol x,E)}\mathcal F_L
    \right|_{(\boldsymbol x,E,\boldsymbol 0)}
    =\begin{pmatrix}
       H_{\mathcal S\mathcal S}-E & -\boldsymbol x\\
       B_{\mathcal S} & 0
     \end{pmatrix}.}
    \label{eq: full_cage_jacobian}
\end{equation}
Allowed variations satisfy
\(\boldsymbol x^{\mathsf T}\delta\boldsymbol x=0\).  The displayed matrix acts
on ambient amplitude coordinates, but the radial normalization direction is
already in its kernel at a solution, so this restriction does not change its
image.

A coupling displacement \(\delta\boldsymbol g\) produces the residual
\begin{equation*}
    \delta_{\boldsymbol g}\mathcal F_L
    \equiv
    \left.D_{\boldsymbol g}\mathcal F_L\right|_{(\boldsymbol x,E,\boldsymbol0)}
    \delta\boldsymbol g.
\end{equation*}
If this residual lies in \(\operatorname{Im}\mathcal J_{\rm cage}\), the state
and energy can cancel it to first order.  What matters is therefore only the
component outside that repairable image.  With \(P_\perp\) denoting a projector
onto a complementary residual subspace (the orthogonal complement for the
standard inner product), define
\begin{equation}
    \begin{aligned}
    \mathfrak O_{{\rm cage},L}
      &\equiv P_\perp
      \left.D_{\boldsymbol g}\mathcal F_L\right|_{(\boldsymbol x,E,\boldsymbol0)},\\
    \mathfrak O_{{\rm cage},L}\delta\boldsymbol g
      &=P_\perp\,\delta_{\boldsymbol g}\mathcal F_L,\\
    c_{\rm cage}(L)
      &=\operatorname{rank}\mathfrak O_{{\rm cage},L}.
    \end{aligned}
    \label{eq: full_cage_obstruction}
\end{equation}
The map \(\mathfrak O_{{\rm cage},L}\) is the first-order cage obstruction: it
keeps precisely the deformation-induced residual that cannot be repaired by
readjusting the caged eigenstate.

\begin{proposition}[First-order continuation criterion for a caged eigenstate]
\label{prop:first_order_cage_continuation}
At a normalized reference caged eigenstate on fixed support \(\mathcal S\), a declared
direction \(\delta\boldsymbol g\) admits corrections
\((\delta\boldsymbol x,\delta E)\), with
\(\boldsymbol x^{\mathsf T}\delta\boldsymbol x=0\), that solve the linearized
caging equations if and only if
\(\mathfrak O_{{\rm cage},L}\delta\boldsymbol g=0\).  Consequently,
\(\ker\mathfrak O_{{\rm cage},L}\) is the preserving tangent space, and
\(c_{\rm cage}(L)=\operatorname{rank}\mathfrak O_{{\rm cage},L}\) counts the
independent first-order tuning conditions within the declared independent
coupling coordinates.
\end{proposition}

This criterion is deliberately only first order and fixed support.  To promote a
tangent solution to a nearby nonlinear family, the normalized full residual
map must also be regular---in particular, its differential must have locally
constant rank---so that its zero set forms a smooth local solution surface.
Even then, the continuation must persist in the desired coupling directions;
a tangent null direction is not by itself a straight finite deformation or
robustness on an open coupling neighborhood.\footnote{Here ``open'' is understood relative to the declared compatible deformation family: the allowed couplings vary over a nonzero neighborhood within that family, with a radius that remains nonzero as $L\to\infty$. It need not be an open neighborhood in the full space of local Hamiltonians.}  Appendix~\ref{ssec: appendix_exact_cage_continuation}
gives the tangent-lift equation, the nonlinear regularity details, the complex
extension, and the metric-independent interpretation of the exact
preserving kernel and obstruction rank.

\subsection{Does the bounded local dark rule survive?}
\label{ssec: branchwise_local_reduction}
\label{ssec: local_q_stability}

A caged eigenstate may continue even after the local diagnostic constructed in
Sec.~\ref{sec: interference_caging} has disappeared.  The second layer therefore
tracks the same bounded physical construction together with the state.  We keep
the same region \(R\) and the same route: a selected boundary row for the
single-row kinetic construction, the same local potential region for the
diagonal construction, or the same finite multi-row operator chart.

For the first two routes, the deformed microscopic Hamiltonian fixes the local
operator.  A multi-row construction may also contain finitely many local
combination coefficients that readjust smoothly.  These are coordinates of the
local operator, not additional Hamiltonian couplings, and they must be
continued jointly rather than refitted independently at each size.  We denote
the resulting operator by \(\widehat L_R(\boldsymbol g)\).

On the fixed-support family,
\(\ket{\delta\psi_L}=(\delta\boldsymbol x,0)^{\mathsf T}\).  Differentiating the
darkness condition
\(\widehat L_R(\boldsymbol g)\ket{\psi_L(\boldsymbol g)}=0\) gives
\begin{equation}
    \boxed{
    \widehat L_R\ket{\delta\psi_L}
    +\delta\widehat L_R\ket{\psi_L}=0.}
    \label{eq: local_channel_tangent_equation}
\end{equation}
The first term is the response of the state to the deformation; the second is
the response of the local operator itself.  Both must cancel for the same
coupling displacement.  In particular, the operator cannot be reconstructed
post hoc after the deformed state is known.  For the kinetic route, the selected
boundary process changes with the Hamiltonian; for the diagonal route, the local
potential and the potential value selected by the caged eigenstate change together.  The
multi-row and constrained-space equations are collected in
Appendix~\ref{ssec: appendix_joint_local_channel_continuation}.

This continuation follows one bounded region and one declared local rule.  Its
failure does not exclude a reorganized caged eigenstate with another local construction.
Conversely, a collective cancellation that grows with system size does not
supply deformation-stable locality: along the thermodynamic sequence, at least
one size-compatible \(\widehat L_R(\boldsymbol g)\) must retain uniformly
bounded support and norm.

If that local operator remains dark, so does the positive witness
\(\widehat Q_R(\boldsymbol g)=\widehat L_R^\dagger(\boldsymbol g)
\widehat L_R(\boldsymbol g)\).  The final question is then entirely thermal:
does the surrounding spectrum continue to activate the same local process?

\subsection{Does the local ETH mismatch survive?}
\label{ssec: thermal_activity_margin}

Exact continuation of the caged eigenstate and of its local dark rule still does not control
the surrounding spectrum.  The outside-support block can change substantially
without acting on the caged eigenstate, so the microcanonical expectation of
\(\widehat Q_R\) may move independently.  Deformation-stable local ETH violation
therefore requires one common coupling neighborhood in which the caged eigenstate, the
bounded local operator, and the local ETH mismatch all persist.

Let \(\mathcal U\) be such a size-independent neighborhood within a declared
family of Hermitian local (or sufficiently rapidly decaying) Hamiltonians, and
let \(\mathcal U_L\) denote its realization at size \(L\).  The deformation
itself must come from one size-compatible bounded local rule; a finite-size
obstruction calculation alone does not impose this physical requirement.  At
each \(\boldsymbol g\), reconstruct the energy-matched microcanonical reference
of the same deformed Hamiltonian using the protocol of
Sec.~\ref{sssec: thermal_reference_protocol}.  If a deformation changes the
exact symmetry content, the reference projector changes with it; an open-family
claim is therefore made on a region of fixed symmetry content, with
symmetry-enhanced endpoints treated separately.  Any additional dynamical
fragment must remain independently selected by the microscopic local dynamics
and thermodynamically large.

For the energy of the continued caged eigenstate \(E_L(\boldsymbol g)\), define
\begin{equation}
    \begin{aligned}
    \tau_{Q,L}(\boldsymbol g)
      &=\operatorname{Tr}\!\left[
        \rho_L^{\rm mc}(E_L(\boldsymbol g),\boldsymbol g)
        \widehat Q_R(\boldsymbol g)
        \right],\\
    \tau_*^{\mathcal U}
      &=\liminf_{L\to\infty}
        \inf_{\boldsymbol g\in\mathcal U_L}
        \tau_{Q,L}(\boldsymbol g)>0.
    \end{aligned}
    \label{eq: thermodynamic_icqmbs_stability_conditions}
\end{equation}
Thus \(\tau_*^{\mathcal U}>0\) means that the same locally dark process remains
uniformly active in the surrounding microcanonical ensemble throughout the
neighborhood.  By itself, this proves deformation-stable local diagonal-ETH
violation, not yet a deformation-stable ICQMBS.

For the ICQMBS upgrade, the raw reference windows must also satisfy the
thermal-background conditions of Definition~\ref{def: icqmbs} uniformly
throughout the same coupling neighborhood: the energy-density width must
vanish, the window must retain positive entropy density, the declared caged
fraction must vanish, and every bounded local observable must concentrate.
Appendix~\ref{ssec: appendix_uniform_thermal_background} records these four
uniform limits explicitly.  A finite coupling grid at a few sizes is evidence,
not a substitute for them.

\begin{proposition}[Three-layer criterion for deformation-stable ICQMBS]
\label{prop:deformation_stable_icqmbs}
Under the locality assumptions above, suppose that, for every sufficiently
large \(L\) and every \(\boldsymbol g\in\mathcal U_L\),
\begin{enumerate}[label=(\roman*),leftmargin=*]
    \item the declared size-compatible Hamiltonian family admits a continued
    caged eigenpair
    \(\bigl(\ket{\psi_L(\boldsymbol g)},E_L(\boldsymbol g)\bigr)\);
    \item one size-compatible local caging operator
    \(\widehat L_R(\boldsymbol g)\), with uniformly bounded physical support and
    norm, obeys
    \(\widehat L_R(\boldsymbol g)\ket{\psi_L(\boldsymbol g)}=0\); and
    \item its witness
    \(\widehat Q_R(\boldsymbol g)=\widehat L_R^\dagger(\boldsymbol g)
    \widehat L_R(\boldsymbol g)\) satisfies \(\tau_*^{\mathcal U}>0\) in
    Eq.~\eqref{eq: thermodynamic_icqmbs_stability_conditions}.
\end{enumerate}
Then the continued family exhibits a deformation-stable local diagonal-ETH
violation throughout \(\mathcal U\).  If the uniform background conditions in
Appendix~\ref{ssec: appendix_uniform_thermal_background} also hold in the
reference spaces selected by the thermal protocol throughout that neighborhood,
the continued family is a deformation-stable ICQMBS.
\end{proposition}

The order of the infimum and thermodynamic limit in
Eq.~\eqref{eq: thermodynamic_icqmbs_stability_conditions} is essential: one
microscopic rule and one size-independent neighborhood must work for all
sufficiently large systems.  At fixed \(L\), \(c_{\rm cage}(L)=0\) means only
that every direction in the declared deformation chart is first-order
unobstructed.  When \(c_{\rm cage}(L)>0\), generic continuation within that
chart requires \(c_{\rm cage}(L)\) independent tuning conditions, although
preserving directions may remain in
\(\ker\mathfrak O_{{\rm cage},L}\).

Within a declared physical deformation family, we reserve \emph{robust} for a
size-compatible continued family that persists on an open neighborhood whose
radius remains nonzero as \(L\to\infty\).  We call it \emph{fragile} when
persistence requires tuning or when no such size-independent neighborhood
survives.  Higher-order obstructions or a shrinking continuation radius can
therefore invalidate a robustness claim even when several finite-size tangent
calculations are unobstructed.

\FloatBarrier

\section{Graph-guided chiral-shell search and finite-size certification}
\label{sec: practical_search}

The preceding sections assume that a candidate caged eigenstate is already known.
Finding such a state is a separate problem.  In a single-particle flat-band
system, the lattice often suggests a compact motif by inspection; in a
many-body configuration graph, searching over arbitrary subsets of product
configurations is instead combinatorial.  We therefore use two simple graph
labels to narrow the search before imposing the exact support--boundary
condition: a chiral grading of the kinetic configuration graph and a common
value of the diagonal potential.

\begin{figure}[!t]
    \centering
    \includegraphics[width=\columnwidth]{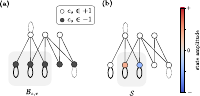}
    \caption{Graph-guided chiral-shell search.
    \textbf{(a)} A chiral grading value \(c\) and common potential value \(v\)
    select the candidate set \(\mathcal B_{c,v}\); its outgoing kinetic
    couplings define the boundary map \(B_{c,v}\).
    \textbf{(b)} A boundary-kernel vector fixes the amplitudes that cancel the
    outgoing rows.  Only its nonzero entries form the reconstructed support
    \(\mathcal S\subseteq\mathcal B_{c,v}\).  The displayed colors indicate
    relative signs for this real-valued schematic.}
    \label{fig:chiral_shell_search_schematic}
\end{figure}

The physical idea is shown in Fig.~\ref{fig:chiral_shell_search_schematic}: 
we assume the configuration graph of \(\widehat K\) is bipartite to demonstrate the idea. However, the full Hamiltonian does not need to respect chiral symmetry or to be a bipartite graph.
The chiral grading selects one vertex class of the bipartite kinetic graph, so
no kinetic edge remains inside the candidate set.  The potential label then fixes the
same diagonal energy on every selected configuration.  The internal
Schr\"odinger equation is therefore automatic, and the search reduces to one
question: can the outgoing amplitudes cancel on the opposite-grading
boundary?  This is a candidate-generation device, not a definition or a
classification of interference caging.

\subsection{Chiral-shell reduction}
\label{ssec: type1_uniform_shell_search}

Suppose the kinetic configuration graph admits a two-coloring that flips under
every allowed kinetic move.  For the search used here, this graph statement is
equivalent to a product-basis diagonal grading: let \(\widehat C\) satisfy
\begin{equation}
\begin{aligned}
    \widehat C^2&=\widehat{\mathbf1},
    &\{\widehat K,\widehat C\}&=0,\\
    \widehat C|s\rangle&=c_s|s\rangle,
    &c_s&\in\{+1,-1\}.
\end{aligned}
\label{eq: acomm_bipartite}
\end{equation}
Every nonzero kinetic edge then joins configurations of opposite grading, so
the kinetic configuration graph is bipartite.  Conversely, a bipartition of
that graph defines such a diagonal \(\widehat C\) by assigning eigenvalues
\(\pm1\) to its two vertex classes.  This equivalence concerns the
configuration graph of \(\widehat K\), not the physical lattice.
In a basis ordered by the grading sign,
\begin{equation}
    K=
    \begin{pmatrix}
        0 & K_{+-}\\
        K_{+-}^{\dagger} & 0
    \end{pmatrix}.
    \label{eq: main_chiral_kinetic_block_form}
\end{equation}
A state restricted to one grading sector therefore has no internal kinetic
motion.  The individual outgoing transitions still exist, and caging requires
their amplitudes to cancel on the opposite-grading boundary.

We next use the diagonal potential as a second vertex label.  For a chosen
value \(v\), combine one grading sign with one uniform-potential shell,
\begin{equation}
\begin{aligned}
    \mathcal B_{c,v}
    &\equiv
    \bigl\{|s\rangle\in\mathcal B:
    c_s=c,\ \langle s|\widehat V(\boldsymbol\lambda)|s\rangle=v\bigr\},\\
    \widehat P_{c,v}
    &\equiv
    \sum_{|s\rangle\in\mathcal B_{c,v}}|s\rangle\langle s|.
\end{aligned}
\label{eq: type1_joint_shell}
\end{equation}
No commutation between \(\widehat K\) and \(\widehat V\) is assumed: kinetic
edges may connect different potential shells.\footnote{Accordingly, no spectral reflection symmetry of the Hamiltonian is assumed \cite{schecter_many-body_2018}. The full configuration graph is still nonbipartite due to the presence of self-loops.}  
What matters is only that
\(\widehat V\) acts as a scalar on the selected one. The selected uniform shell may be structural or tuned in the sense of
Sec.~\ref{ssec: potential_shell_constraints}.  That distinction matters for
deformation stability, not for the fixed-parameter kernel search; shell mergers
and related special cases are collected in
Appendix~\ref{ssec: appendix_uniform_shell_structure}.  
Consequently,
\begin{equation}
    \widehat P_{c,v}\widehat K\widehat P_{c,v}=0,
    \qquad
    \widehat V(\boldsymbol\lambda)\widehat P_{c,v}
    =v\widehat P_{c,v}.
    \label{eq: type1_internal_block_zero}
\end{equation}
The internal energy is fixed at \(E=v\), so only leakage from the candidate set
remains.

Let \(\partial\mathcal B_{c,v}\) be the product configurations outside the
candidate set that receive kinetic amplitude from it.  Their outgoing
couplings form the rectangular boundary map
\begin{equation}
    B_{c,v}
    \equiv
    \widehat P_{\partial\mathcal B_{c,v}}
    \widehat K\widehat P_{c,v}.
    \label{eq: type1_boundary_matrix}
\end{equation}
The search is then simply
\begin{equation}
\boxed{
    B_{c,v}\boldsymbol x=0,
    \qquad
    E(\boldsymbol\lambda)=v.}
    \label{eq: type1_boundary_nullspace}
\end{equation}
A nonzero kernel vector supplies amplitudes whose outgoing contributions cancel
row by row.  Its \emph{actual support} \(\mathcal S\) is the subset of
\(\mathcal B_{c,v}\) on which those amplitudes are nonzero, and is reconstructed
before certification.

At fixed parameters, a complete kernel solve is exhaustive for this declared
class.\footnote{This is a physical search reduction; no generic runtime or memory advantage over exact diagonalization is claimed.}
Any exact caged eigenstate with
\(\mathcal S\subseteq\mathcal B_{c,v}\) can be extended by zero over the rest of
the candidate set and lies in \(\ker B_{c,v}\); no omitted internal kinetic
equation is generated because one grading sector contains no kinetic edges.
This is only \emph{relative} completeness: the search says nothing about caged eigenstates
whose support spans both grading sectors or more than one potential value.
Dropping the chiral restriction gives the more general uniform-shell
eigenspace search of Appendix~\ref{ssec: general_uniform_shell_search}, where
the internal kinetic eigenspace must also be solved explicitly.  Accordingly,
the chiral restriction is a search simplification rather than a condition for
interference caging; Appendix~\ref{ssec: spin1_xy_five_site_nonchiral_cage}
gives an explicit caged eigenstate on a nonbipartite resolved kinetic graph.

\subsection{What the finite-size solution certifies}

The kernel solve generates candidates; certification is a separate step.  For
each candidate we reconstruct its actual support \(\mathcal S\), re-evaluate
the internal eigenvalue and outgoing-boundary equations on that support,
verify \(B_{\mathcal S}\neq0\), and finally embed the state in the full resolved
sector to check the full-Hamiltonian residual.  This separates candidate generation
from verification in the full resolved sector.

For an analytic or exact-arithmetic vector, exact vanishing of these equations
establishes the finite-size statement of interference caging.  For a numerical
nullspace, the certificate is conditional on the stated rank tolerance and
precision; Appendix~\ref{ssec: finite_size_search_certification} gives the
corresponding singular-value and residual reporting conventions.

These checks establish only what they directly test.  A complete
\(B_{c,v}\)-kernel solve is exhaustive only within the declared chiral shell,
and a finite-size residual certificate does not by itself supply a bounded
local caging operator, a thermodynamic local-ETH distinction, or deformation stability.
The implementation details, degenerate-subspace handling, and optional graph
reductions are collected in Appendix~\ref{appendix: supplementary_search_methods}; every reduced candidate is ultimately lifted
back to the full product basis for certification.

The search therefore returns a finite-size caged eigenstate together with an
explicit support--boundary certificate.  It does not by itself establish a
bounded real-space caging rule, a thermodynamic local-ETH distinction, or
deformation stability.  The two model sections next identify the repeated local
mechanism behind the selected solutions and test those additional requirements
independently.

\FloatBarrier

\section{Spin-1 \texorpdfstring{$XY$}{XY} chain: local caging and compatible deformation}
\label{sec: spin1_xy}

In the spin-1 $XY$ chain, the framework recovers the known staggered
$\pi$-bimagnon QMBS tower, which also admits quasiparticle and
projector-embedding descriptions \cite{schecter_weak_2019}.  The
support--boundary construction exposes the repeated local interference behind
the tower's exactness.  We use this mechanism to identify Hamiltonian
deformations that preserve the tower and its bounded local caging operators.
We then prove
that one fluctuation-scale raw-window sequence realizes the thermodynamic
ICQMBS criteria and use narrower windows as complementary finite-size diagnostics of
the surrounding many-body spectrum.

\subsection{Model and benchmark QMBS tower}
\label{ssec: spin1_xy_model}

\subsubsection{Hamiltonian and chiral grading}

Consider a chain of $L$ spin-1 degrees of freedom with local $S^z$
eigenstates $\ket{+}$, $\ket{0}$, and $\ket{-}$.  The Hamiltonian is
\begin{equation}
\begin{aligned}
    \widehat H_{\mathrm{XY}}
    &=\widehat K+\widehat V,\\
    \widehat K
    &=\frac{J}{2}\sum_{(r,r+1)}
    \left(
        \widehat S_r^+\widehat S_{r+1}^-
        +\widehat S_r^-\widehat S_{r+1}^+
    \right),\\
    \widehat V
    &=h\sum_{r=1}^{L}\widehat S_r^z
      +D\sum_{r=1}^{L}(\widehat S_r^z)^2,
\end{aligned}
\label{eq: xy_hamiltonian}
\end{equation}
where $J$ is the conventional $XY$ exchange coupling, $h$ the longitudinal field, and $D$ the
single-ion anisotropy.  The bond sum runs over nearest neighbors for open
boundary conditions (OBC), and includes $(L,1)$ for periodic boundary
conditions (PBC).  The conserved total magnetization
\begin{equation}
    \widehat M=\sum_{r=1}^{L}\widehat S_r^z
\end{equation}
allows us to work in a fixed-$M$ product basis, where $\widehat K$ gives the
configuration-graph edges and $\widehat V$ the diagonal vertex energies.  For
generic $D\neq0$, these two pieces do not commute,
$[\widehat K,\widehat V]\neq0$; at $D=0$ the remaining field term is the
sectorwise constant $hM\,\widehat{\mathbf1}$.

For OBC, and for even $L$ under PBC, the physical chain is bipartite,
$\Omega=\Omega_A\sqcup\Omega_B$.  Choosing one alternating sublattice defines
\begin{equation}
    \widehat C_A
    =\prod_{r\in\Omega_A}\exp(i\pi\widehat S_r^z),
    \label{eq: xy_chiral_operator}
\end{equation}
which satisfies $\widehat C_A^2=\widehat{\mathbf1}$ and
\begin{equation}
    \{\widehat K,\widehat C_A\}=0.
    \label{eq: xy_chiral_anticommutation}
\end{equation}
Thus the kinetic configuration graph is bipartite,\footnote{The chiral 
grading is a property of the kinetic graph; 
it need not originate from a bipartition of the physical lattice. In some
models a bipartition of the physical lattice provides the grading, whereas in
others the required configuration parity has a different microscopic origin.}
and the grading selects a
vertex class for the chiral-shell search below.  This grading is a search aid,
not the origin of the caged eigenstate: exactness is decided by the boundary
cancellation.

For odd $L$ under PBC, the physical chain is nonbipartite, so the alternating
real-space construction of $\widehat C_A$ is unavailable.  Independently, the
kinetic configuration graph is also nonbipartite, as shown explicitly in
Appendix~\ref{ssec: spin1_xy_five_site_nonchiral_cage}.
The chiral-shell search therefore ceases to apply, although interference caging itself need 
not disappear; Appendix~\ref{ssec: spin1_xy_five_site_nonchiral_cage} gives an explicit 
odd-periodic caged eigenstate.

\begin{figure}[!t]
    \centering
    \includegraphics[width=\columnwidth]{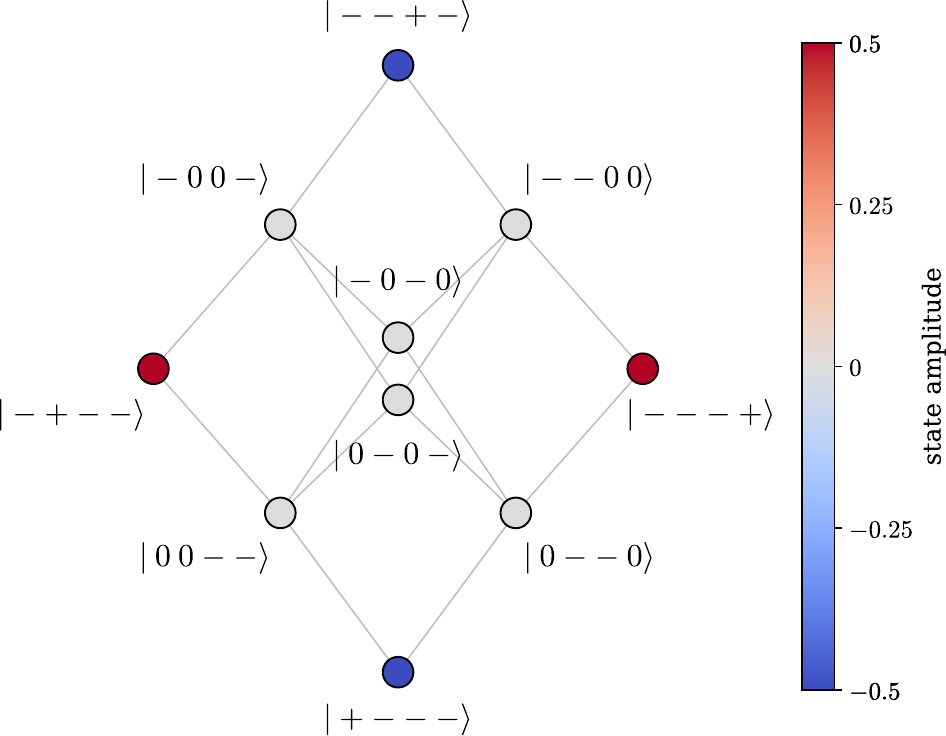}
    \caption{Configuration graph of a representative $L=4$ spin-1 $XY$
    tower member under PBC in the sector of total magnetization $M=-2$.  Vertices are
    product states and edges are local exchange processes; warm and cool colors
    denote opposite real amplitudes.  Each gray boundary vertex is reached from
    two support configurations with opposite amplitudes, displaying the local
    destructive interference that underlies the tower.  Self-loops are omitted.}
    \label{fig:spin1_xy_hamiltonian_graph_L4}
\end{figure}

\subsubsection{Known staggered QMBS tower}

We next recall the exact staggered $\pi$-bimagnon QMBS tower introduced in
Ref.~\cite{schecter_weak_2019}.  For OBC, and for even $L$ under PBC, start from
the fully polarized state
\begin{equation}
    \ket{\mathrm{F}_{-}}=\bigotimes_{r=1}^{L}\ket{-}_r.
    \label{eq: xy_fully_polarized_reference}
\end{equation}
The $n$th tower member is
\begin{equation}
    \ket{\Psi_n}
    =\mathcal N_n(\widehat J^+)^n\ket{\mathrm{F}_{-}},
    \qquad
    \widehat J^+
    =\frac12\sum_{r=1}^{L}(-1)^r(\widehat S_r^+)^2,
    \label{eq: xy_scar_a}
\end{equation}
with $n=0,\ldots,L$, where $\mathcal N_n$ normalizes the state and
$\widehat J^+$ is the staggered $\pi$-bimagnon raising operator.  Equivalently,
\begin{equation}
    \ket{\Psi_n}
    =\binom{L}{n}^{-1/2}
      \sum_{\substack{\Xi\subseteq\Omega\\|\Xi|=n}}
      (-1)^{\sum_{r\in\Xi}r}\ket{\boldsymbol s_\Xi},
    \label{eq: xy_scar_product_expansion}
\end{equation}
where $\Xi$ is the set of sites carrying $\ket{+}$ and
$\ket{\boldsymbol s_\Xi}$ has $\ket{-}$ elsewhere.  The algebraic form connects
to the known tower, while the product-basis form exposes the alternating
amplitudes responsible for the local interference below.  Figure~\ref{fig:spin1_xy_hamiltonian_graph_L4}
previews that pattern: each active boundary configuration is reached from two
support configurations with opposite amplitudes.

  Every support
configuration contains no $\ket0$ state and therefore has
\begin{equation}
    M_n=-L+2n,
    \qquad
    E_n=h(-L+2n)+DL.
    \label{eq: xy_uniform_shell_energy}
\end{equation}
Thus the potential is uniform on the support of every tower member for every
uniform $D$.  Moreover, within a fixed-$M$ sector the field term is simply
$hM\,\widehat{\mathbf1}$, so changing $h$ only shifts the resolved spectrum.
Neither uniform $h$ nor uniform $D$ is required for the exact caging mechanism;
the special choice used later concerns the convenience of the thermal ensemble
bridge.

\subsection{Local origin of the exact tower}

\subsubsection{Two-parent interference from the chiral-shell search}
\label{ssec: xy_sn_state}

We first ask whether the benchmark tower can be recovered from the search
construction rather than inserted from its closed form.  Taking $D\neq0$
temporarily isolates the uniform-potential shell with no local $\ket0$ state,
\begin{equation}
    \mathcal B_n^{\mathrm{XY}}
    =\left\{\ket{\boldsymbol s_\Xi}:\ \Xi\subseteq\Omega,\ |\Xi|=n\right\}.
    \label{eq: xy_type1_shell}
\end{equation}
This use of $D$ selects the candidate set; it does not create the eigenstate,
which remains exact after setting $D=0$.

Inside this shell there is only one way to leak out locally: a neighboring
$\ket{+-}$ or $\ket{-+}$ pair is converted into the same $\ket{00}$ state,
\begin{equation}
\begin{aligned}
    \widehat k_{r,r+1}
      &=\frac{J}{2}\left(
        \widehat S_r^+\widehat S_{r+1}^-
        +\widehat S_r^-\widehat S_{r+1}^+
        \right),\\
    \widehat k_{r,r+1}\ket{+-}
      &=\widehat k_{r,r+1}\ket{-+}=J\ket{00},
\end{aligned}
    \label{eq: xy_local_kinetic_action}
\end{equation}
while $\ket{++}$ and $\ket{--}$ are dark separately.  Each active boundary
configuration therefore has exactly two parents with equal matrix elements.
Writing a candidate vector in this shell as
$\sum_{\sigma\in\mathcal B_n^{\mathrm{XY}}}x_n(\sigma)\ket{\sigma}$,
where $x_n(\sigma)$ is the amplitude of configuration $\sigma$, destructive
interference requires
\begin{equation}
    x_n(\ldots+-\ldots)=-x_n(\ldots-+\ldots).
    \label{eq: xy_adjacent_swap_relation}
\end{equation}
Here the omitted entries are identical on the two sides: only the neighboring
$\ket{+-}$ and $\ket{-+}$ pair is exchanged.  This local rule is precisely the
boundary-kernel equation
\begin{equation}
    B_n^{\mathrm{XY}}\boldsymbol x_n=0,
    \qquad
    B_n^{\mathrm{XY}}
    =\widehat P_{\partial\mathcal B_n^{\mathrm{XY}}}\widehat K
      \widehat P_{\mathcal B_n^{\mathrm{XY}}}.
    \label{eq: xy_type1_boundary_kernel}
\end{equation}
Adjacent swaps connect all binary configurations at fixed $n$, so for
$0<n<L$ the kernel is one-dimensional for OBC and even-$L$ PBC and is generated
by
\begin{equation}
    x_n(\Xi)=(-1)^{\sum_{r\in\Xi}r}.
    \label{eq: xy_unique_kernel_vector}
\end{equation}
The polarized endpoints are trivially one-dimensional.  Appendix~\ref{ssec:
xy_boundary_kernel_proof} gives the connectivity and uniqueness proof.

The same cancellation occurs independently on every physical bond
$R_r=\{r,r+1\}$.  After fixing the outside environment, the only active
conditional component is proportional to $\ket{+-}-\ket{-+}$, while
$\ket{++}$ and $\ket{--}$ are annihilated separately.  Hence
\begin{equation}
    \widehat k_{r,r+1}\ket{\Psi_n}=0
    \label{eq: xy_frustration_free}
\end{equation}
for every bond.  The known global tower is therefore reconstructed from a
repeated bounded two-parent cancellation.

\subsubsection{From cancellation to bounded local caging operators}
\label{ssec: xy_three_local_witnesses}

Because the cancellation is local, the tower admits bounded operators that
encode the same darkness.  On $R_r$ define the directed incoming amplitude and
its Hermitian counterpart,
\begin{equation}
\begin{aligned}
    \widehat A_{R_r}
      &=J\ket{00}\left(\bra{+-}+\bra{-+}\right),\\
    \widehat Z_{R_r}
      &=\widehat A_{R_r}+\widehat A_{R_r}^{\dagger},
\end{aligned}
    \label{eq: xy_directed_q_witness}
\end{equation}
with normalized positive witnesses
\begin{equation}
\begin{aligned}
    \widehat Q^{A}_{R_r}
      &=\frac{\widehat A_{R_r}^{\dagger}\widehat A_{R_r}}{2J^2},\\
    \widehat Q^{Z}_{R_r}
      &=\frac{\widehat Z_{R_r}^{2}}{2J^2}.
\end{aligned}
    \label{eq: xy_reduced_iz}
\end{equation} 
The uniform shell gives an independent diagonal detector,
\begin{equation}
\begin{aligned}
    D\widehat Y_r
      &=D\left[(\widehat S_r^z)^2-\widehat{\mathbf1}_r\right],\\
    \widehat Y_r&=(\widehat S_r^z)^2-\widehat{\mathbf1}_r,\\
    \widehat Q^{Y}_r&=\widehat Y_r^2=\ket0_r\!\bra0_r.
\end{aligned}
    \label{eq: xy_local_potential_darkness}
\end{equation}
Factoring out $D$ separates the coupling from the dimensionless shell operator
$\widehat Y_r$, which remains well defined when the thermal comparison is specialized
to $D=0$.  At that reference point $\widehat A_{R_r}$ and $\widehat Z_{R_r}$ are the primary kinetic
caging operators, whereas $\widehat Y_r$ should be read as the normalized continuation
of the diagonal-shell observable from $D\neq0$.  All three bounded operators annihilate every tower member,
\begin{equation}
    \widehat A_{R_r}\ket{\Psi_n}
    =\widehat Z_{R_r}\ket{\Psi_n}
    =\widehat Y_r\ket{\Psi_n}=0.
    \label{eq: xy_three_channel_darkness}
\end{equation}
The conditional-environment-independent proof and normalization details are given in
Appendix~\ref{ssec: xy_local_operator_blocks_appendix}.

The known QMBS tower has therefore been reduced to repeated bounded local
caging operators.  We can now ask the stronger question of Sec.~\ref{sec:
deformation_stability}: can the caged eigenstate and its local caging operators
be continued together under a genuine Hamiltonian deformation?

\subsection{Exact continuation under Hamiltonian deformation}

\subsubsection{Compatibility of exchange phases}
\label{sssec: xy_sn_break_h3}

Within the phase-pattern tower ansatz below, a generic complex exchange spoils
the two-parent cancellation because the paths into $\ket{00}$ acquire
different phases.  This cancellation is preserved exactly when the exchange
phase compensates the relative phase of the two parents.  Allowing
\begin{equation}
    \widehat J_\eta^+
    =\frac12\sum_r\eta_r(\widehat S_r^+)^2,
    \qquad |\eta_r|=1,
    \label{eq: xy_generalized_tower_phase}
\end{equation}
and a complex Hermitian bond exchange
\begin{equation}
    \widehat k_{rs}(t_{rs})
    =\frac12\left(
      t_{rs}\widehat S_r^+\widehat S_s^-
      +t_{rs}^*\widehat S_r^-\widehat S_s^+
    \right),
    \label{eq: xy_general_exchange}
\end{equation}
the two-parent cancellation is preserved exactly when
\begin{equation}
    t_{rs}^*\eta_r+t_{rs}\eta_s=0.
    \label{eq: xy_exchange_phase_condition}
\end{equation}
For the staggered tower $\eta_r=(-1)^r$ and a translation-invariant range-$d$
exchange,
\begin{equation}
    t_d^*+(-1)^dt_d=0,
    \qquad
    \begin{cases}
      t_d\in\mathbb R, & d\ \text{odd},\\
      t_d\in i\mathbb R, & d\ \text{even}.
    \end{cases}
    \label{eq: xy_complex_exchange_phase_line}
\end{equation}
Thus real odd-range exchange and purely imaginary even-range exchange preserve
the full QMBS tower within this ansatz and preserve interference caging
for its interior members $0<n<L$.  The polarized endpoints $n=0,L$ remain
exact but have empty active boundary.

For concise notation, define
\begin{equation}
    \widehat K_d(t_d)
    =\frac12\sum_r\left[
       t_d\widehat S_r^+\widehat S_{r+d}^-
      +t_d^*\widehat S_r^-\widehat S_{r+d}^+
    \right].
    \label{eq: xy_range_d_exchange}
\end{equation}
A size-compatible family that changes the surrounding spectrum while leaving
the tower exact is
\begin{equation}
\begin{aligned}
    \widehat H_\kappa
      &=\widehat K_1(J)+\widehat K_3(J_3^{\rm ref})
        +\widehat K_2(i\kappa),\\
    \frac{J_3^{\rm ref}}{J}&=0.1,
    \qquad
    \frac{\kappa_\star}{J}=0.1.
\end{aligned}
    \label{eq: xy_reference_family}
\end{equation}
Here $J_3^{\rm ref}/J=0.1$ fixes a representative nonzero point on the exact
real odd-range compatible direction, while $\kappa_\star/J=0.1$ selects the
representative point along the independent imaginary even-range direction.
Neither value is optimized or fine-tuned.  The omission of $h$ and $D$ from
Eq.~\eqref{eq: xy_reference_family} is likewise not a caging requirement.  In
the fixed-$M$ sector, $h$ is an overall energy shift, while a uniform $D$ leaves
the tower and the bounded local annihilators exact by
Eq.~\eqref{eq: xy_uniform_shell_energy}.  We choose the $D=0$ slice below
because it restores an exact spectral-center bridge and makes the thermodynamic
thermal target analytically controllable; Appendix~\ref{ssec:
xy_generic_microcanonical_appendix} gives a matched finite-temperature control
at nonzero $D$.
The scan in Appendix~\ref{ssec: xy_joint_deformation_appendix} shows positive
finite-size microcanonical witness values over $0\le J_3/J\le0.2$.
For $J\kappa\neq0$, nearest- and next-nearest-neighbor bonds form triangular
loops carrying a gauge-invariant phase $\pm\pi/2$, so the imaginary even-range
term cannot be removed by local $S^z$ rephasing.  Equation~\eqref{eq:
xy_exchange_phase_condition} classifies exact continuation within the tower
ansatz; Appendix~\ref{ssec: xy_joint_deformation_appendix} compares it with the
broader first-order repair problem.

After removing the inessential fixed-$M$ field shift and choosing $D=0$, the
same family retains an antiunitary spectral reflection
\cite{schecter_many-body_2018} even though nonzero $\kappa$ breaks ordinary
inversion and the unitary $\widehat C_A$ anticommutation:
\begin{equation}
    \widehat\Theta=\widehat C_A\mathcal K,
    \qquad
    \widehat\Theta\widehat H_\kappa\widehat\Theta^{-1}
    =-\widehat H_\kappa,
    \label{eq: xy_antiunitary_spectral_reflection}
\end{equation}
where $\mathcal K$ is complex conjugation in the $S^z$ product basis.  On the
selected $M=-2$ tower sequence with translation momentum $k=0$ or $\pi$, $\widehat\Theta$ acts within
the resolved sector (Appendix~\ref{ssec: xy_thermal_ensembles_appendix}), so
the tower remains pinned to the spectral center,
\begin{equation}
    e_{\Psi,L}(\kappa)
    =e_{\beta=0,(M,k),L}(\kappa)
    =e_{\beta=0,M,L}(\kappa)=0.
    \label{eq: xy_family_beta_zero_energy_match}
\end{equation}

\subsubsection{Continuation of the local caging operators}

The same phase condition that preserves the caged eigenstate also preserves a
bounded local annihilator.  For every compatible bond,
\begin{equation}
\begin{aligned}
    \widehat A_{rs}(t_{rs})
      &=\ket{00}_{rs}
        \left(t_{rs}^{*}\bra{+-}+t_{rs}\bra{-+}\right)_{rs},\\
    \widehat Z_{rs}(t_{rs})
      &=\widehat A_{rs}(t_{rs})+\widehat A_{rs}^{\dagger}(t_{rs}).
\end{aligned}
    \label{eq: xy_deformed_directed_channel}
\end{equation}
Whenever Eq.~\eqref{eq: xy_exchange_phase_condition} holds, these operators
annihilate the tower and retain two-site support.  Along
Eq.~\eqref{eq: xy_reference_family}, the original nearest-neighbor
$\widehat A_{R_r},\widehat Z_{R_r}$ and $\widehat Y_r$ therefore remain fixed and exactly annihilate every
tower member, while the
added $K_2(i\kappa)$ term has corresponding bounded second-neighbor operators.
Their explicit form and the diagonal-shell extension are given in
Appendix~\ref{ssec: xy_joint_deformation_appendix}.

The exact caged eigenstate and its local caging operators thus survive the same
physical deformation.  The remaining question is whether this exact local
annihilation stays exceptional inside the many-body background of the deformed
Hamiltonian.

\subsection{Thermodynamic closure and finite-size local ETH diagnostics}
\label{ssec: xy_cage_thermal_contrast}

The selected tower-member sequence and its bounded local dark rule continue
exactly along the compatible family.  For the thermal part of the ICQMBS
test, a raw microcanonical witness value must stay bounded away from zero,
and the background must concentrate locally.  We first test these conditions
numerically in a narrow window, using the complete magnetization-preserving
two-site algebra for the concentration test.  We then prove the thermodynamic
criterion for every fixed bounded region on a fluctuation-scale window.

\begin{figure*}[!t]
    \centering
    \includegraphics[width=0.98\textwidth]{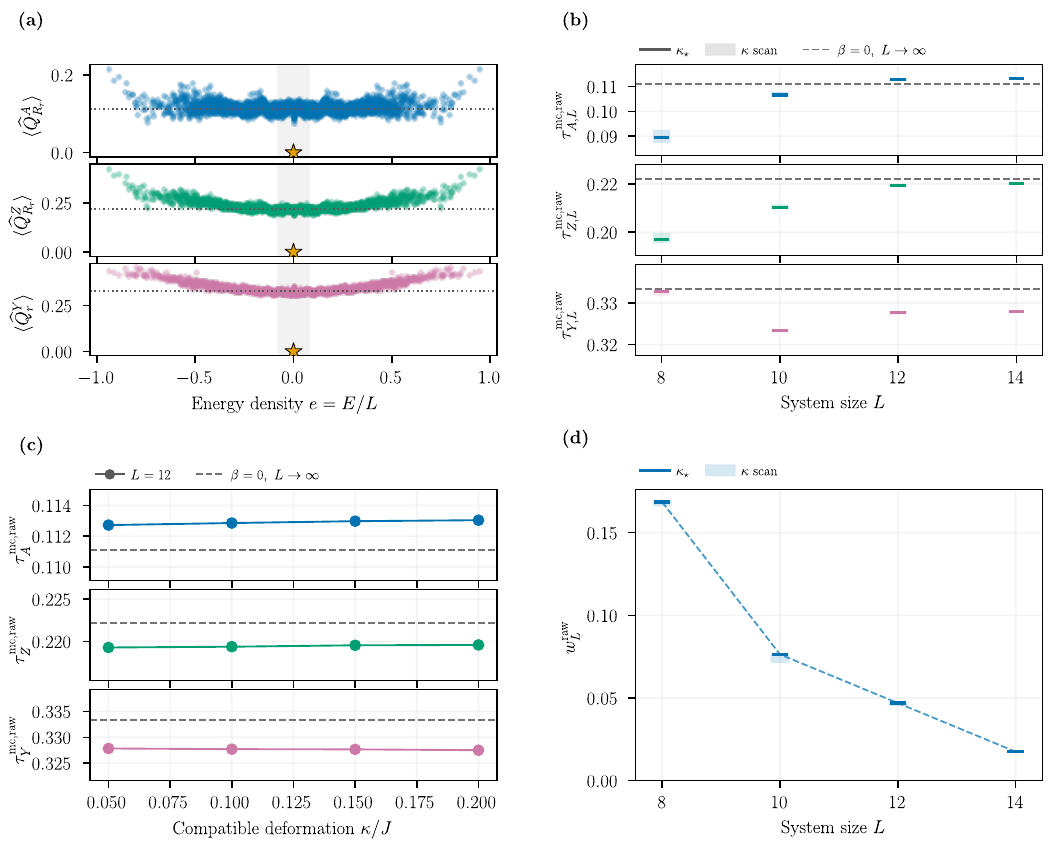}
    \caption{Finite-size diagnostics of the third, thermal layer of the
    compatible spin-1 construction, using the common
    energy half-width $\Delta E=(J/2)L^{1/4}$.  \textbf{(a)} \emph{Local exception:}
    resolved $L=12$ ETH scatter at $\kappa_\star/J=0.1$; the star is the
    exactly dark tower member, the shaded band the comparison window, and the
    horizontal segments the raw microcanonical values.  \textbf{(b)} \emph{Size and sampled
    deformation range:} raw microcanonical values of the witnesses from $A,Z,Y$
    at $\kappa_\star$ for $L=8,10,12,14$.  The pale range glyphs span the sampled positive-$\kappa$
    values where a deformation scan is available ($L\le12$); the dashed lines
    are the thermodynamic fixed-$M$, $\beta=0$ limits
    $(1/9,2/9,1/3)$, rather than exact finite-size traces.  \textbf{(c)}
    \emph{Deformation scan:} the same witness values across the compatible $\kappa$
    grid at $L=12$, with the same thermodynamic guides.  The exact
    sector-resolved $\beta=0$ traces are also $\kappa$-independent;
    the small variation of the raw microcanonical values is a finite-window effect.  \textbf{(d)} \emph{Background
    concentration:} worst-case raw width over the complete
    space of magnetization-preserving Hermitian two-site operators versus $L$; the pale
    range glyphs again span the sampled positive-$\kappa$ values for
    $L\le12$, while $L=14$ shows only the representative point.  Dashed
    segments connecting system sizes are guides to the eye, not scaling fits.
    Appendix~\ref{ssec: xy_beta_zero_overlap_appendix} gives the exact finite-size
    trace comparisons; Appendix~\ref{ssec: xy_finite_size_thermal_diagnostics_appendix}
    gives cleaning and window controls.}
    \label{fig: xy_analytical_numerical_support}
\end{figure*}

For $\alpha\in\{A,Z,Y\}$, we use the shorthand
\begin{equation}
    \tau_{\alpha,L}^{\rm mc,raw}
    \equiv
    \operatorname{Tr}\!\left(\rho_L^{\rm mc,raw}\widehat Q^{\alpha}\right),
    \label{eq: xy_raw_microcanonical_witness_shorthand}
\end{equation}
where $\widehat Q^{\alpha}$ denotes $\widehat Q^A_{R_r}$,
$\widehat Q^Z_{R_r}$, or $\widehat Q^Y_r$, respectively.
Figure~\ref{fig: xy_analytical_numerical_support} compares the raw
microcanonical witness values and background widths across sizes and
compatible couplings.

\subsubsection{Resolved reference and finite-size evidence}

We use even $L$ under PBC, set $h=D=0$, and follow the tower member with
$n_L=(L-2)/2$, hence $M=-2$ and $k=n_L\pi\pmod{2\pi}$.\footnote{At fixed
$M$, the field $h$ only shifts energies.  The choice $D=0$ supplies the
analytic spectral-center bridge; the exact tower and its local operators do
not require it.  Appendix~\ref{ssec: xy_generic_microcanonical_appendix}
gives finite-$D$ matched-canonical controls.}
The numerical comparison uses the half-width $\Delta E=(J/2)L^{1/4}$ and
representative coupling $\kappa_\star/J=0.1$.
The raw microcanonical values of all three witnesses are positive through
$L=14$.  At that size,
\begin{equation}
    (\tau_A,\tau_Z,\tau_Y)_{L=14}^{\rm mc,raw}
    \simeq(0.113,0.220,0.328),
    \label{eq: xy_L14_direct_microcanonical}
\end{equation}
compared with zero for all three witnesses on the caged eigenstate.

To interpret this local mismatch as an ICQMBS diagnostic, we also test
concentration of the surrounding background beyond these three observables.
We evaluate the raw concentration width $w_L^{\rm raw}$ of
Eq.~\eqref{eq: appendix_raw_concentration_width} over the complete space of
magnetization-preserving Hermitian two-site operators.  The local
Hilbert--Schmidt normalization sets the sum of the squared magnitudes of
the two-site matrix elements to one.  The width narrows monotonically from
\begin{equation}
    w_8^{\rm raw}\simeq0.169
    \quad\text{to}\quad
    w_{14}^{\rm raw}\simeq0.0175,
    \label{eq: xy_representative_concentration_sequence}
\end{equation}
with the intermediate sizes following the same decrease.  Independent window
choices, joint-dark cleaning, and a complete three-site operator-space test
show the same qualitative narrowing.  These checks demonstrate that the trend
is not tied to the three witnesses or to one window convention, but they
remain finite-size diagnostics; Appendix~\ref{ssec: xy_finite_size_thermal_diagnostics_appendix}
collects the operator-space dimensions, secondary sequences, and systematics.

The spectral center supplies a natural infinite-temperature comparator.
Thermodynamic fixed-$M$ counting gives the analytic limits
\begin{equation}
    (\tau_A,\tau_Z,\tau_Y)_{\beta=0}
    \longrightarrow\left(\frac19,\frac29,\frac13\right),
    \label{eq: xy_beta_zero_witness_limits}
\end{equation}
close to the observed raw values.  The resolved $\beta=0$ trace is exactly
$\kappa$-independent, explaining why the narrow-window means vary weakly
across the compatible family in Fig.~\ref{fig: xy_analytical_numerical_support}(c).
Their residual variation is a finite-window effect, not exact spectral
invariance.  Appendix~\ref{ssec: xy_beta_zero_overlap_appendix} gives the
sampled grid and quantitative ensemble comparison.

\subsubsection{Thermodynamic ICQMBS realization}
\label{sssec: xy_thermodynamic_realization}

For a sufficiently large fixed $c>0$, choose the fluctuation-scale raw
half-width in the same resolved sector,
\begin{equation}
\begin{aligned}
    \delta E_L^{\rm th}&=c|J|L^{1/2},\\
    \delta e_L^{\rm th}
      &\equiv\frac{\delta E_L^{\rm th}}{L}
       =c|J|L^{-1/2}\longrightarrow0.
\end{aligned}
    \label{eq: xy_thermodynamic_proof_window}
\end{equation}
Thus the total-energy window grows on the fluctuation scale while its
energy-density width shrinks.  Let $\rho_L^{\rm mc,th}$ be the normalized
projector onto this raw window.  Appendix~\ref{ssec: xy_thermodynamic_closure_appendix}
proves the following result.

\begin{proposition}[Thermodynamic spin-1 ICQMBS family]
\label{prop: xy_thermodynamic_icqmbs_main}
For the compatible Hamiltonians in Eq.~\eqref{eq: xy_reference_family}, the
selected staggered-bimagnon tower sequence is an ICQMBS for the raw windows in
Eq.~\eqref{eq: xy_thermodynamic_proof_window}, and it is deformation-stable
throughout $0.05<\kappa/J<0.20$ in the sense of
Proposition~\ref{prop:deformation_stable_icqmbs}.
\end{proposition}

The open interval is within the exactly compatible exchange family of
Eq.~\eqref{eq: xy_reference_family}; arbitrary local perturbations need not
preserve it.
Spectral reflection fixes the resolved energy center, and an $O(L)$
energy-variance bound ensures that the window retains a size-independent
positive fraction of the exponentially large sector.  This gives positive
window entropy density and a vanishing fraction for the selected tower member.
Translation averaging \cite{mori_weak_2016} and fixed-$M$ covariance control
the effect of conditioning on the raw window: every fixed bounded-local mean
changes by $O(L^{-1/2})$, and its background variance is $O(L^{-1})$.
Energy-block compression makes the concentration statement independent of
the basis chosen inside an exact degeneracy, and the bounds are uniform along
the compatible interval.  In particular,
$\operatorname{Tr}(\rho_L^{\rm mc,th}\widehat Q^A_{R_r})\to1/9$.
The witness thus vanishes on the selected sequence and has a nonzero
limiting raw microcanonical value for the same Hamiltonian. The background
concentrates on every fixed bounded region throughout the compatible interval.
The full bounds are in Appendix~\ref{ssec: xy_thermodynamic_closure_appendix};
Appendix~\ref{appendix: relations_other_formalisms} compares this thermal
closure with the known tower and local-parent descriptions.

The proof window differs from the narrower
$\Delta E=(J/2)L^{1/4}$ protocol in Fig.~\ref{fig: xy_analytical_numerical_support}.
It requires only a size-independent positive sector fraction, which need not
approach one.  The numerical protocol probes a sharper spectral neighborhood;
its microcanonical--$\beta=0$ bridge remains a finite-size diagnostic and is
not assumed in the theorem.  The thermodynamic theorem is therefore proved on the $D=0$
slice because that slice admits the analytic $\beta=0$ closure, not because the exact tower or its
local caging operators require $D=0$.  Finite-$D$ matched-canonical controls are
given in Appendix~\ref{ssec: xy_generic_microcanonical_appendix}.

The spin-1 $XY$ example therefore realizes both the local caging mechanism and the
thermodynamic hierarchy of Secs.~\ref{sec: interference_caging} and~\ref{sec: deformation_stability}.
Proposition~\ref{prop: xy_thermodynamic_icqmbs_main} establishes the ICQMBS
classification in the local diagonal-ETH sense of
Definition~\ref{def: icqmbs}; the distinction from stronger chaos and
dynamical-thermalization diagnostics is discussed there.  As an independent
check at the representative interior point $\kappa_\star/J=0.1$,
symmetry-resolved adjacent-gap ratios for $L=8,10,12$ are consistent with the
Gaussian orthogonal ensemble (GOE) value and remain well separated from
Poisson statistics
\cite{oganesyan2007,atas2013}.  This supports a nonintegrable many-body
background but is not used in the proposition; Appendix~\ref{ssec:
xy_thermal_ensembles_appendix} gives the symmetry-class audit and numerical
values.

\section{Square-lattice quantum dimer model: fixed-width local caging and compatible deformation}
\label{sec: square_qdm}

We consider the square-lattice quantum dimer model (QDM) on finite periodic
$L_x\times L_y$ lattices, where $L_x$ and $L_y$ count vertices along the
two lattice directions.  The construction starts on a $4\times4$ torus and
repeats a four-column motif along $x$, giving $4N\times4$ tori with
$L_x=4N$, $L_y=4$, and $N$ a positive integer.
This provides the constrained counterpart of the spin-chain benchmark.
A plaquette flip is local in real
space, but close packing determines which moves are admissible.  The extra task
is therefore to reduce configuration-space cancellation to a bounded operator
on the physical links. Its positive square gives the witness.

We follow the same three stages as in Sec.~\ref{sec: spin1_xy}.  First, the
$4\times4$ boundary equation yields a compact four-covering motif and bounded
kinetic caging operators.  Second, repeating that motif gives an exact
$4N\times4$ periodic-product eigenstate and the same bounded kinetic caging
operators for every positive integer $N$ along a size-independent checkerboard
phase deformation.  Third, we test its finite-temperature background in
the spatial symmetry sector containing the caged eigenstate at $L_x=8,12$,
using complete raw microcanonical windows and energy-matched canonical
comparators.  The thermal comparison remains finite-size evidence rather
than a thermodynamic ICQMBS classification.

The thermodynamic scope is deliberately quasi-one-dimensional: we fix $L_y=4$
and increase $L_x$, so locality and any thermodynamic target
refer to this fixed-width sequence; no true two-dimensional claim is made here.\footnote{An operator may wrap the fixed
circumference and remain bounded.  A limit in which $L_y$ also grows requires
a separate locality and thermal analysis.}

\subsection{Model and chiral-shell candidate set}
\label{ssec: square_qdm_model}
\label{ssec: square_qdm_definition}

\begin{figure*}[!t]
    \centering
    \includegraphics[width=\textwidth]{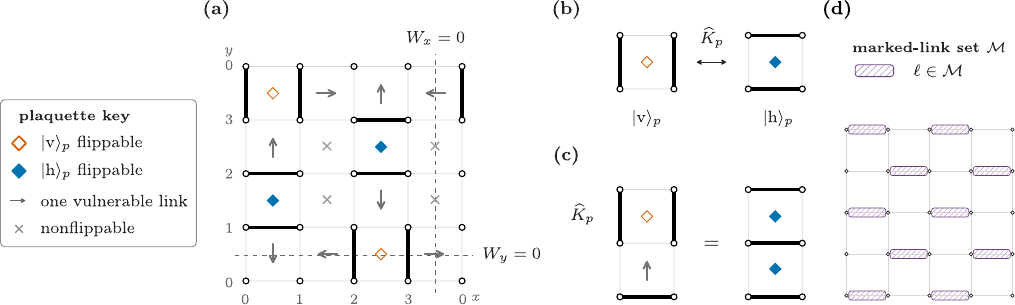}
    \caption{Microscopic ingredients and chiral grading of the square-lattice
    QDM.  \textbf{(a)} Representative close-packed dimer covering in the
    $(W_x, W_y)=(0,0)$ sector, where $W_x$ and $W_y$ are the signed dimer
    crossings of the two noncontractible cuts relative to a reference covering.
    Thick dark segments denote occupied dimers;
    duplicated boundary rows and columns show periodic images, and dashed lines
    mark the two winding cuts.  The plaquette symbols distinguish horizontal
    and vertical flippable configurations, nonflippable plaquettes, and a
    nonflippable plaquette with one vulnerable link.  \textbf{(b)} Elementary
    kinetic move: $\widehat K_p$ exchanges the vertical and horizontal
    flippable arrangements on plaquette $p$, generating an edge of the
    constrained configuration graph.  \textbf{(c)} A flip (on the upper plaquette) 
    can also change the
    flippability of a neighboring plaquette: when the latter has a single
    vulnerable link shared with $p$, flipping $p$ completes its parallel dimer
    pair.  \textbf{(d)} Marked-link construction for the grading operator
    $\widehat C$.  Choosing exactly one marked link on every plaquette ensures
    that each plaquette flip reverses the marked-link parity, giving
    $\{\widehat K,\widehat C\}=0$.  Together with the flippability label
    $N_{\mathrm{fp}}$, this grading defines the chiral-shell candidate sets.}
    \label{fig:qdm_basis_flip_chiral_4x4}
\end{figure*}

\subsubsection{Constrained dynamics and winding sector}

A close-packed square-lattice dimer covering occupies links so that exactly
one occupied link touches each site.  We denote a covering by $\ket D$ and its
link occupations by
\begin{equation}
    \widehat n_\ell\ket D=n_\ell(D)\ket D,
    \qquad
    \sum_{\ell\ni\mathbf r}n_\ell(D)=1.
    \label{eq: qdm_dimer_constraint}
\end{equation}
The second equality is the close-packing constraint.  It is the source of the
many-body character of the example: a plaquette is not an independent local
variable, because its allowed dimer arrangements are coupled to neighboring
plaquettes through shared sites.

On a torus, local plaquette flips preserve two winding numbers $(W_x, W_y)$,
defined by the signed dimer crossings of two noncontractible cuts relative to
a reference covering.  We work in the largest sector,
$(W_x,W_y)=(0,0)$, unless stated otherwise.  Figure~\ref{fig:qdm_basis_flip_chiral_4x4}
shows the two winding cuts and summarizes the constrained basis, elementary
flip, and chiral grading used below.  In the finite-size calculations we
enumerate the complete set of close-packed coverings in the selected winding
sector before any subsequent symmetry resolution; the ordering of those basis
configurations is only a coordinate convention.

On an $L_x\times L_y$ torus, $p_{(x,y)}$ denotes the unit plaquette with
lower-left vertex $(x,y)$.  Site coordinates are taken modulo $L_x$ and $L_y$;
the fixed-width sequence below has $L_y=4$.
A plaquette $p$ is flippable when two parallel dimers occupy opposite edges.
Writing the two local arrangements as $\ket{{\rm h}}_p$ and
$\ket{{\rm v}}_p$, define
\begin{equation}
\begin{aligned}
    \widehat U_p&=\ket{{\rm v}}_p\!\bra{{\rm h}}_p,\\
    \widehat K_p&=\widehat U_p+\widehat U_p^\dagger,
    \qquad
    \widehat F_p=\widehat K_p^2.
\end{aligned}
    \label{eq: qdm_local_flip_flippability}
\end{equation}
Both $\widehat U_p$ and $\widehat U_p^\dagger$ annihilate a nonflippable
covering.  The operator $\widehat K_p$ performs the flip, while
$\widehat F_p$ projects onto the two flippable configurations.  The
Rokhsar--Kivelson Hamiltonian is \cite{rokhsar_superconductivity_1988,moessner2010quantum}
\begin{equation}
\begin{aligned}
    \widehat H_{\mathrm{QDM}}
    &=\widehat K+\lambda\widehat V,\\
    \widehat K&=-\sum_p\widehat K_p,
    \qquad
    \widehat V=\sum_p\widehat F_p.
\end{aligned}
    \label{eq: RK_Hamiltonian}
\end{equation}
Here $\lambda$ is the uniform flippability-potential coupling.
The plaquette-flip amplitude sets the energy unit, $J=1$; it is restored
explicitly in the phase-deformed Hamiltonian below.  The operator
$\widehat V$ counts flippable plaquettes,
\begin{equation}
    \widehat V\ket D=N_{\mathrm{fp}}(D)\ket D.
    \label{eq: qdm_uniform_potential_label}
\end{equation}
Coverings with the same $N_{\mathrm{fp}}$ form a uniform-potential shell.
The kinetic and potential pieces are
genuinely noncommuting, $[\widehat K,\widehat V]\neq0$, because a plaquette
flip can change the flippability of neighboring plaquettes.

The interference mechanism is the constrained analogue of the two-parent
process in the spin chain.  Distinct support coverings can be flipped into the
same boundary covering.  Their individual moves are allowed, but their signed
amplitudes can cancel at that common image.  Because the close-packing
constraint controls which flips exist, this cancellation must be checked in
the complete constrained basis.

\subsubsection{Chiral grading and boundary equation}
For the even-by-even periodic lattices used below, choose a marked-link set
$\mathcal M$ such that every plaquette contains exactly one marked link, as in Fig.~\ref{fig:qdm_basis_flip_chiral_4x4}(d), and
define
\begin{equation}
    \widehat C
    =\prod_{\ell\in\mathcal M}
    \left(\widehat{\mathbf1}-2\widehat n_\ell\right),
    \qquad
    \widehat C^2=\widehat{\mathbf1}.
    \label{eq: qdm_chiral_operator}
\end{equation}
Every plaquette flip changes the occupation of the unique marked link in that
plaquette, so
\begin{equation}
    \{\widehat K,\widehat C\}=0.
    \label{eq: qdm_chiral_anticommutation}
\end{equation}
The kinetic configuration graph is therefore bipartite.  A covering $\ket D$ has grading
sign $c_D=\pm1$, and every kinetic edge connects opposite signs.

For fixed grading sign $c$ and flippability $n$, define the chiral-shell
candidate set
\begin{equation}
    \mathcal B_{c,n}
    =\bigl\{\ket D:c_D=c,\ N_{\mathrm{fp}}(D)=n\bigr\},
    \label{eq: qdm_type1_shell}
\end{equation}
with projector $\widehat P_{c,n}$.  On this candidate set,
\begin{equation}
    \widehat P_{c,n}\widehat K\widehat P_{c,n}=0,
    \qquad
    \widehat V\widehat P_{c,n}=n\widehat P_{c,n}.
    \label{eq: qdm_type1_internal_zero}
\end{equation}
The internal eigenvalue problem is automatic, and the search reduces to
\begin{equation}
    B_{c,n}\boldsymbol x=0,
    \qquad
    B_{c,n}=\widehat P_{\partial\mathcal B_{c,n}}\widehat K\widehat P_{c,n},
    \qquad
    E=\lambda n.
    \label{eq: qdm_type1_boundary_solve}
\end{equation}
This is the QDM realization of the chiral-shell search ansatz in
Sec.~\ref{ssec: type1_uniform_shell_search}.  Every kernel vector used below
is restricted to its actual support and satisfies the full Hamiltonian equation
together with the stated nontriviality conditions.

\subsection{Recovery of the compact local caging construction}
\label{ssec: qdm_finite_size_boundary_solve}

\begin{figure*}[!t]
    \centering
    \includegraphics[width=\textwidth]{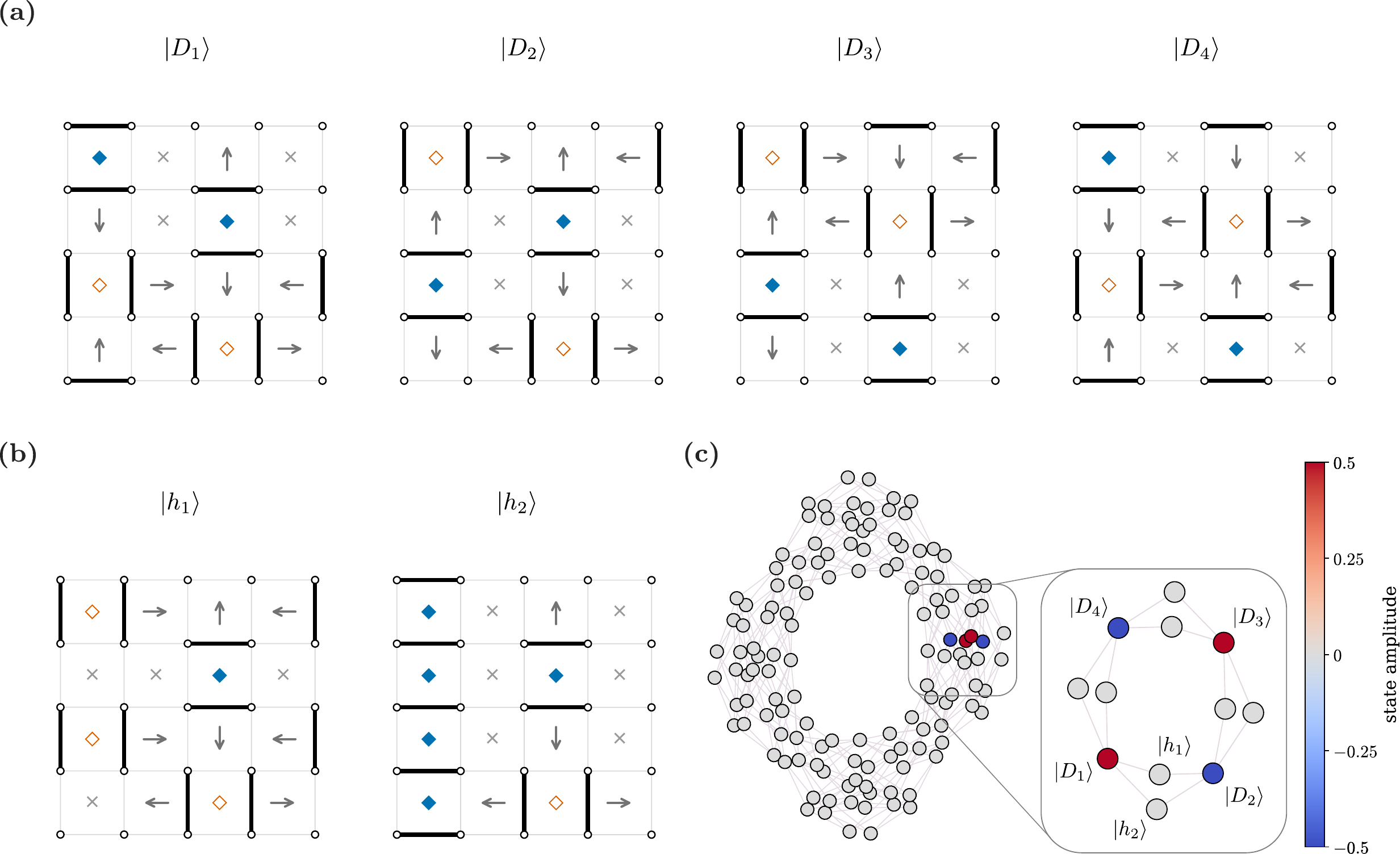}
    \caption{Compact caged eigenstate and its paired boundary cancellation for the $4\times4$ torus in the zero-winding sector.  \textbf{(a)} The four dimer coverings
    in the normalized support, $\ket{D_1},\ldots,\ket{D_4}$.  \textbf{(b)} Two
    representative active-boundary configurations reached from pairs of
    support coverings.  \textbf{(c)} The bounded real-space motif embedded in
    the finite-size configuration graph.  Warm and cool colors denote opposite
    amplitude signs of the caged eigenstate; gray vertices have zero amplitude.
    The inset in (c) isolates the configuration support $\mathcal S$ and its active
    boundary; edges from those boundary vertices to farther-away configurations
    are omitted for clarity. Self-loops are also omitted.}
    \label{fig:qdm_type1_local_interference_4x4}
\end{figure*}

\subsubsection{Boundary equation and compact representative}

For the $4\times4$ torus in the zero-winding sector, the constrained Hilbert
space has dimension $132$.  Let $\mathcal E_{(0,n)}$ denote the caged
subspace with zero kinetic eigenvalue and flippability $n$.  The chiral-shell
boundary matrix has finite-size nullities
\begin{equation}
    \dim\mathcal E_{(0,4)}=9,
    \qquad
    \dim\mathcal E_{(0,6)}=1,
    \label{eq: qdm_main_cage_dimensions}
\end{equation}
These nullities characterize the complete finite-size chiral-shell kernel;
the exactness of the specific state used below does not rely on the numerical
nullity.  Rather than
beginning with the full nine-dimensional subspace, we select a compact
four-covering representative whose signed boundary matrix gives an explicit
algebraic boundary relation and carries the bounded local mechanism.
Figure~\ref{fig:qdm_type1_local_interference_4x4} shows its support and active
boundary.

A normalized compact-motif caged eigenstate is
\begin{equation}
    \ket{\psi_{\mathrm{motif}}}
    =\frac12
    \left(
      \ket{D_1}-\ket{D_2}+\ket{D_3}-\ket{D_4}
    \right).
    \label{eq: qdm_compact_cage_state}
\end{equation}
All four coverings have the same grading sign and $N_{\mathrm{fp}}=4$.
Pairs of plaquette flips reach the same boundary covering with opposite
amplitudes.  For one cancellation unit,
\begin{equation}
    \left(
    \widehat K_{p_{(0,1)}}+\widehat K_{p_{(0,3)}}
    \right)
    \left(\ket{D_1}-\ket{D_2}\right)=0,
    \label{eq: qdm_type1_cancellation}
\end{equation}
and a translated relation cancels the leakage from $\ket{D_3}$ and
$\ket{D_4}$.  Hence
\begin{equation}
    \widehat K\ket{\psi_{\mathrm{motif}}}=0,
    \qquad
    \widehat V\ket{\psi_{\mathrm{motif}}}
    =4\ket{\psi_{\mathrm{motif}}}.
    \label{eq: qdm_local_state_KV}
\end{equation}
The complete sparse representatives and support--boundary matrices are given
in Appendix~\ref{appendix: square_qdm_cage_manifold}.  This four-covering
state is the elementary motif.  The fixed-width sequence used later repeats
that motif: its global configuration support grows with $L_x$ even though the
local cancellation rule and the physical support of its caging operators stay
bounded.  We therefore reserve ``compact motif'' for the local building block
and refer to the larger states as the periodic-product caged sequence.

\subsubsection{Bounded kinetic local caging operators}
\label{ssec: qdm_local_collective_interference}
\label{sssec: qdm_compact_reduced_iz_decomposition}

A boundary row initially carries a projector onto the outside dimer
environment.  The constrained conditioning analysis shows that every environment
with nonzero conditional weight of the caged eigenstate obeys the same paired
cancellation, so the outside projector can be removed.  The detailed conditioning proof is in
Appendix~\ref{ssec: qdm_complementary_sector_certificate}.

At longitudinal coordinate $x$, choose one opposite plaquette pair and define
the physical operator region
\begin{equation}
    R_x^{(\eta)}=\partial p_{(x,\eta)}\cup\partial p_{(x,\eta+2)},
    \qquad \eta\in\{0,1\},
    \label{eq: qdm_eight_link_region}
\end{equation}
where $\partial p$ is the set of four links bounding plaquette $p$,
and transverse coordinates are understood modulo four.  This region contains
eight links: four horizontal links between vertex columns $x,x+1$ and the
four vertical links on the two selected plaquette boundaries.  Its geometric
envelope has longitudinal width two and wraps the fixed circumference; the
full plaquette column would instead contain twelve links.  We write $R$ for
one fixed placement $R_x^{(\eta)}$, without taking the union of both pair
choices.  The resulting directed and Hermitian kinetic operators are
$\widehat A_R$ and
$\widehat Z_R=\widehat A_R+\widehat A_R^{\dagger}$, normalized here to
$\|\widehat A_R\|=\|\widehat Z_R\|=1$.
Appendix~\ref{ssec: qdm_complementary_sector_certificate} gives their local
matrix elements and the link coordinates.  Let
$\ket{\psi_{\mathrm{comp}}}$ denote either the compact motif
$\ket{\psi_{\mathrm{motif}}}$ or a selected symmetry-resolved translation
superposition built from it.  Then
\begin{equation}
\begin{aligned}
    \widehat A_R\ket{\psi_{\mathrm{comp}}}&=0,
    &\widehat Q_R^A&=\widehat A_R^\dagger\widehat A_R,\\
    \widehat Z_R\ket{\psi_{\mathrm{comp}}}&=0,
    &\widehat Q_R^Z&=\widehat Z_R^2.
\end{aligned}
\label{eq: qdm_reduced_iz_annihilation}
\end{equation}
This environment-independence step is the essential constrained-model addition to the general local-caging
construction of Sec.~\ref{sec: interference_caging}.

For the compact motif, choose $R=R_x^{(0)}$; the shifted shell
\begin{equation}
    \widehat Y_R^{\mathrm{shell}}
    =\frac12\left[
      \widehat F_{p_{(x,0)}}+\widehat F_{p_{(x,2)}}
      -2\widehat{\mathbf1}
    \right]
    \label{eq: qdm_intended_shifted_potential}
\end{equation}
is also dark; its translated counterpart uses the odd-$y$ pair.  A fixed-pair
operator is not dark after translation projection because that pair is not
flippable on every translated component.  The primary
symmetry-resolved fixed-width route therefore uses $\widehat A_R$ and
$\widehat Z_R$; the localized diagonal route is kept in Appendix~\ref{appendix: square_qdm_cage_manifold}.

\subsubsection{Locality split in the finite-size caged subspace}
\label{ssec: qdm_collective_limitation}

Relative to the bounded stripe-motif generator class of Appendix~\ref{appendix: square_qdm_cage_manifold},
\begin{equation}
\begin{aligned}
    \dim\mathcal E_{(0,4)}&=9,\\
    \dim\mathcal E_{\mathrm{loc}}&=8,\\
    \dim\!\left(\mathcal E_{(0,4)}/\mathcal E_{\mathrm{loc}}\right)&=1.
\end{aligned}
    \label{eq: qdm_main_local_collective_decomposition}
\end{equation}
The eight-dimensional locally generated subspace contains the compact route
used below.  The one-dimensional quotient is also exactly caged.  Within the
plaquette-linear kinetic class considered here, its first null combination on
$4\times4$ uses all plaquettes, so this class supplies no smaller-support
annihilator at that size.  This is a class-relative limitation, not a proof
that no bounded operator of any form exists.  The quotient, radius search, and deformation ranks are therefore kept
in Appendix~\ref{ssec: qdm_collective_locality_appendix} and are not used in
the fixed-width thermal comparison.

\subsection{Compatible continuation of the local caging construction}
\label{sssec: qdm_deformation_stability}
\label{ssec: qdm_kinetic_obstruction_main}

\subsubsection{Checkerboard compatibility and torus holonomy}

Let $\boldsymbol\phi=\{\phi_p\}$ collect one real phase for each
plaquette.  A general Hermitian plaquette-phase deformation is
\begin{equation}
    \widehat K(\boldsymbol\phi)
    =-J\sum_p\left(
      e^{i\phi_p}\widehat U_p
      +e^{-i\phi_p}\widehat U_p^\dagger
    \right).
    \label{eq: qdm_phase_deformation_family}
\end{equation}
For the fixed compact motif, a paired boundary cancellation survives when its
two parent flips acquire the same complex coefficient.  Thus
\begin{equation}
    \phi_{p_{(x,1)}}=\phi_{p_{(x,3)}}
    \pmod{2\pi},
    \label{eq: qdm_pair_phase_compatibility}
\end{equation}
with translated equal-phase constraints repeated along the strip.  A phase
difference between the two parents leaves an uncanceled boundary amplitude
for this fixed-state construction.  This is a sufficient exact compatibility
rule for the motif, not a completeness statement for all state-readjusting
continuations of caged eigenstates; the broader finite-size cage Jacobian is analyzed in
Appendix~\ref{ssec: qdm_deformation_maps}.

The relevant local gauge quotient can be stated directly in the dimer basis.
A local link rephasing
\begin{equation}
    \widehat G(\boldsymbol\theta)
    =\exp\!\left(i\sum_\ell\theta_\ell\widehat n_\ell\right)
    \label{eq: qdm_link_rephasing_unitary}
\end{equation}
changes plaquette-flip phases by the corresponding link--plaquette incidence
map.  Phase patterns in that image are local basis rephasings and are not used
as evidence for a physical kinetic deformation.  Within this declared local
gauge class, the compatible phase space contains an analytically nontrivial
direction.  If $\mathsf M_{\rm lg}$ denotes the link-to-plaquette phase map,
the checkerboard vector obeys
$\boldsymbol\chi^{\mathsf T}\mathsf M_{\rm lg}=0$ on the even periodic
lattices considered here because the two plaquettes sharing any link have
opposite checkerboard signs.  Since
$\boldsymbol\chi^{\mathsf T}\boldsymbol\chi>0$, it cannot belong to the
linear link-rephasing image.  In plaquette coordinates it is
\begin{equation}
    \chi_{p_{(x,y)}}=(-1)^{x+y}.
    \label{eq: qdm_checkerboard_phase_pattern}
\end{equation}
The paired plaquettes differ by two sites around the circumference and hence
have the same checkerboard sign.  This obstruction concerns the linear
link-rephasing image on the periodic lattice.  A link rephasing can nevertheless
move all checkerboard phases onto a single closing seam.  The non-gauge content
is therefore a torus boundary holonomy, rather than a distinct bulk phase
coupling.  Appendix~\ref{ssec: qdm_boundary_holonomy} derives this representation
and its boundary norm bound.  The gauge qualification does not concern
arbitrary configuration-dependent diagonal unitaries.

\subsubsection{Exact fixed-width repetition and resolved sector}

A size-compatible continuation keeps the local phase amplitude fixed rather
than normalizing the full plaquette-phase vector as the strip grows.  We denote
this real checkerboard phase amplitude by $\varphi$ and, for an arbitrary
uniform flippability coupling $\lambda$, use
\begin{equation}
\begin{aligned}
    \widehat H_{\varphi,\lambda}
    ={}&-J\sum_p\left[
      e^{i\varphi\chi_p}\widehat U_p
      +e^{-i\varphi\chi_p}\widehat U_p^\dagger
    \right]\\
    &+\lambda\sum_p\widehat F_p.
\end{aligned}
\label{eq: qdm_checkerboard_reference_family}
\end{equation}
Thus the analytic caging construction below does not require a special value of
$\lambda$.

For this checkerboard family the circumference-four continuation is exact.  In
each four-column longitudinal motif only $x=0,2$ are active and $x=1,3$ are
inactive.  Within an active column, the cancellation pairs $y$ with $y+2$
($y=0$ or $1$, depending on the support component), so the transverse pair
choice does not alter the longitudinal pattern.  Hence no plaquette flip couples
neighboring motifs.  The paired parents carry opposite exact amplitudes and
have equal checkerboard phases because
$\chi_{p_{(x,y+2)}}=\chi_{p_{(x,y)}}$.  Their kinetic contributions therefore cancel
identically cell by cell.  Since every support configuration contains exactly
four flippable plaquettes per motif, the $N$-fold periodic product is an exact
eigenstate on every $4N\times4$ torus, with $N$ a positive integer, for
arbitrary real $\varphi$:
\begin{equation}
    \widehat H_{\varphi,\lambda}\ket{\psi_{L_x}}
    =\lambda L_x\ket{\psi_{L_x}},
    \qquad L_x=4N.
    \label{eq: qdm_exact_periodic_product_family}
\end{equation}
Appendix~\ref{ssec: qdm_fixed_width_appendix} gives the local seam
construction.  For $L_x=4,8,12$, the checkerboard phase vector has unit relative
distance from the declared local link-rephasing image.

For $\varphi>0$, the checkerboard Hamiltonian is invariant under every
translation $(a,b)$ with $a+b$ even.  Writing $\widehat T_x$ and
$\widehat T_y$ for the one-site lattice translations, we resolve this subgroup
using
\begin{equation}
    \widehat T_{\rm diag}=\widehat T_x\widehat T_y,
    \qquad \widehat T_y^2,
    \label{eq: qdm_checkerboard_translation_generators}
\end{equation}
and select the translation character of the caged eigenstate,
\begin{equation}
    k_{\rm diag}=\frac{\pi}{2},
    \label{eq: qdm_checkerboard_selected_irrep}
\end{equation}
within the $+1$ eigensector of $\widehat T_y^2$.
A transverse reflection $\widehat S_y$ about a row of plaquette centers
preserves the checkerboard pattern and also acts within this translation
sector, so its parity must be resolved.
We select its eigenvalue on the caged eigenstate: $s_y=-1,+1,-1$ for
$L_x=4,8,12$, respectively.  The resulting sector dimensions are
$1$, $875$, and $39667$.  In particular, the $4\times4$ target sector
contains no thermal background and is excluded from the thermal size
comparison.  Appendix~\ref{ssec: qdm_microcanonical_activity_appendix}
gives the symmetry and window details; the $\varphi=0$ endpoint has enhanced
one-site translation symmetry and is kept as a separate control.

\subsubsection{Continuation of the kinetic local caging operators}
\label{ssec: qdm_witness_continuation_main}

The periodic-product state has growing many-body support, but the operator
region $R$ remains the same eight-link opposite-plaquette region in
Eq.~\eqref{eq: qdm_eight_link_region}, within two adjacent vertex columns.
The local annihilation also continues analytically.  For any fixed outside dimer
environment of $R$ with nonzero amplitude in the repeated state, the
conditional state on $R$ is either inactive or contains the same two support
parents with opposite amplitudes as in the elementary motif.  The
environment-independent operator $\widehat A_R$ therefore annihilates every
conditional component separately.  Its boundary child has zero amplitude in
the periodic-product support, so $\widehat A_R^\dagger$ annihilates the state
as well.  Because the paired checkerboard phases are equal, the same local
identity is unchanged along the full compatible phase family.  Hence, for every
positive integer $N$ and arbitrary real $\varphi$,
\begin{equation}
\begin{aligned}
    \widehat A_R\ket{\psi_{4N}}&=0,\\
    \widehat Z_R\ket{\psi_{4N}}&=0,
\end{aligned}
\qquad N\in\mathbb N,
\label{eq: qdm_joint_cage_operator_continuation}
\end{equation}
with $R$ of fixed longitudinal width two.  Appendix~\ref{ssec:
qdm_fixed_width_appendix} gives the conditional-state repetition proof.  The
$L_x=4,8,12$ residual calculations are independent finite-size checks of this
all-size identity, while the target-energy and joint-dark ranks at those sizes
characterize the surrounding exceptional subspace.  Equation~\eqref{eq:
qdm_exact_periodic_product_family} independently fixes the energy density for
arbitrary uniform $\lambda$,
\begin{equation}
    \frac{e_\psi}{J}
    =\frac{E_{\psi,L_x}}{4L_xJ}=\frac{\lambda}{4J}.
    \label{eq: qdm_periodic_product_energy_density}
\end{equation}
For the thermal diagnostics below we specialize only now to the order-one
nonzero reference $\lambda=\lambda_\star$ with $\lambda_\star/J=1$, and
quote numerical energies in units $J=1$.  This value is not required by the
caging construction and is not optimized.  We deliberately do not set
$\lambda=0$: that purely kinetic limit restores the spectral-reflection
structure and moves the caged eigenstate to zero energy, making a $\beta=0$ bridge
the natural special-case candidate.  Retaining $\lambda_\star\ne0$ instead
tests the energy-matched finite-temperature continuation introduced in
Sec.~\ref{sssec: thermodynamic_implementation}.  Although
$\lambda_\star/J=1$ coincides with the conventional Rokhsar--Kivelson coupling
at $\varphi=0$, no RK ground-state property is used in the interference-caging
proof.  The present thermal data are at this fixed $\lambda_\star$ and do not
by themselves establish stability in the $\lambda$ direction.

At nonzero checkerboard phase, the eight compact $(0,4)$ representatives of
the $4\times4$ chiral-shell caged subspace remain exact, whereas the collective
ninth $(0,4)$ representative and the $(0,6)$ caged eigenstate are lifted.

The joint-dark ranks in the resolved raw windows through $L_x=12$ are given
in Appendix~\ref{ssec: qdm_microcanonical_activity_appendix}.
They characterize the exceptional subspace but do not replace the raw
microcanonical-window comparison below.

\subsection{Fixed-width thermal comparison for the checkerboard family}
\label{sssec: qdm_fixed_width_continuation}
\label{ssec: qdm_fixed_width_thermal_main}

\begin{figure*}[!t]
    \centering
    \includegraphics[width=0.98\textwidth]{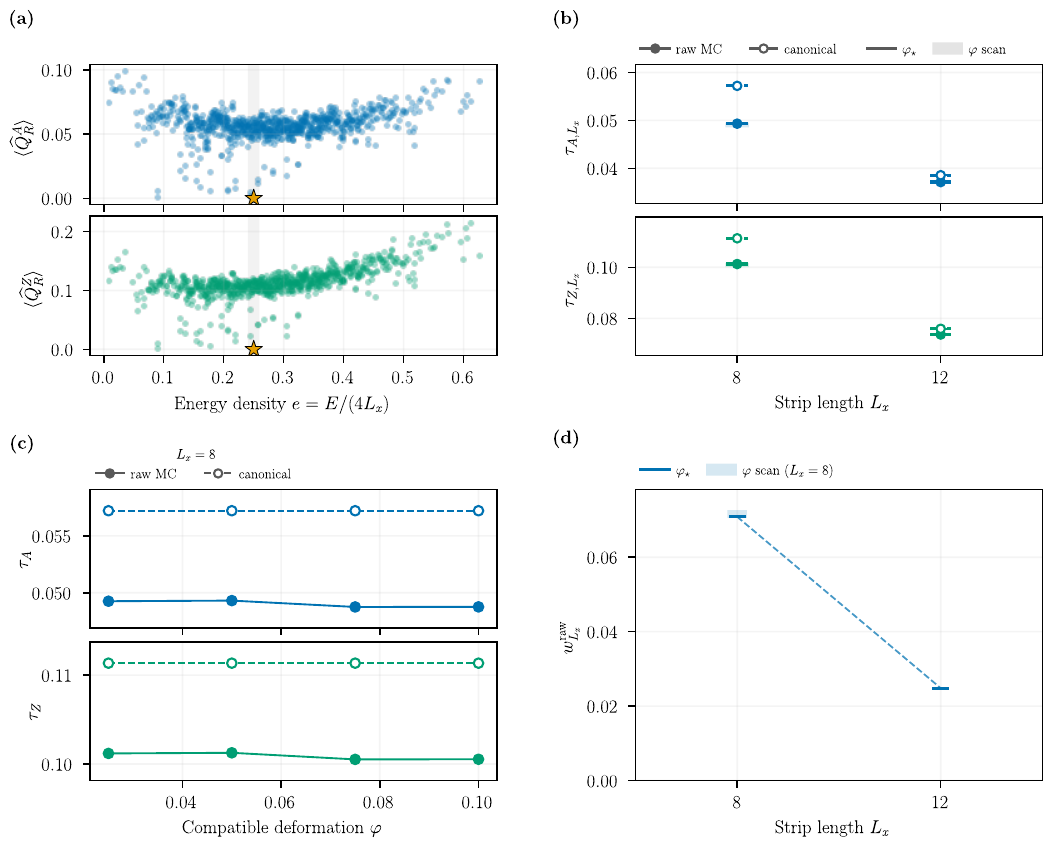}
    \caption{Finite-size thermal diagnostics at $\lambda_\star/J=1$ and
    $L_y=4$ in the spatial symmetry sector containing the caged eigenstate,
    including its transverse reflection parity $s_y$.
    The caged family and bounded $A_R,Z_R$ rule are analytic for all
    $4N\times4$ strips.  Blue and green denote the witnesses from $\widehat A_R$
    and $\widehat Z_R$, respectively.
    \textbf{(a)} \emph{Local exception:} $L_x=8$, $s_y=+1$ eigenstate scatter at
    $\varphi_\star=0.05$; stars mark the caged eigenstate at $e_\psi/J=1/4$,
    and the shaded strip marks the raw window $|E-E_\psi|\le J/4$.
    \textbf{(b)} \emph{Size and sampled
    deformation range:} Filled and open markers are raw microcanonical and
    energy-matched canonical means at $L_x=8,12$, with $s_y=+1,-1$,
    respectively.  Both raw windows are complete; only the $L_x=12$
    canonical means use stochastic trace estimates from random state vectors
    (typicality), with conditional standard errors
    below marker size.  \textbf{(c)} \emph{Deformation scan:} Raw and canonical means versus
    $\varphi$ at $L_x=8$; their weak variation is not asserted to be exact
    invariance.  \textbf{(d)} \emph{Background
    concentration:} Raw fluctuation width
    $w_{L_x}^{\rm raw}$ for the complete constrained algebra on the
    eight-link region $R$ of Eq.~\eqref{eq: qdm_eight_link_region}, using
    the local Hilbert--Schmidt quotient metric at $\varphi_\star$:
    operator directions whose sector projection vanishes are identified with zero.
    In (b,d), horizontal ticks mark the representative phase;
    the positive-phase scan at $L_x=8$ varies weakly.  Only the
    representative phase is available at $L_x=12$.  The dashed segment in (d)
    is a guide to the eye, not a scaling fit.  The one-dimensional $4\times4$ target sector
    is omitted.  Appendix~\ref{appendix: square_qdm_cage_manifold} gives
    window, uncertainty, and covariance details.}
    \label{fig:qdm_analytical_numerical_support}
\end{figure*}

The first two layers are analytic: the repeated state and its bounded
$A_R,Z_R$ dark rule continue on every $4N\times4$ torus along the compatible
checkerboard family.  We now test the third layer, asking whether that exact
darkness remains locally exceptional in the same-Hamiltonian thermal
background.  Figure~\ref{fig:qdm_analytical_numerical_support} combines the
raw microcanonical witness values, the canonical comparison, and
concentration on the declared eight-link region $R$.

\subsubsection{Raw reference and energy-matched comparator}

At the numerical reference point $\lambda_\star/J=1$, the repeated caged eigenstate has
energy density $e_\psi/J=1/4$.
We use the spatial sector selected in
Sec.~\ref{sssec: qdm_deformation_stability}, including the caged eigenstate's
reflection parity, and the same total-energy half-width $J/4$ at $L_x=8,12$.
The corresponding energy-density half-width is $J/(16L_x)$.
Both raw windows are complete and retain the caged eigenstate, 
containing $47$ and $1959$ states, respectively.  At $L_x=8$ the full
eigensystem is used; at $L_x=12$ the partial spectrum covers the complete
window, with residual and subspace-size convergence checks reported in
Appendix~\ref{ssec: qdm_microcanonical_activity_appendix}.
The one-dimensional $4\times4$ target sector supplies no thermal background
and is omitted from this comparison.

For a canonical comparator, choose $\beta_{L_x}$ so that the mean energy in
the same resolved sector equals $E_{\psi,L_x}$. The ensemble bridge is deliberate at finite
temperature. This provides a separate
check of the raw local means at each length. A canonical sequence
could extend the thermal comparison to larger sizes only after its
bounded-local equivalence to the defining raw microcanonical sequence is
controlled.  Appendix~\ref{ssec: qdm_microcanonical_activity_appendix} gives
the ensemble, spectral-coverage, and typicality checks.
For $\alpha\in\{A,Z\}$, let $\tau_{\alpha,L_x}^{\rm mc,raw}$ and
$\tau_{\alpha,L_x}^{\rm can}$ be the raw microcanonical and matched canonical
expectations of $\widehat Q_R^\alpha$.  Define
\begin{equation}
\begin{aligned}
    \delta_{\alpha,L_x}
    &=\left|
      \tau_{\alpha,L_x}^{\rm mc,raw}
      -\tau_{\alpha,L_x}^{\rm can}
      \right|,\\
    \Delta_{L_x}
    &=\max_{\alpha\in\{A,Z\}}\delta_{\alpha,L_x}.
\end{aligned}
\label{eq: qdm_local_ensemble_mismatch}
\end{equation}
At the representative phase $\varphi_\star=0.05$, this mismatch decreases from
$\Delta_8\simeq0.0101$ to $\Delta_{12}\simeq0.00225$.
Figure~\ref{fig:qdm_analytical_numerical_support}(b) displays the microcanonical
and canonical witness values.  Agreement of these values is a diagnostic
for canonical continuation, not a replacement for the raw-window criterion.

\subsubsection{Finite-size evidence and thermodynamic scope}

At both nontrivial lengths, the two witnesses vanish on the caged eigenstate
and have positive raw microcanonical values.  To test the background more
broadly, we evaluate the raw concentration width $w_{L_x}^{\rm raw}$ of
Eq.~\eqref{eq: appendix_raw_concentration_width} over the complete space of
allowed Hermitian local operators on the eight-link region $R$, using the
local Hilbert--Schmidt quotient metric of
Eq.~\eqref{eq: qdm_local_hs_quotient_metric}.
The width decreases from $w_8^{\rm raw}\simeq0.0709$ to
$w_{12}^{\rm raw}\simeq0.0248$.  This is a complete test on $R$;
concentration on other bounded regions remains open.  Neighboring windows
and joint-dark cleaning support the same finite-size trend, with all values
and systematics in Appendix~\ref{ssec: qdm_microcanonical_activity_appendix}.

The phase response at $L_x=8$ is weak compared with the change between the
two lengths.  This is consistent with a phase-only deformation that preserves
flip magnitudes, connectivity, the flippability term, and $e_\psi$, while the
matched temperatures remain near infinite temperature.  Neither argument
implies exact phase independence.  Only the representative phase is
available at $L_x=12$, so the scan tests stability at one length rather than
an open thermodynamic interval.

Thus the checkerboard family has an exact eigenstate and the same bounded
local caging mechanism at every $4N\times4$ size, together with positive raw
microcanonical witness values and a narrowing background at $L_x=8,12$.  The
raw-window thermodynamic conditions in Definition~\ref{def: icqmbs} also
require a microcanonical witness value bounded away from zero as the strip
length grows and concentration on every bounded fixed-width region.  The fixed-width ICQMBS
classification and its uniform thermal continuation on an open checkerboard
interval therefore remain open.  The true two-dimensional
limit, in which $L_y$ also grows, is a separate open problem.
Appendix~\ref{ssec: qdm_microcanonical_activity_appendix}
records the remaining thermodynamic requirements alongside the full numerical
comparison.

\section{Conclusions and outlook} \label{sec: conclusions}

Interference caging connects an exact many-body eigenstate to a physical
test of nonthermal behavior \emph{from first principle}.  
Destructive interference confines the state to
a set of configurations; when that cancellation follows a bounded local rule,
it supplies a measurement on a fixed region of the physical lattice.  We give
two sufficient constructions, based on local kinetic cancellation and a
locally resolved uniform potential.  In each, the positive witness
$\widehat Q_R=\widehat L_R^\dagger\widehat L_R$ vanishes on the caged
eigenstate.  A nonzero thermodynamic microcanonical value for the same
Hamiltonian at the same energy density establishes a local violation of
diagonal ETH.  The ICQMBS classification additionally requires a genuine
locally thermal background in the diagonal-ETH sense defined in
Sec.~\ref{ssec: thermodynamic_eth_criterion}: an unmodified symmetry-resolved
window with positive entropy density, concentration of bounded local observables,
and a vanishing fraction of caged eigenstates.  These constructions are sufficient rather than
necessary; other local measurements may diagnose nonthermal states when the
present reductions fail.

The deformation framework asks whether the exact state, its local mechanism,
and its thermal distinction survive changes of the interactions.  It
identifies coupling directions that can be accommodated by readjusting the
state and energy, and those that obstruct the caged eigenstate.  A thermodynamic stability
claim requires these changes to follow one size-compatible local interaction
rule, preserve the bounded caging operator, and keep the microcanonical witness
value bounded away from zero uniformly on a size-independent coupling
neighborhood as the system grows.  The surrounding spectrum must remain
thermodynamic and locally concentrated on that same neighborhood.  This gives a precise meaning to stability within a
declared compatible family and completes the framework with predictive power.

The spin-1 $XY$ chain provides an analytic thermodynamic realization of the
full framework.  The construction recovers the staggered $\pi$-bimagnon tower
and its bounded local caging operators, identifies a compatible exchange
deformation that preserves both the exact states and their local mechanism,
and closes the local thermodynamic comparison on a fluctuation-scale
$O(L^{1/2})$ microcanonical window.  A size-independent positive fraction of
the resolved sector remains in this window, the background is locally
concentrated, the selected tower-member fraction vanishes, and the kinetic
witness has a positive limiting microcanonical value throughout an open
compatible deformation interval, establishing a deformation-stable ICQMBS
family in the sense of Definition~\ref{def: icqmbs}.
The tower, local-parent structure, and translation-averaging tools are
established ingredients; the additional conclusion is the raw-window thermal
diagnosis, with symmetry resolution and degeneracy control uniform along the
same compatible family.

The square-lattice quantum dimer model tests the same framework in a constrained
Hilbert space.  A compact four-covering motif generates an exact repeated
circumference-four family under a checkerboard phase deformation, and the
same bounded kinetic caging operators annihilate the repeated state
analytically for every $4N\times4$ strip.  Finite-size calculations in the
spatial symmetry sector containing the caged eigenstate further show nonzero
raw microcanonical witness values together
with a narrowing background on the declared eight-link region.  Thus, the exact-state and bounded-local
mechanism layers are established analytically, while the thermal layer has
strong finite-size support but is not yet closed thermodynamically.  The true
two-dimensional limit also remains open.

The local caging operators found in these models raise the question of how
interference caging connects to other QMBS constructions. In the spin-1
$XY$ chain, the quasiparticle creator relates successive
members of the staggered $\pi$-bimagnon QMBS tower, and the local caging
operators identify the leakage processes that cancel in those eigenstates.
The witnesses also annihilate the tower and can supply local constraints
for a projector embedding. Appendix~\ref{appendix: relations_other_formalisms}
gives the additional conditions for such an embedding, for identifying
the intended subspace, and for recovering the algebra generated by the
local Hamiltonian terms. The ICQMBS diagnosis starts from a bounded local
reduction of interference and applies the thermal and spectral tests above,
independently of a tower construction or a projector embedding.
Whether bounded local operators can characterize an
entire caged subspace through their common kernel remains open in general,
as does the role of any algebraic or topological obstructions.
Interference caging can also occur within a
Hilbert-space fragment identified from the local dynamics independently of
the caged eigenstate. An ICQMBS diagnosis within that fragment requires an
energy window with positive entropy density and local concentration.

The repeated local motifs in both models suggest that relations between
product configurations could help narrow the search for caged eigenstates.
This is a possible extension of the restricted chiral-shell search in
Sec.~\ref{sec: practical_search}, which finds candidate eigenstates without
full-spectrum diagonalization.\footnote{
The chiral-shell search uses a bipartite kinetic configuration graph. 
It is a practical simplification rather than a requirement for interference caging. 
Appendix~\ref{ssec: spin1_xy_five_site_nonchiral_cage} gives caged eigenstates
in a non-bipartite kinetic configuration graph.}
A weighted graph automorphism permutes product configurations while
preserving the kinetic matrix elements
(Appendix~\ref{ssec: graph_symmetry_candidate_reduction}). Such a permutation
can act nonlocally in real space and relate configurations beyond those
connected by lattice or on-site symmetries. A permutation that also
preserves the diagonal potential commutes with the full Hamiltonian.
The orbits, or sets of configurations related by these permutations, can
expose repeated boundary constraints. An open question is whether those
orbits can systematically identify supports that admit solutions of both the
internal eigenvalue equation and the boundary cancellation condition.

By linking interference, local measurements, and stability under changes of the interactions, the framework provides a common way to construct and test exceptional many-body states across constrained and unconstrained systems.

\section*{Data Availability}

The \texttt{qlinks} software used for constrained-basis construction,
interference-cage searches, and numerical analyses is publicly available in
Ref.~\cite{qlinks_repository_2026}. The numerical data supporting the findings
of this work are not currently deposited in a public repository because the
model-specific numerical records have not been consolidated into a standalone,
reusable dataset. These data are available from the authors upon request.

\section*{Acknowledgments}

This work was supported by the MOST Young Scholar Fellowship (Grants
No.~112-2636-M-007-008-, No.~113-2636-M-007-002-, and
No.~114-2636-M-007-001-), the National Center for Theoretical Sciences (Grant
No.~113-2124-M-002-003-) under the Ministry of Science and Technology (MOST),
Taiwan, and the Yushan Young Scholar Program (Administrative Support Grant
Fellowship) of the Ministry of Education, Taiwan.

We thank Debasish Banerjee, Akira Furusaki, Chisa Hotta, Ying-Jer Kao, Arnab
Sen, Chin-Hung Lin, Jun Takahashi, and Zhen-Sheng Yuan, listed alphabetically, for insightful
discussions.  We especially thank Debasish Banerjee for independently
cross-checking part of our numerical data by exact diagonalization.

OpenAI ChatGPT (GPT-5.6 Sol and GPT-6) was used during manuscript preparation for
editorial revision, consistency checks, scientific-argument review, and
assistance with parts of the numerical analysis and figure preparation.  The
authors directed these uses and independently verified the AI-assisted text,
scientific arguments, numerical results, figures, and references against the
underlying calculations, source data, and cited literature.  The authors take
full responsibility for the scientific content and final manuscript.

\FloatBarrier

\appendix

\section{Local caging operators and local ETH-violation witnesses}
\label{appendix: local_iz_technical_details}

This appendix collects the technical statements behind Sec.~\ref{sec: interference_caging}: constrained-space locality, bounded kinetic and diagonal reductions, normalization, and the resolved thermal estimators used by the model appendices.  Throughout, ``local'' means bounded support on the physical lattice.

\subsection{What locality means in a constrained Hilbert space}
\label{ssec: appendix_conditional_blocks}
\label{ssec: appendix_complementary_sector_proof}

A boundary-row cancellation initially remembers the outside configuration.
A witness must come from one microscopic rule on a
bounded region $R$, with the constraints selecting the allowed local
block.  This is the content of the environment-independence criterion used in
Proposition~\ref{prop: complementary_sector_reduction}.

Let $R\subset\Omega$ be a bounded physical region and
$\overline R=\Omega\setminus R$.  For constraints diagonal in the microscopic
product basis, the physical Hilbert space decomposes as
\begin{equation}
    \mathcal H_{\rm phys}
    =\bigoplus_{\alpha}
      \ket\alpha_{\overline R}\otimes\mathcal H_{R|\alpha},
    \label{eq: appendix_conditional_hilbert_space}
\end{equation}
where $\alpha$ labels an outside environment and $\mathcal H_{R|\alpha}$ is the
set of local configurations compatible with that environment.  Let
$\Pi_{R|\alpha}$ project onto this conditional local space.

The key restriction is that these conditional blocks are not independent
operators.  They must all descend from one microscopic operator
$\widehat L_R^{\rm phys}$ supported on $R$, whose block in environment
$\alpha$ is
\begin{equation}
    \widehat L_R^{(\alpha)}
    =\Pi_{R|\alpha}\widehat L_R^{\rm phys}\Pi_{R|\alpha}.
    \label{eq: appendix_conditional_block_restriction}
\end{equation}
Thus the apparent dependence on $\alpha$ enters only through the constraint
projection of the same microscopic rule.  We write $\widehat L_R$ for this
microscopic operator restricted to $\mathcal H_{\rm phys}$.

Resolve a normalized physical state as
\begin{equation}
\begin{aligned}
    \ket\psi
    &=\sum_\alpha c_\alpha
      \ket\alpha_{\overline R}\otimes\ket{\psi_{R|\alpha}}_R,\\
    \ket{\psi_{R|\alpha}}&\in\mathcal H_{R|\alpha}.
\end{aligned}
    \label{eq: appendix_conditional_state_resolution}
\end{equation}
Orthogonality of different outside environments gives
\begin{equation}
    \bigl\|\widehat L_R\ket\psi\bigr\|^2
    =\sum_\alpha |c_\alpha|^2
      \bigl\|\widehat L_R^{(\alpha)}\ket{\psi_{R|\alpha}}\bigr\|^2.
    \label{eq: appendix_complement_sector_norm}
\end{equation}
Consequently,
\begin{equation}
    \boxed{
    \widehat L_R\ket\psi=0
    \quad\Longleftrightarrow\quad
    c_\alpha\,
    \widehat L_R^{(\alpha)}\ket{\psi_{R|\alpha}}=0
    \quad\text{for every }\alpha.}
    \label{eq: appendix_environment_independence_criterion}
\end{equation}
This is the constrained-space form of
Proposition~\ref{prop: complementary_sector_reduction}.  A block with
$c_\alpha=0$ is absent and imposes no condition; whenever $c_\alpha\neq0$, the
same microscopic local rule must annihilate the corresponding conditional
state.  Different outside environments are orthogonal and therefore cannot
cancel one another after the extensive conditioning projector is removed.

\subsection{From boundary rows to bounded kinetic operators}
\label{ssec: appendix_rank_one_active_channel}
\label{ssec: appendix_directed_positive_constraint}
\label{ssec: appendix_multirow_kinetic_reduction}
\label{ssec: appendix_minimal_reduction_radius}

Once the outside condition has been removed, a boundary row can be read as a
local transition rule.  We first write the directed and Hermitian operators for
one row, and then state what must remain true when several nearby rows are
combined.

For one selected boundary row $h$, define the incoming local superposition
\begin{equation}
    \ket{v_h}_{\Gamma_h}
    =\sum_{u\in N_{\mathcal S}(h)}K_{hu}^{*}\ket u_{\Gamma_h}.
    \label{eq: appendix_incoming_boundary_vector}
\end{equation}
The directed and Hermitian kinetic operators are
\begin{equation}
    \widehat A_{h;\Gamma_h}=\ket h\!\bra{v_h},
    \qquad
    \widehat Z_{h;\Gamma_h}
    =\widehat A_{h;\Gamma_h}+\widehat A_{h;\Gamma_h}^{\dagger}.
    \label{eq: appendix_directed_transition_map}
\end{equation}
Because $h$ lies outside the caged eigenstate's configuration support while the components of $v_h$ lie
inside it, $\langle h|v_h\rangle=0$.  Hence
\begin{align}
    \widehat A_{h;\Gamma_h}^{\dagger}\widehat A_{h;\Gamma_h}
    &=\ket{v_h}\!\bra{v_h},
    \label{eq: appendix_directed_positive_constraint}\\
    \widehat Z_{h;\Gamma_h}^{2}
    &=\ket{v_h}\!\bra{v_h}
      +\|v_h\|^2\ket h\!\bra h,
    \label{eq: appendix_z_square_directed_decomposition}
\end{align}
and
\begin{equation}
    \ker\widehat Z_{h;\Gamma_h}
    =\ker\widehat A_{h;\Gamma_h}
     \cap\ker\widehat A_{h;\Gamma_h}^{\dagger}.
    \label{eq: appendix_directed_kernel_inclusion}
\end{equation}
Thus $\widehat A^\dagger\widehat A$ measures the coherent incoming-transition
weight, whereas $\widehat Z^2$ also tests occupation of the boundary
configuration.  The directed operator can therefore yield a witness even
when the stronger Hermitian reduction fails.

The same construction extends to finitely many nearby boundary rows contained
in one bounded region $R$.  With complex coefficients $\eta_h$, define
\begin{equation}
    \widehat A_{\mathcal C;R}
    =\sum_{h\in\mathcal C}\eta_h\widehat A_{h;R},
    \qquad
    \widehat Z_{\mathcal C;R}
    =\widehat A_{\mathcal C;R}+\widehat A_{\mathcal C;R}^{\dagger}.
    \label{eq: appendix_multirow_directed_channel}
\end{equation}
The Hermitian definition remains valid for complex hopping amplitudes.  A
\emph{bounded multi-row kinetic reduction} is established when this common
microscopic rule is dark in every occupied environment according to
Eq.~\eqref{eq: appendix_environment_independence_criterion}, with the physical
diameter of $R$ bounded uniformly along the thermodynamic sequence.

The construction must fit inside a region of uniformly bounded diameter along the size sequence; a cancellation whose required support grows with size does not establish this local route.  We do not assume a minimum-support optimization, and bounded multi-row reducibility remains open for the exceptional square-QDM collective representative.

\subsection{When a potential shell yields a witness}
\label{ssec: appendix_uniform_shell_structure}
\label{ssec: appendix_projected_potential_reduction}

The diagonal route separates a global statement from a local one.  The caged eigenstate's
configuration support must first have one common potential value; then a bounded part of that
potential must remain sharp in every occupied environment.  Only the second
statement produces the local operator $\widehat Y_R$.

Write a diagonal potential in terms of an independent coupling vector,
\begin{equation}
    \widehat V(\boldsymbol\lambda)=\sum_{\mu=1}^{m}\lambda_\mu\widehat N_\mu,
\end{equation}
and let $\boldsymbol n(s)$ be the corresponding diagonal eigenvalue vector on a
support configuration $s\in\mathcal S$.  Choosing one reference
$s_0\in\mathcal S$, define the support-variation space
\begin{equation}
    \mathcal D_{\mathcal S}
    =\operatorname{span}\{\boldsymbol n(s)-\boldsymbol n(s_0):s\in\mathcal S\}.
    \label{eq: appendix_shell_variation_space}
\end{equation}
Then the support lies in one uniform-potential shell precisely when
\begin{equation}
    \boxed{\boldsymbol\lambda\in\mathcal D_{\mathcal S}^{\perp}.}
    \label{eq: appendix_shell_orthogonality}
\end{equation}
If $\mathcal D_{\mathcal S}=\{0\}$, the shell is structural in the declared
coupling space; otherwise the orthogonality condition identifies the required
linear tuning relations.  The dimension of $\mathcal D_{\mathcal S}$ therefore
counts the independent linear conditions, but we do not introduce a separate
diagnostic symbol for it.

The global shell condition
\begin{equation}
    \widehat V\widehat P_{\mathcal S}=v_{\mathcal S}\widehat P_{\mathcal S}
    \label{eq: appendix_global_shell_condition}
\end{equation}
fixes the potential contribution to the energy of the caged eigenstate. A
witness from the potential additionally requires a bounded region
$R$ for which
\begin{equation}
    \widehat V_R\ket\psi=v_R\ket\psi,
    \qquad
    \widehat Y_R=\widehat V_R-v_R\widehat{\mathbf1}_R,
    \qquad
    \widehat Y_R\ket\psi=0.
    \label{eq: appendix_local_shifted_potential_channel}
\end{equation}
In a constrained model, $\widehat Y_R$ must again be the restriction of one
microscopic $R$-supported rule and must satisfy
Eq.~\eqref{eq: appendix_environment_independence_criterion} in every occupied
environment.  None of these shell statements requires
$[\widehat K,\widehat V]=0$; they concern the scalar action of the diagonal
operator on the selected caged eigenstate's configuration support.  On the selected state one has
$[\widehat K,\widehat V]\ket\psi=0$, but this does not imply a global
commutation relation.  If the caged eigenstate's configuration support lies on one chiral subset of the
bipartite kinetic graph, the internal kinetic block vanishes; together with the
boundary equation this gives $\widehat K\ket\psi=0$ and hence
$E=v_{\mathcal S}$.

For the positive witness $\widehat Q_R=\widehat Y_R^2$,
\begin{equation}
    \langle\widehat Q_R\rangle_{\rm mc}
    =\operatorname{Var}_{\rm mc}(V_R)
     +\bigl(\langle\widehat V_R\rangle_{\rm mc}-v_R\bigr)^2.
    \label{eq: appendix_shell_positive_witness_decomposition}
\end{equation}
Thus, an accidental first-moment equality does not remove the witness when
thermal local-potential fluctuations remain nonzero.

\subsection{Shifted local operators from internal kinetic equations}
\label{ssec: appendix_shifted_kinetic_construction}

The shifted-potential construction has a kinetic counterpart when the
\emph{internal} eigenvalue equation admits a bounded local reduction.  On the
uniform-potential shell of Eq.~\eqref{eq: uniform_shell_scalar_action}, let
$K_{\mathcal S\mathcal S}=\widehat P_{\mathcal S}\widehat K
\widehat P_{\mathcal S}$ denote the internal kinetic block.  The
support--boundary equations become
\begin{equation}
    K_{\mathcal S\mathcal S}\boldsymbol x=\varepsilon\boldsymbol x,
    \qquad B_{\mathcal S}\boldsymbol x=0,
    \label{eq: appendix_internal_kinetic_eigenvalue}
\end{equation}
so that $\widehat K\ket\psi=\varepsilon\ket\psi$ and
$E=\varepsilon+v_{\mathcal S}$.  The kinetic eigenvalue $\varepsilon$ need
not vanish.  Its sharpness is a global statement and does not by itself make
any bounded part of the kinetic Hamiltonian sharp.

\subsubsection{A sharp local kinetic value}

Fix the microscopic decomposition $\widehat K=\sum_X\widehat K_X$.
On a bounded physical region $R$, define the Hermitian operator
\begin{equation}
    \widehat K_R=\sum_{X\subseteq R}\widehat K_X.
    \label{eq: appendix_local_kinetic_operator}
\end{equation}
If this local operator has a sharp value $\kappa_R$ in the caged eigenstate,
then
\begin{equation}
    \widehat K_R\ket\psi=\kappa_R\ket\psi,
    \qquad
    \widehat L_R^{\rm kin}
    =\widehat K_R-\kappa_R\widehat{\mathbf1}_R
    \label{eq: appendix_local_shifted_kinetic_operator}
\end{equation}
gives $\widehat L_R^{\rm kin}\ket\psi=0$.
The same statement holds for a declared bounded Hermitian combination of
microscopic kinetic processes on $R$.  The local eigenvalue $\kappa_R$ is
real and satisfies $|\kappa_R|\le\|\widehat K_R\|$; it need not equal the
total kinetic eigenvalue $\varepsilon$, which can be extensive.  Along a
thermodynamic sequence, the operator and shift must follow one
size-compatible local rule with uniformly bounded support and norm.

The locality requirement is exactly the environment-independence test of
Eq.~\eqref{eq: appendix_environment_independence_criterion}.  With the
conditional states of Eq.~\eqref{eq: appendix_conditional_state_resolution},
it reads
\begin{equation}
    c_\alpha\bigl(\widehat K_R^{(\alpha)}
        -\kappa_R\widehat{\mathbf1}_{R|\alpha}\bigr)
        \ket{\psi_{R|\alpha}}=0
    \quad\text{for every }\alpha,
    \label{eq: appendix_shifted_kinetic_environment_condition}
\end{equation}
where $\widehat K_R^{(\alpha)}$ is the restriction of the same microscopic
operator to $\mathcal H_{R|\alpha}$, and
$\widehat{\mathbf1}_{R|\alpha}$ is that block's identity.  Thus the shift
must be common to all occupied environments.  In a tensor-product Hilbert
space, this is equivalently
$\operatorname{supp}\rho_R^\psi\subseteq\ker\widehat L_R^{\rm kin}$,
where $\rho_R^\psi$ is the reduced density matrix of the caged eigenstate on
$R$.  A full-rank $\rho_R^\psi$ permits only the trivial shifted operator.

The positive square is
\begin{equation}
    \widehat Q_R^{\rm kin}
    =(\widehat L_R^{\rm kin})^\dagger\widehat L_R^{\rm kin}
    =(\widehat K_R-\kappa_R\widehat{\mathbf1}_R)^2.
    \label{eq: appendix_shifted_kinetic_positive_witness}
\end{equation}
It vanishes exactly on the caged eigenstate, while
\begin{equation}
\begin{aligned}
    \langle\widehat Q_R^{\rm kin}\rangle_{\rm mc}
    &=\operatorname{Var}_{\rm mc}(K_R)\\
    &\quad+\bigl(\langle\widehat K_R\rangle_{\rm mc}-\kappa_R\bigr)^2.
\end{aligned}
    \label{eq: appendix_shifted_kinetic_thermal_decomposition}
\end{equation}
After the common normalization specified below, the positive-liminf
criterion of Eq.~\eqref{eq: appendix_positive_witness_activity} supplies the
same sufficient local diagonal-ETH test as for the boundary and potential
constructions.

\subsubsection{A selected internal row}

An individual internal row generally requires a projector shift.  Let
$u\in\mathcal S$ and
$N_{\mathcal S}(u)=\{v\in\mathcal S:K_{uv}\neq0\}$, with
$K_{uv}=\bra u\widehat K\ket v$.  Suppose all occupied incoming kinetic
processes for this row are supported inside one bounded region $R$.  Their
parents then agree with $u$ on a common outside configuration $\alpha_u$.
Define the local row operator and target projector by
\begin{equation}
\begin{aligned}
    \widehat A_{u;R}
    &=\ket u_R\sum_{v\in N_{\mathcal S}(u)}K_{uv}\,{}_R\!\bra v,\\
    \widehat P_{u;R}&=\ket u_R\,{}_R\!\bra u.
\end{aligned}
    \label{eq: appendix_internal_kinetic_row_operators}
\end{equation}
Here the subscript $R$ denotes restriction of a product configuration to $R$.
The row $\sum_vK_{uv}x_v=\varepsilon x_u$ gives
\begin{equation}
    \Bigl[\widehat P_{\alpha_u;\overline R}\otimes
      \bigl(\widehat A_{u;R}-\varepsilon\widehat P_{u;R}\bigr)\Bigr]
      \ket\psi=0,
    \label{eq: appendix_conditioned_internal_kinetic_row}
\end{equation}
where $\widehat P_{\alpha_u;\overline R}
=\ket{\alpha_u}\!\bra{\alpha_u}_{\overline R}$ selects that outside
configuration.  If the outside conditioning can be removed by
Eq.~\eqref{eq: appendix_environment_independence_criterion}, then
\begin{equation}
    \widehat L_R^{\rm row}
    =\widehat A_{u;R}-\varepsilon\widehat P_{u;R},
    \qquad \widehat L_R^{\rm row}\ket\psi=0
    \label{eq: appendix_shifted_internal_kinetic_row}
\end{equation}
is another local caging operator, provided its norm remains uniformly bounded
under the same normalization convention.  The shift acts on the local target
projector, rather than the identity on all of $R$.  Since $u$ is occupied,
adjoint darkness is not automatic; the positive witness is always
$(\widehat L_R^{\rm row})^\dagger\widehat L_R^{\rm row}$, whereas Hermitian
symmetrization additionally requires
$(\widehat L_R^{\rm row})^\dagger\ket\psi=0$.  Finite combinations of
internal rows obey the same locality and normalization requirements.

\subsection{Fixing the witness normalization}
\label{ssec: appendix_caging_operator_normalization}

The darkness condition fixes a kernel but not an overall scale:
\begin{equation}
    \widehat L_R\ket\psi=0
    \quad\Longleftrightarrow\quad
    a\widehat L_R\ket\psi=0,
    \qquad a\neq0.
    \label{eq: appendix_caging_operator_scale_equivalence}
\end{equation}
We therefore factor any bare local construction as
\begin{equation}
    \widehat L_R^{\rm bare}(\boldsymbol\lambda)
    =\nu_R(\boldsymbol\lambda)\widehat L_R(\boldsymbol\lambda),
    \qquad \nu_R\neq0,
    \label{eq: appendix_local_scale_factorization}
\end{equation}
where only the common local energy scale is removed; relative magnitudes, signs, and phases remain part of the interference rule.  The normalized positive witness is
\begin{equation}
    (\widehat L_R^{\rm bare})^\dagger\widehat L_R^{\rm bare}
    =|\nu_R|^2\widehat Q_R,
    \qquad
    \widehat Q_R\equiv\widehat L_R^\dagger\widehat L_R.
    \label{eq: appendix_positive_witness_scale}
\end{equation}
We use a size-independent convention, usually $0\le\widehat Q_R\le\widehat{\mathbf1}$.  Two conventions are thermodynamically equivalent when their relative factors satisfy
\begin{equation}
    0<a_-\le |a_L(\boldsymbol\lambda)|\le a_+<\infty
    \label{eq: appendix_admissible_normalization_bounds}
\end{equation}
uniformly in size and throughout the declared coupling neighborhood.  Vanishing or diverging $L$-dependent rescalings are inadmissible.  If a chosen common scale crosses zero, one instead keeps a nonsingular coupling-independent local generator or chooses another local caging operator.  The spin-1 and QDM sections use these dimensionless conventions.

\subsection{Constructing the thermal reference}
\label{ssec: appendix_energy_matching_and_canonical_role}

All thermal comparisons use the same Hamiltonian and the resolved reference space specified in Sec.~\ref{sssec: thermal_reference_protocol}.  Let $\widehat P_{\rm ref, L}$ project onto that invariant space and let $I_L$ be the window centered on $E_L$.  Retaining complete exact-energy blocks gives
\begin{equation}
    \widehat P_{\mathcal W_L}
    =\widehat P_{\rm ref,L}\,\mathbf 1_{I_L}(\widehat H_L),
    \qquad
    \rho_L^{{\rm mc,raw}}
    =\frac{\widehat P_{\mathcal W_L}}
           {\operatorname{Tr}\widehat P_{\mathcal W_L}}.
    \label{eq: microcanonical_state_definition}
\end{equation}
This is the defining main-text state $\rho_L^{\rm mc}$; in this appendix we
write it as $\rho_L^{{\rm mc, raw}}$ only to distinguish it from the cleaned
finite-size diagnostic below.  A partial eigensolver can represent this window only when completeness of the contained interval is independently certified.  The window rule is fixed before inspecting the witness and is tested against reasonable width or target-count variations.

\subsubsection{Joint-dark cleaning}
Cleaning is a finite-size diagnostic, not the defining ensemble.  For a declared translated family of positive witnesses, define
\begin{equation}
    \widehat Q_{\rm all}=\sum_{a,R}\widehat Q_R^a.
    \label{eq: appendix_joint_dark_operator}
\end{equation}
Within each resolved exact-energy block $\widehat P_{E, L}$, the kernel of $\widehat P_{E, L}\widehat Q_{\rm all}\widehat P_{E, L}$ defines the joint-dark subspace.  Summing these projectors over the window gives $\widehat P_{{\rm jd},L}$ and
\begin{equation}
    \rho_L^{{\rm mc,clean}}
    =\frac{\widehat P_{\mathcal W_L}-\widehat P_{{\rm jd},L}}
           {\operatorname{Tr}(\widehat P_{\mathcal W_L}-\widehat P_{{\rm jd},L})}.
    \label{eq: appendix_energy_matching_microcanonical}
\end{equation}
This construction is basis independent inside exact degeneracies.  With
\begin{equation}
    f_{{\rm jd},L}
    =\frac{\operatorname{Tr}\widehat P_{{\rm jd},L}}
           {\operatorname{Tr}\widehat P_{\mathcal W_L}},
    \label{eq: appendix_cage_excised_fraction}
\end{equation}
and $\rho_L^{\rm jd}=\widehat P_{{\rm jd},L}/\operatorname{Tr}\widehat P_{{\rm jd},L}$ when the rank is nonzero,
\begin{equation}
    \rho_L^{{\rm mc,raw}}
    =(1-f_{{\rm jd},L})\rho_L^{{\rm mc,clean}}
     +f_{{\rm jd},L}\rho_L^{\rm jd},
    \label{eq: appendix_raw_clean_convex_decomposition}
\end{equation}
so every bounded Hermitian local observable satisfies
\begin{equation}
    \left|\operatorname{Tr}[(\rho_L^{{\rm mc,raw}}-\rho_L^{{\rm mc,clean}})\widehat O_R]\right|
    \le 2f_{{\rm jd},L}\|\widehat O_R\|,
    \label{eq: appendix_general_raw_clean_bound}
\end{equation}
with the sharper bound $f_{{\rm jd},L}$ for $0\le\widehat O_R\le\widehat{\mathbf1}$.  Thus $f_{{\rm jd}, L}\to0$ makes raw and cleaned bounded-local limits identical whenever either exists.

\subsubsection{Canonical and \texorpdfstring{$\beta=0$}{beta=0} bridges}
A canonical comparator in the same resolved space is
\begin{equation}
    \rho^{\rm can}_{\beta,L}
    =\frac{\widehat P_{\rm ref,L}e^{-\beta\widehat H_L}}
           {\operatorname{Tr}(\widehat P_{\rm ref,L}e^{-\beta\widehat H_L})},
    \label{eq: appendix_resolved_canonical_state}
\end{equation}
with $\beta_L$ chosen, when possible, from
\begin{equation}
    \operatorname{Tr}(\rho^{\rm can}_{\beta_L,L}\widehat H_L)=E_L.
    \label{eq: appendix_canonical_energy_matching}
\end{equation}
Using this as a surrogate for the raw microcanonical sequence requires local equivalence on every fixed bounded region,
\begin{equation}
    \operatorname{Tr}(\rho_L^{{\rm mc,raw}}\widehat O_R)
    -\operatorname{Tr}(\rho^{\rm can}_{\beta_L,L}\widehat O_R)
    \longrightarrow0,
    \label{eq: appendix_local_ensemble_equivalence}
\end{equation}
under hypotheses appropriate to the resolved sector or independently verified in the model \cite{brandao2015ensembleequivalence,tasaki2018localequivalence}.  At $\beta=0$,
\begin{equation}
    \rho^{\rm can}_{0,L}
    =\frac{\widehat P_{\rm ref,L}}{\operatorname{Tr}\widehat P_{\rm ref,L}},
    \qquad
    E_{0,L}^{\rm ref}
    =\frac{\operatorname{Tr}(\widehat P_{\rm ref,L}\widehat H_L)}
           {\operatorname{Tr}\widehat P_{\rm ref,L}},
    \label{eq: appendix_beta_zero_resolved_trace}
\end{equation}
and
\begin{equation}
    \tau^{\rm can}_{Q,L}(0)
    =\frac{\operatorname{Tr}(\widehat P_{\rm ref,L}\widehat Q_R)}
           {\operatorname{Tr}\widehat P_{\rm ref,L}}.
    \label{eq: appendix_beta_zero_witness_trace}
\end{equation}
The $\beta=0$ trace is a valid shortcut only after energy matching and a local bridge to the raw resolved microcanonical sequence; a still coarser symmetry-sector trace requires one additional bounded-local equivalence check.

\subsection{From finite-size estimators to the thermodynamic criterion}
\label{ssec: appendix_thermodynamic_witness_details}

The finite-size estimators above enter the thermodynamic claim through three independent controls.

\subsubsection{Bulk window and local mismatch}
Use the positive-entropy, shrinking-energy-density raw windows of
Eq.~\eqref{eq: main_microcanonical_window_scaling}.  A normalized positive
witness dark on the caged sequence must obey
\begin{equation}
    \tau_*^{\rm raw}
    =\liminf_{L\to\infty}
      \operatorname{Tr}(\rho_L^{{\rm mc,raw}}\widehat Q_R)>0.
    \label{eq: appendix_positive_witness_activity}
\end{equation}
For $0\le\widehat Q_R\le\widehat{\mathbf1}$, Eq.~\eqref{eq: appendix_general_raw_clean_bound} reduces to
\begin{equation}
    \left|\operatorname{Tr}(\rho_L^{{\rm mc,raw}}\widehat Q_R)
      -\operatorname{Tr}(\rho_L^{{\rm mc,clean}}\widehat Q_R)\right|
    \le f_{{\rm jd},L}.
    \label{eq: exceptional_fraction_activity_bound}
\end{equation}
Thus, cleaned values have the same limit only when $f_{{\rm jd}, L}\to0$.
Let $\rho_{R,L}^{\psi}$ and $\rho_{R,L}^{{\rm mc,raw}}$ be the reduced
density matrices on $R$ of the caged eigenstate and raw microcanonical state.
Their trace distance is $d_{\rm tr}(\rho,\sigma)=\tfrac12\|\rho-\sigma\|_1$,
where $\|\cdot\|_1$ is the trace norm.  When $\widehat Q_R$ has an ordinary
$R$-supported effect representative, the variational characterization of trace
distance gives
\begin{equation}
    \liminf_{L\to\infty}
    d_{\rm tr}(\rho_{R,L}^{\psi},\rho_{R,L}^{{\rm mc,raw}})
    \ge\tau_*^{\rm raw}.
    \label{eq: trace_distance_lower_bound}
\end{equation}
For constrained-space operators defined only by conditional blocks, the bounded-local expectation mismatch itself is the ETH statement.

\subsubsection{Raw background concentration}
Use the exact-energy-block outlier fraction of
Eq.~\eqref{eq: main_background_eth_concentration}.  Concentration on one
fixed region can be checked on a finite spanning Hermitian basis of its
symmetry-allowed projected local operator space; it does not by itself
establish concentration on larger regions.

If cleaned blocks are used numerically, Cauchy interlacing gives, for $\epsilon>d_L$ and $d_L=|\langle\widehat O_R\rangle_{\rm raw}-\langle\widehat O_R\rangle_{\rm clean}|$,
\begin{equation}
    f_{L,\epsilon}^{\rm raw}[\widehat O_R]
    \le 2f_{{\rm jd},L}
      +(1-f_{{\rm jd},L})
       f_{L,\epsilon-d_L}^{\rm clean}[\widehat O_R],
    \label{eq: appendix_raw_from_clean_concentration_bound}
\end{equation}
while $d_L\le2f_{{\rm jd},L}\|\widehat O_R\|$ follows from Eq.~\eqref{eq: appendix_general_raw_clean_bound}.  Hence, cleaned concentration plus $f_{{\rm jd}, L}\to0$ recovers raw concentration on the same region.  Level statistics are complementary and do not replace this local test.

The root-mean-square spread of the compressed values for one observable is
\begin{equation}
    w_L(\widehat O_R)^2
    =\frac{1}{N_{\rm win,L}}\sum_{E,j}
    \left(o_{E,j}^{(R)}-\langle\widehat O_R\rangle_{{\rm mc},L}\right)^2,
    \label{eq: appendix_local_observable_spread}
\end{equation}
where the sum includes all the directions in each exact-energy block of the
raw window.  Counting deviations larger than \(\epsilon\) gives
\(f_{L,\epsilon}[\widehat O_R]\le w_L(\widehat O_R)^2/\epsilon^2\).
For a fixed region \(R\), the raw concentration width is
\begin{equation}
    w_L^{\rm raw}=\max_{\|\widehat O_R\|_{\rm loc}=1}w_L(\widehat O_R),
    \label{eq: appendix_raw_concentration_width}
\end{equation}
where the maximum runs over the allowed Hermitian operator space on \(R\)
with the local normalization \(\|\cdot\|_{\rm loc}\) specified in each
model.  In a basis orthonormal in this metric, the squared spread is the
quadratic form of the covariance matrix in
Eq.~\eqref{eq: xy_block_invariant_covariance}, and \(w_L^{\rm raw}\) is the
square root of its largest eigenvalue.  A vanishing width therefore controls
concentration uniformly over the normalized operators on \(R\).

\subsubsection{Infinite-temperature shortcut}
The resolved trace in Eq.~\eqref{eq: appendix_beta_zero_resolved_trace}
supplies the thermal target only when its limiting energy density matches
the caged sequence and the local-equivalence condition
\eqref{eq: appendix_local_ensemble_equivalence} holds.  A coarser-sector
trace requires its additional bounded-local bridge.

\section{Technical criteria for stability under Hamiltonian deformations}
\label{appendix: deformation_stability_details}
\label{appendix: thermodynamic_continuation_stability}

Section~\ref{sec: deformation_stability} gives the physics-first
three-layer criterion: continue the caged eigenstate, continue one bounded local dark rule,
and then ask whether the same local mismatch survives thermodynamically.  Here
we collect the technical conditions behind those three steps.  We first treat
the fixed-support tangent problem, including complex-Hermitian coordinates,
nonlinear continuation, and degenerate caged subspaces.  We then continue the
local operator jointly with the state and close by stating the uniform thermal
background conditions needed for the ICQMBS upgrade.  Throughout, a
thermodynamic deformation means one size-compatible Hermitian local family;
the non-Hermitian remarks below concern only the finite-size algebraic scope.

\subsection{From tangent compatibility to a finite continuation of a caged eigenstate}
\label{ssec: appendix_exact_cage_continuation}
\label{ssec: technical_deformation_maps}

The first question is finite dimensional: after the Hamiltonian changes,
can the amplitudes and energy readjust so that the same support remains an exact
caged eigenstate?  The Jacobian answers this only to first order.  We therefore separate the
tangent test from the extra chart, regularity, and size-compatibility conditions
needed for a finite physical family.

\subsubsection{Scope and size-independent deformation neighborhoods}

The continuation coordinates presuppose one fixed microscopic Hilbert space and
constraint structure.  If the constraint itself changes, the allowed
configuration basis and conditional local projectors change, and a new residual
map must be constructed rather than differentiated through the present chart.

Algebraically, a right caged eigenvector is insensitive to an independent
incoming block and to
$\delta H_{\overline{\mathcal S}\overline{\mathcal S}}$ in the general
non-Hermitian extension of Eq.~\eqref{eq: undeformed_cage_block_stability}, as
long as $\delta H_{\mathcal S\mathcal S}=\delta B_{\mathcal S}=0$.
This freedom may even be nonlocal or non-Hermitian, but it is not the physical
ETH deformation considered here.  Our thermodynamic deformation family is
Hermitian; a genuinely non-Hermitian continuation would additionally require
complex energies, left--right eigenvectors, and exceptional-point control.

Let the chosen Hermitian perturbation family be
$\delta\widehat H_L=\sum_a\delta g_a\widehat W_{a,L}$.  Locality is not
required for the finite-size linear-algebra test itself.  It enters when the
finite-size tangent is promoted to a physical thermodynamic deformation.  For
finitely many uniform couplings, any fixed normalized coupling-space norm
defines a size-independent neighborhood.  If instead
$\delta\widehat H_L=\sum_X\delta\widehat h_X$ contains an extensive set of
independently perturbed bounded local terms, where $X$ denotes the support of
one term, a convenient local interaction norm on the finite geometry
$\Omega_L$ is
\begin{equation}
    \|\delta\widehat H_L\|_{\rm loc}
    =\sup_{x\in\Omega_L}\sum_{X\ni x}\|\delta\widehat h_X\|.
    \label{eq: appendix_local_deformation_norm}
\end{equation}
Here $\|\cdot\|$ is the operator norm, and the sum refers to the declared
microscopic interaction decomposition.  For a finite-range interaction with
bounded local coordination, this norm keeps an $O(1)$ change of every local
term inside an $O(1)$ neighborhood as the system grows.  Measuring
$\boldsymbol g$ from the reference Hamiltonian, we denote either the
finite-dimensional coupling norm or the induced local interaction norm by
$\|\boldsymbol g\|_{\rm def}$ and write
\begin{equation}
    \mathcal U_L(g_0)
    =\{\boldsymbol g:\|\boldsymbol g\|_{\rm def}\le g_0\}.
    \label{eq: appendix_uniform_deformation_neighborhood}
\end{equation}
The thermodynamic stability criterion requires a radius $g_0$ that is
independent of $L$.

\subsubsection{Real coordinates for complex-Hermitian families}

Complex amplitudes do not make the physical tangent problem complex: a
Hermitian Hamiltonian is still varied by independent real couplings.  For fixed
configuration support $\mathcal S$, start from the residual map
$\mathcal F_L$ in Eq.~\eqref{eq: full_cage_equation_map} and take
\begin{equation}
    \boldsymbol x\in\mathbb C^{|\mathcal S|},
    \qquad E\in\mathbb R,
    \qquad \boldsymbol g\in\mathbb R^{m_L},
    \label{eq: appendix_real_complex_parameter_spaces}
\end{equation}
where the real coordinates $\boldsymbol g$ parametrize Hermitian local
Hamiltonians.  Any coordinate chart used for a rank or codimension statement
must be nonsingular at the reference point.  Thus a nonzero complex hopping
$t=J e^{i\phi}$ may be represented locally by $(J,\phi)$, whereas at a
reference point with $J=0$ one should instead use Cartesian coordinates
$(\operatorname{Re}t,\operatorname{Im}t)$, equivalently independent Hermitian
cosine and sine directions.  If $q$ is the number of complex components of the residual, then
$\mathcal F_L\in\mathbb C^q$ is regarded as the real map
\begin{equation}
    \mathcal F_L^{\mathbb R}
    =\begin{pmatrix}\operatorname{Re}\mathcal F_L\\
                     \operatorname{Im}\mathcal F_L\end{pmatrix}
    :\mathbb R^{2|\mathcal S|-2}\times\mathbb R\times\mathbb R^{m_L}
      \longrightarrow\mathbb R^{2q},
    \label{eq: appendix_realified_residual_map}
\end{equation}
after normalization and global phase are removed.  Equivalently, one may retain compact complex notation while using the real inner
product
\begin{equation}
    (\boldsymbol u,\boldsymbol v)_{\mathbb R}
    =\operatorname{Re}\langle\boldsymbol u|\boldsymbol v\rangle.
    \label{eq: appendix_real_inner_product}
\end{equation}
All kernels, orthogonal complements, ranks, and codimensions in this complex
extension are understood in this real sense.  Thus the unqualified rank used
in the main text becomes the physical real rank after realification.  In particular, $\mathcal J_{\rm cage}^\dagger$
denotes the adjoint for these underlying real inner products; the dagger is
retained only to avoid duplicating the main-text notation.

At a normalized reference state, choose
$\langle\boldsymbol x|\delta\boldsymbol x\rangle=0$ to remove the normalization
and phase directions.  This identifies the state variation with the real tangent space of the ray
$[\boldsymbol x]$ in the complex projective space
$\mathbb{CP}^{|\mathcal S|-1}$.
Differentiating the two components of $\mathcal F_L$ gives
\begin{align}
    (H_{\mathcal S\mathcal S}-E)\delta\boldsymbol x
      -\boldsymbol x\,\delta E
      &=-\delta H_{\mathcal S\mathcal S}\boldsymbol x,
      \label{eq: appendix_internal_linear_response}\\
    B_{\mathcal S}\,\delta\boldsymbol x
      &=-\delta B_{\mathcal S}\,\boldsymbol x.
      \label{eq: appendix_boundary_linear_response}
\end{align}
Stacking these equations gives the tangent-lift equation
\begin{equation}
    \mathcal J_{\rm cage}
    \begin{pmatrix}\delta\boldsymbol x\\ \delta E\end{pmatrix}
    =-D_{\boldsymbol g}\mathcal F_L\,\delta\boldsymbol g,
    \label{eq: cage_tangent_lift_equation}
\end{equation}
with the Jacobian $\mathcal J_{\rm cage}$ displayed in
Eq.~\eqref{eq: full_cage_jacobian}.  Its image is the residual subspace
that can be repaired by changing the state and energy.  The lower block alone
gives the intuitive necessary test:
$P_{\ker B_{\mathcal S}^\dagger}\delta B_{\mathcal S}\,\boldsymbol x$ cannot be canceled by changing
the support amplitudes.  The full problem is stricter because the same
$\delta\boldsymbol x$ must also satisfy
Eq.~\eqref{eq: appendix_internal_linear_response}.

A deformation direction is first-order compatible precisely when its generated
residual lies in the range of $\mathcal J_{\rm cage}$.  Equivalently, its
component orthogonal to that range must vanish.  In the main text this
projection is denoted by $P_\perp$, and the generated residual by
$\delta_{\boldsymbol g}\mathcal F_L$; explicitly,
$P_\perp=P_{\ker\mathcal J_{\rm cage}^{\dagger}}$ and
\begin{equation}
    \delta_{\boldsymbol g}\mathcal F_L
    =D_{\boldsymbol g}\mathcal F_L\,\delta\boldsymbol g
    =\begin{pmatrix}
       \delta H_{\mathcal S\mathcal S}\boldsymbol x\\
       \delta B_{\mathcal S}\boldsymbol x
     \end{pmatrix}.
    \label{eq: appendix_full_cage_parameter_derivative}
\end{equation}
Its repairable and unrepairable parts are
\begin{equation}
    D_{\boldsymbol g}\mathcal F_L\,\delta\boldsymbol g
    =P_{\operatorname{Im}\mathcal J_{\rm cage}}
      D_{\boldsymbol g}\mathcal F_L\,\delta\boldsymbol g
     +P_{\ker\mathcal J_{\rm cage}^{\dagger}}
      D_{\boldsymbol g}\mathcal F_L\,\delta\boldsymbol g,
    \label{eq: cage_residual_tangent_normal_decomposition}
\end{equation}
and the second term is exactly the obstruction in
Eq.~\eqref{eq: full_cage_obstruction}.  The orthogonal projector is only a
convenient representative of the cokernel map, which identifies residual
vectors differing by an amplitude-and-energy repair: replacing the residual vector
by any invertible linear recombination, or changing the positive-definite inner
product used to choose a complement, leaves the condition
$D_{\boldsymbol g}\mathcal F_L\,\delta\boldsymbol g\in
\operatorname{Im}\mathcal J_{\rm cage}$ unchanged.  Hence the preserving
kernel and obstruction rank are independent of the residual-space metric and of
invertible residual-coordinate changes; a nonsingular reparameterization of the
coupling coordinates only changes the coordinate representation of the same
preserving subspace.  Individual singular values used in numerical conditioning
diagnostics are not invariant in this sense.
Notice that the two ranks play different roles: $\operatorname{rank}_{\mathbb R}\mathcal J_{\rm cage}$
gives the dimension of the repairable residual space, whereas
$c_{\rm cage}=\operatorname{rank}_{\mathbb R}\mathfrak O_{\rm cage}$ counts
the independent \emph{real} first-order tuning conditions within the declared
Hamiltonian perturbation family.  A complex residual direction can therefore
supply two independent real constraints when its real and imaginary parts are
independently accessible by physical coupling variations.

\subsubsection{From a finite-size nullspace to a physical deformation family}

\paragraph*{Finite-size preserving directions.}
For a physical local search,
choose an independent real basis
$\{\widehat W_a\}_{a=1}^{m}$ of Hermitian perturbations allowed by the
declared real-space range, microscopic constraint, and exact symmetries that
are to be retained.  If $\{\boldsymbol u_\mu\}$ is a real-orthonormal basis
of $\ker\mathcal J_{\rm cage}^{\dagger}$ and $\boldsymbol e_a$ is the
unit vector selecting the perturbation $\widehat W_a$, the obstruction
matrix is
\begin{equation}
    [\mathfrak O_{\rm cage}]_{\mu a}
    =\bigl(\boldsymbol u_\mu,
      D_{\boldsymbol g}\mathcal F_L\,\boldsymbol e_a\bigr)_{\mathbb R},
    \qquad
    \delta\boldsymbol g\in\ker\mathfrak O_{\rm cage},
    \label{eq: constructive_cage_preserving_kernel}
\end{equation}
with $(\cdot,\cdot)_{\mathbb R}$ from
Eq.~\eqref{eq: appendix_real_inner_product}.  Its real nullspace returns all
first-order preserving combinations in that deformation basis.  This converts
stability from a test of a guessed perturbation into a linear nullspace search.
Repeating the calculation after increasing the allowed real-space range gives
a progressive local-deformation search.

\paragraph*{Size compatibility.}
A finite-size nullspace becomes a thermodynamic deformation direction only if
the perturbations are inherited from one microscopic rule as the system grows.
Concretely, take one finite-range
microscopic interaction rule and let $\delta\widehat H_L$ denote its restriction
to the finite geometry $\Omega_L$.  We call the sequence size compatible when
its interaction range is bounded, $\sup_L\|\delta\widehat H_L\|_{\rm loc}<\infty$,
and the local coefficients are inherited from the same rule rather than chosen
independently after solving each finite-size nullspace.  First-order compatibility of the caged eigenstate
 along the sequence then requires
\begin{equation}
    \delta\widehat H_L\in\ker\mathfrak O_{{\rm cage},L}
    \qquad\text{for all sufficiently large }L,
    \label{eq: appendix_size_compatible_cage_direction}
\end{equation}
where membership means that the coefficient vector of the restricted
interaction lies in the corresponding finite-size kernel.  For a uniform
finite-dimensional coupling family this reduces to following the same coupling
direction at every size.  For an extensive inhomogeneous deformation basis it is a
nontrivial compatibility condition and prevents an arbitrary sequence of
finite-size null vectors from being interpreted as one physical deformation.

Because $\mathcal S$ is the exact nonzero support of the reference caged eigenstate, a
fixed-support chart is valid only while every tracked support amplitude remains
nonzero.  Continuity guarantees such a neighborhood around the reference
point, but if a component crosses zero, the support chart has reached its
boundary and the residual map should be rebuilt with the reorganized support
rather than continued through that crossing with the old Jacobian.

The same distinction applies beyond first order.  A constant-rank theorem at
each fixed $L$ may provide a local continuation radius $g_{0,L}>0$, but these
radii can shrink to zero with system size.  Robustness on an open thermodynamic neighborhood within the declared family
requires a common lower bound $\inf_{L\ge L_0}g_{0,L}>0$, or an explicit exact
family known to exist on such a common neighborhood.  Likewise, when
$c_{\rm cage}(L)>0$, a tuned preserving manifold has thermodynamic meaning only
if the finite-size tuned manifolds arise from one size-compatible microscopic
parameterization.

A useful restricted slice keeps the state vector itself fixed.  Setting
$\delta\boldsymbol x=0$ and allowing only an energy shift gives
\begin{equation}
    \mathcal A_{\psi}(\delta\widehat H)
    =\bigl(\widehat{\mathbf 1}-\ket\psi\!\bra\psi\bigr)
      \delta\widehat H\ket\psi,
    \qquad
    \delta\widehat H\in\ker\mathcal A_{\psi}.
    \label{eq: appendix_fixed_state_preserving_slice}
\end{equation}
Thus $\ker\mathcal A_{\psi}$ is exactly the fixed-state-preserving slice of
the general construction, up to an energy shift.  It is generally more
restrictive than $\ker\mathfrak O_{\rm cage}$ because it forbids the caged eigenstate
amplitudes from readjusting.

A zero obstruction rank means that no first-order tuning is required within
the declared coupling coordinates, but it does not exclude higher-order
obstructions.  More generally, the existence of a tangent lift at one point is
not yet an exact straight-line continuation in coupling space.  At each fixed
size, promoting tangent data to a nearby solution family requires smoothness and
the appropriate constant-rank hypotheses for the normalized full map
$\mathcal F_L$.  When its zero set is a smooth manifold, every admitted tangent
vector integrates to a local curve on that manifold; obtaining an open family
parameterized by freely chosen couplings additionally requires the projection
from the solution manifold to those coupling coordinates to have the required
local rank.  Standard operator and matrix perturbation theory gives the broader
continuation context \cite{kato1995,stewart_sun1990}.  Numerically, a regular
solution permits predictor--corrector continuation: take a small tangent step,
correct back to $\mathcal F_L=0$, recompute the tangent space, and iterate.  The
size-compatible and uniform-radius requirements above are what promote such
finite-size continuations to a thermodynamic statement.

\subsubsection{Degenerate caged subspaces}

For a degenerate caged eigenspace, an individual eigenvector is not a basis-independent
continuation object.  Let
$\widehat P_{\rm cage}(\boldsymbol g)$ be a smooth rank-preserving
projector whose range remains inside the declared fixed-support subspace.  The
existence of such a smooth projector is an additional continuation hypothesis:
a spectrally isolated cluster provides one by standard perturbation theory,
whereas at an unresolved level crossing it must be selected by an exact
symmetry, an analytic construction, or another basis-independent rule.  Then
\begin{equation}
    [1-\widehat P_{\rm cage}]\widehat H(\boldsymbol g)\widehat P_{\rm cage}=0
    \label{eq: appendix_invariant_caged_manifold_residual}
\end{equation}
preserves an invariant caged subspace.  Preserving an exactly degenerate
eigenspace at one energy additionally requires
\begin{equation}
    \widehat P_{\rm cage}\widehat H(\boldsymbol g)\widehat P_{\rm cage}
    =E(\boldsymbol g)\widehat P_{\rm cage}.
    \label{eq: appendix_degenerate_caged_eigenspace_condition}
\end{equation}
This distinction is relevant whenever a finite-size caged eigenstate is embedded in a
degenerate eigenspace rather than followed as an isolated eigenvector.

\subsection{Joint continuation of the bounded local dark rule}
\label{ssec: appendix_joint_local_channel_continuation}

Preserving the eigenpair is only the first layer.  The same deformation
must also preserve one bounded microscopic rule that remains dark in every
relevant environment.  The state and that local rule are therefore continued
with one common correction.

\subsubsection{Fixed environment and local-operator charts}

Appendix~\ref{ssec: appendix_conditional_blocks} gives the fixed-Hamiltonian
locality criterion in a constrained Hilbert space.  Under deformation, the same
condition must hold throughout one common coupling neighborhood.  Let $R$ be a
bounded region containing the tracked single-row or finite multi-row process.
Choose once and for all a finite-volume index set $\mathcal I_R$ containing
every outside environment that can acquire nonzero conditional amplitude
anywhere in the declared deformation neighborhood, and resolve the continued
state as
\begin{equation}
    \ket{\psi(\boldsymbol g)}
    =\sum_{\alpha\in\mathcal I_R}
      \ket\alpha_{\overline R}\otimes
      \ket{\psi_{R|\alpha}(\boldsymbol g)}_R.
    \label{eq: appendix_branchwise_environment_resolution}
\end{equation}
The unnormalized conditional vector $\ket{\psi_{R|\alpha}}$ is allowed to vanish
at the reference point.  Using a fixed index set is essential for a smooth
tangent problem: an environment absent at $\boldsymbol g=0$ can become occupied
at first order and must then be constrained by the same local microscopic rule.

For the kinetic or shifted-potential constructions used in this work, let
$\widehat L_R^{\rm phys}(\boldsymbol g,\boldsymbol\ell)$ denote one declared
local-operator chart applied to the deformed Hamiltonian.  The auxiliary
coordinates $\boldsymbol\ell\in\mathbb R^p$ are absent for a canonical single
boundary row or shifted potential.  For a bounded multi-row kinetic
combination, they form a finite real coordinate chart for the normalized local
coefficients, with real and imaginary parts included separately when needed.
Their dimension and physical support are fixed independently of system size.
The superscript $\rm phys$ emphasizes the microscopic operator; its restriction
to the constrained physical Hilbert space is the main-text operator.  Its total
first-order response is
\begin{equation}
    \delta\widehat L_R^{\rm phys}
    =\sum_a\delta g_a\,
      \partial_{g_a}\widehat L_R^{\rm phys}
     +\sum_b\delta\ell_b\,
      \partial_{\ell_b}\widehat L_R^{\rm phys},
    \label{eq: appendix_local_channel_parameter_derivative}
\end{equation}
with the second sum absent when the microscopic construction fixes the local
operator uniquely.  The coefficients $\boldsymbol\ell$ are auxiliary local
repair variables, not Hamiltonian couplings: they must be solved jointly with
the continued state inside one smooth, size-compatible chart and cannot be
chosen independently after each finite-size eigenstate is known.
After removing the overall operator scale, choose independent
$\boldsymbol\ell$ coordinates in the fixed normalization convention of
Appendix~\ref{ssec: appendix_caging_operator_normalization}, smoothly on the
coupling neighborhood.  A different smooth nonzero representative
$\widehat L'_R=a(\boldsymbol g,\boldsymbol\ell)\widehat L_R$
changes the derivative by
\begin{equation}
    \delta\widehat L'_R\ket\psi
    =a\,\delta\widehat L_R\ket\psi
     +(\delta a)\widehat L_R\ket\psi
    =a\,\delta\widehat L_R\ket\psi,
    \label{eq: appendix_rescaling_invariant_linearized_darkness}
\end{equation}
because the reference state is dark.  Hence the linearized annihilation
condition and the joint preserving directions are unchanged by a smooth
nonzero rescaling.  Before imposing the fixed normalization, the rescaling
multiplies the witness and its microcanonical value by $|a|^2$.
The operator chart is singular at $a=0$.

For a single kinetic boundary row,
$\delta\widehat A_h=\ket h\!\bra{\delta v_h}$, where
$\ket{\delta v_h}$ is read directly from the local boundary row of
$\delta H$.  For a Hermitian kinetic construction,
$\delta\widehat Z_R=\delta\widehat A_R+
\delta\widehat A_R^{\dagger}$; for the shifted-potential route,
$\delta\widehat Y_R=\delta\widehat V_R-
\delta v_R\widehat I_R$.  Thus the canonical single-row and shifted-potential
operators change directly with the same microscopic coupling variation.  A
multi-row operator may also move in its finite local coefficient chart, but that
motion is part of the joint solve rather than a post-hoc refit to the continued
state.  The corresponding positive witness follows as
\begin{equation}
    \widehat Q_R(\boldsymbol g)
    =\widehat L_R^{\dagger}(\boldsymbol g)\widehat L_R(\boldsymbol g),
    \qquad
    \delta\widehat Q_R
    =\delta\widehat L_R^{\dagger}\widehat L_R
     +\widehat L_R^{\dagger}\delta\widehat L_R.
    \label{eq: appendix_deformed_positive_witness}
\end{equation}
The reference constructions are $\widehat A_R$, $\widehat Z_R$,
$\widehat Y_R$, or a declared bounded multi-row kinetic combination.

\subsubsection{Conditional-block darkness in constrained spaces}

For an outside environment $\alpha$, let $\Pi_{R|\alpha}$ denote the
projector onto the locally allowed states in $R$.  The conditional action of
the microscopic operator is
\begin{equation}
    \widehat L_R^{(\alpha)}(\boldsymbol g,\boldsymbol\ell)
    =\Pi_{R|\alpha}
      \widehat L_R^{\rm phys}(\boldsymbol g,\boldsymbol\ell)
      \Pi_{R|\alpha}.
    \label{eq: constrained_local_block_restriction}
\end{equation}
Continuation as one physical local operator requires
\begin{equation}
    \widehat L_R^{(\alpha)}(\boldsymbol g,\boldsymbol\ell)
    \ket{\psi_{R|\alpha}(\boldsymbol g)}=0
    \qquad\forall\,\alpha\in\mathcal I_R.
    \label{eq: appendix_joint_environment_darkness}
\end{equation}
For an environment whose conditional vector vanishes, this equation is
trivial at that parameter point; retaining the block in the fixed residual is
what imposes the correct first-order condition if that environment begins to
carry amplitude.  This test follows a fixed bounded region $R$ and a fixed microscopic
construction of $\widehat L_R$.  Failure on this chart does not exclude a
family with reorganized configuration support, a different bounded region, or
another physically motivated local caging operator.
At the reference point, this condition also gives a simple exact parent
construction: if $\widehat Q_R=\widehat L_R^{\dagger}\widehat L_R$ annihilates
the caged eigenstate, then
\begin{equation}
    \widehat H(\lambda)=\widehat H+\lambda\widehat Q_R
    \label{eq: witness_generated_exact_deformation}
\end{equation}
retains that eigenstate for every real $\lambda$.  This proves the existence
of bounded local eigenstate-preserving perturbations, but such a state-adapted
term need not belong to the microscopic Hamiltonian class under study and does
not by itself establish thermal stability.

\subsubsection{Joint tangent equation and obstruction}

The same state correction must solve the caging equation and every conditional
local-annihilation equation.  Collect these conditions in a joint residual with
one local-annihilation block for every $\alpha\in\mathcal I_R$,
\begin{equation}
    \mathcal F_L^{\rm joint}
    =\begin{pmatrix}
       \mathcal F_L(\boldsymbol x,E,\boldsymbol g)\\
       \{\widehat L_R^{(\alpha)}(\boldsymbol g,\boldsymbol\ell)
          \ket{\psi_{R|\alpha}(\boldsymbol x,\boldsymbol g)}\}_{\alpha}
     \end{pmatrix}=0.
    \label{eq: augmented_sectorwise_continuation_map}
\end{equation}
At first order this gives
\begin{equation}
    \delta\widehat L_R^{(\alpha)}\ket{\psi_{R|\alpha}}
    +\widehat L_R^{(\alpha)}\ket{\delta\psi_{R|\alpha}}=0
    \qquad\forall\alpha,
    \label{eq: appendix_joint_linearized_darkness}
\end{equation}
where, for a fixed microscopic constraint,
$\delta\widehat L_R^{(\alpha)}=\Pi_{R|\alpha}
(\delta\widehat L_R^{\rm phys})\Pi_{R|\alpha}$.  The first term is the
operator response to the Hamiltonian deformation; the second is the response
of the same conditional state amplitudes that enter the cage Jacobian,
with every $\delta\psi_{R|\alpha}$ generated by the same
$\delta\boldsymbol x$ that solves the continuation problem for the caged eigenstate.
Checking the caged eigenstate and local operator independently with unrelated state
corrections would therefore be insufficient.

The same projection used for the caged eigenstate alone now yields the constructive local
filter.  Define
\begin{equation}
    \mathcal J_{\rm joint}
    =D_{(\boldsymbol x,E,\boldsymbol\ell)}\mathcal F_L^{\rm joint},
    \qquad
    \mathfrak O_{\rm joint}
    =P_{\ker\mathcal J_{\rm joint}^{\dagger}}
      D_{\boldsymbol g}\mathcal F_L^{\rm joint}.
    \label{eq: appendix_joint_obstruction_map}
\end{equation}
The cokernel and coordinate-invariance argument of
Appendix~\ref{ssec: technical_deformation_maps} also applies here, including
nonsingular normalized $\boldsymbol\ell$ coordinates: preserving kernels
and real codimensions are invariant, while numerical singular values remain
chart-dependent conditioning diagnostics.  Thus
\begin{equation}
    \boxed{\delta\boldsymbol g\in\ker\mathfrak O_{\rm joint}}
    \label{eq: appendix_joint_preserving_kernel}
\end{equation}
selects all first-order directions in the declared deformation basis for which
one joint correction $(\delta\boldsymbol x,\delta E,\delta\boldsymbol\ell)$
continues both the exact caged eigenstate and the chosen local caging operator.  If $\ker\mathfrak O_{\rm cage}$ is larger, the discarded directions
preserve the eigenstate but not that local explanation.  At regular points the
same predictor--corrector continuation described above can be applied to
$\mathcal F_L^{\rm joint}$ to construct a finite preserving path rather than
only its tangent space.

\subsubsection{Uniform locality along the thermodynamic sequence}

For a thermodynamic local deformation, the same size-compatible microscopic
perturbation rule introduced in Eq.~\eqref{eq: appendix_size_compatible_cage_direction}
must in fact satisfy the stronger condition
$\delta\widehat H_L\in\ker\mathfrak O_{{\rm joint},L}$ for all sufficiently
large $L$.  A finite-size joint solution becomes a thermodynamic local statement only when
the physical diameter of $R$ is bounded uniformly for all sufficiently large
systems and all couplings in the declared neighborhood, and any auxiliary
coefficients $\boldsymbol\ell_L(\boldsymbol g)$ arise from the same bounded
local chart with a size-independent normalization rather than from unrelated
finite-size fits.  We do not introduce a
minimum-radius diagnostic: the present claims require only existence of one
such uniformly bounded construction.  In particular, finite-size collective
cancellation does not decide whether the exceptional square-QDM representative
admits a bounded multi-row kinetic reduction.

\subsubsection{Uniform thermal-background upgrade}
\label{ssec: appendix_uniform_thermal_background}

For the ICQMBS upgrade, the same coupling neighborhood must also satisfy the
thermal-background requirements of Definition~\ref{def: icqmbs} uniformly.
Writing $N_{\rm win,L}(\boldsymbol g)$ and
$N_{\rm cage,L}(\boldsymbol g)$ for the raw-window and declared caged-subspace
dimensions, and $f_{L,\epsilon}[O_R;\boldsymbol g]$ for the concentration
fraction of Eq.~\eqref{eq: main_background_eth_concentration}, the conditions
are
\begin{equation}
    \begin{aligned}
    \sup_{\boldsymbol g\in\mathcal U_L}
      \frac{\delta E_L(\boldsymbol g)}{|\Omega_L|}&\longrightarrow0,\\
    \liminf_{L\to\infty}\inf_{\boldsymbol g\in\mathcal U_L}
      \frac{\log N_{\rm win,L}(\boldsymbol g)}{|\Omega_L|}&>0,\\
    \sup_{\boldsymbol g\in\mathcal U_L}
      \frac{N_{\rm cage,L}(\boldsymbol g)}{N_{\rm win,L}(\boldsymbol g)}
      &\longrightarrow0,\\
    \sup_{\boldsymbol g\in\mathcal U_L}
      f_{L,\epsilon}[O_R;\boldsymbol g]&\longrightarrow0.
    \end{aligned}
    \label{eq: appendix_uniform_deformed_background_conditions}
\end{equation}
The last limit is required for every fixed bounded region $R$, bounded
Hermitian $O_R$, and $\epsilon>0$.  These are the deformation-wide versions
of the four background conditions already imposed at a single Hamiltonian; the
uniformity in $\boldsymbol g$ is what upgrades a pointwise ICQMBS statement to
an open-family one.

\section{Search implementation and finite-size certification}
\label{appendix: supplementary_search_methods}

This appendix collects only the implementation details needed to reproduce the
search and finite-size certification of Sec.~\ref{sec: practical_search}.
Candidate labels and graph reductions may simplify the intermediate problem,
but the complete caged subspace is identified before representative
localization and every reported state is finally certified in the full resolved
sector.

\subsection{Declaring the search space}
\label{ssec: search_scope_basis}
\label{ssec: fictitious-particle}
\label{sssec: sym_basis}
\label{sec: labeling}

The search starts from a constrained real-space product basis
\(\mathcal B=\{\ket s\}\).  Conserved quantities that are diagonal in this
basis, such as total magnetization or winding number, can be imposed before the
search matrices are built.  A symmetry that instead acts through coherent
superpositions is resolved only after the search solutions have been collected
into a caged subspace invariant under that symmetry.  If the symmetry maps one
searched shell or seed into a distinct symmetry-related shell, the full
degenerate orbit is collected first and the symmetry is diagonalized only
within the resulting invariant caged subspace.  Thus post-solve symmetry
resolution never assigns a quantum number to a subspace that is not closed
under the symmetry.  Reordering or integer-encoding the product configurations
changes only their coordinate representation, not ranks, nullities, spectra,
or physical states.

The candidate class must be fixed before solving.  Within one declared shell
and numerical rank convention, a complete internal-eigenspace and
boundary-kernel calculation is exhaustive for states supported in that shell.
It says nothing about supports crossing different potential values, grading
classes omitted by the chiral ansatz, or candidate sets not generated by the
chosen labels.  A shell that exists only on a coupling-tuning locus is equally
valid for this fixed-parameter search, but its tuned character must be recorded
and carries no continuation claim.

Candidate generation, finite-size certification, the thermodynamic ETH test,
and deformation stability remain separate layers.  The shell solves below
address only the first; the residual and reporting conventions of
Sec.~\ref{ssec: finite_size_search_certification} provide the second.

\subsection{Chiral-shell kernel solve}
\label{ssec: type1_search_implementation}

For a declared chiral shell $\mathcal B_{c,v}$, Sec.~\ref{sec: practical_search} reduces the candidate problem to a boundary nullspace.  Two implementation details are useful.

First, when the diagonal potential is parameterized as
\(\widehat V(\boldsymbol\lambda)=\sum_\mu
\lambda_\mu\widehat N_\mu\), shell membership should be tracked through the
configuration feature vector \(\boldsymbol n(s)\), not inferred only from
rounded floating-point energies.  Identical feature vectors define structural
degeneracies of the declared coupling family; distinct vectors that coincide
only at the chosen \(\boldsymbol\lambda\) define tuned shell mergers.  If shell
grouping is performed numerically, its tolerance is part of the reported
search convention.

Second, for each nonempty \(\mathcal B_{c,v}\), the implementation constructs
the sparse boundary map \(B_{c,v}\) and computes its complete numerical or
exact nullspace according to the declared rank convention.  The
positive-semidefinite \emph{boundary-overlap matrix}
\(B_{c,v}^{\dagger}B_{c,v}\) can be useful for exposing independent groups of
leakage columns.  Its entries are weighted inner products,
\begin{equation}
    (B_{c,v}^{\dagger}B_{c,v})_{ij}
    =\sum_h (B_{c,v})_{hi}^{*}(B_{c,v})_{hj}.
    \label{eq: boundary_overlap_weighted_inner_product}
\end{equation}
For signed or complex weights, this is not an unweighted shared-neighbor adjacency: shared boundary rows can cancel in the inner product.  Exact block diagonalization may be used computationally, but the original boundary equation remains authoritative.

The chiral kinetic block also gives a useful two-step representation.  With the
notation of Eq.~\eqref{eq: main_chiral_kinetic_block_form},
\begin{equation}
\begin{aligned}
    \ker K_{+-}^{\dagger}
    &=\ker(K_{+-}K_{+-}^{\dagger}),\\
    \ker K_{+-}
    &=\ker(K_{+-}^{\dagger}K_{+-}).
\end{aligned}
\label{eq: bipartite_projection_kernel_identity}
\end{equation}
These two-step matrices can expose same-grading structure; for weighted graphs, their diagonal entries are sums of squared hopping magnitudes rather than weighted degrees.  They do not replace the boundary-nullspace solve.

\subsection{When the internal kinetic block does not vanish}
\label{ssec: general_uniform_shell_search}
\label{ssec: type2_uniform_shell_search}

The chiral-shell reduction is special because the internal kinetic block
vanishes.  A uniform-potential shell can still support a caged eigenstate when internal
kinetic motion remains, but one must then solve the internal eigenproblem and
the boundary condition together.  Let
\begin{equation}
    \mathcal B_v
    \equiv
    \bigl\{|s\rangle\in\mathcal B:
    \langle s|\widehat V|s\rangle=v\bigr\}
\end{equation}
be such a shell, with projector \(\widehat P_v\).  Its internal kinetic block
and outgoing boundary map are
\begin{equation}
    K_v=\widehat P_v\widehat K\widehat P_v,
    \qquad
    B_v=\widehat P_{\partial\mathcal B_v}\widehat K\widehat P_v.
    \label{eq: general_internal_boundary_blocks}
\end{equation}
A shell-supported caged eigenstate with kinetic contribution \(\varepsilon\) lies in
\begin{equation}
    \ker B_v\cap\mathcal E_{\varepsilon}(K_v),
    \label{eq: general_eigenspace_intersection}
\end{equation}
where \(\mathcal E_{\varepsilon}(K_v)\) denotes the complete kinetic eigenspace.
If the columns of \(U_{\varepsilon}\) span that eigenspace, the remaining solve
is
\begin{equation}
    B_vU_{\varepsilon}\boldsymbol y=0.
    \label{eq: general_eigenspace_boundary_solve}
\end{equation}

Using the complete eigenspace is essential.  Testing an arbitrary numerical
eigenbasis vector by vector can hide a nonleaking linear combination or make a
basis-dependent cancellation look intrinsic to one vector.  In exact
arithmetic, \(\mathcal E_{\varepsilon}(K_v)\) is the exact spectral eigenspace.
In floating point, nearby eigenvalues are grouped with a declared tolerance
\(\tau_E\).  This grouping is a numerical candidate convention, not a claim of
exact degeneracy, so \(\tau_E\) and the adjacent spectral separation should be
reported.  If distinct kinetic eigenvalues are accidentally merged, the
boundary solve can return a dark vector in their combined span that is not an
eigenvector of \(K_v\); the independent internal residual of
Sec.~\ref{ssec: finite_size_search_certification} then rejects the candidate.

The solve is exhaustive within the declared shell only when the complete
kinetic eigenspaces are correctly grouped and their intersections with
\(\ker B_v\) are computed completely.  Neither this search nor the chiral-shell
search is exhaustive for supports spanning more than one potential value; in
that case the full internal equation
\(H_{\mathcal S\mathcal S}\boldsymbol x=E\boldsymbol x\) must be solved without
the shell reduction.  The five-site spin-one caged eigenstate of
Appendix~\ref{ssec: spin1_xy_five_site_nonchiral_cage} provides a simple example
with a nonempty internal graph.

\subsection{Subspace first, representative second}
\label{ssec: degenerate_cage_localization}

A search generally returns a caged subspace rather than a preferred set of
vectors.  Let the columns of \(U\) form an orthonormal basis of that certified
subspace.  Existence and dimension are properties of the complete subspace and
must be decided before inspecting whichever basis a nullspace or eigensolver
happens to return.

One can choose sparse representatives afterward for interpretation.  For
\(\ket{\psi(\boldsymbol y)}=U\boldsymbol y\), define the product-basis inverse
participation ratio
\begin{equation}
    \operatorname{IPR}(\boldsymbol y)
    =\sum_{s\in\mathcal B}
    \left|\langle s|U\boldsymbol y\rangle\right|^4,
    \qquad \|\boldsymbol y\|_2=1.
    \label{eq: cage_subspace_ipr}
\end{equation}
Maximizing it, optionally with orthogonality constraints to earlier choices,
favors vectors supported on fewer product configurations.  This postprocessing
neither changes the caged subspace nor tests its existence.  Different bases
of the same subspace can have different IPRs, apparent supports, and
cancellation patterns, and a large product-basis IPR does not establish
real-space localization or a bounded local caging operator.

Local-operator analysis follows the same ordering: a tested bounded operator family may select only part of a certified caged subspace, and failure within that family does not exclude every bounded multi-row reduction.

\subsection{From candidate to finite-size certificate}
\label{ssec: finite_size_search_certification}

Candidate shells, disconnected search blocks, or symmetry-related seeds can
produce repeated vectors.  Before anything is reported as a caged subspace,
embedded solutions at the same energy are concatenated and reduced to an
orthonormal basis with a rank-revealing factorization.  This removes duplicates
and linear dependencies without selecting a preferred sparse basis.

\paragraph*{Support convention.}
For an analytic or exact-arithmetic vector, the actual support
\(\mathcal S=\{s:x_s\neq0\}\) is unambiguous.  In floating-point calculations
we instead report the thresholded \emph{numerical support}
\begin{equation}
    \mathcal S_{\tau_{\rm sup}}
    \equiv
    \{s:|x_s|>\tau_{\rm sup}\},
    \label{eq: numerical_support_threshold}
\end{equation}
where \(\tau_{\rm sup}\) is independent of the eigenspace-grouping, rank, and
residual tolerances.  A physical conclusion that relies on an unusually sparse
numerical representative should be stable under reasonable changes of
\(\tau_{\rm sup}\), unless the relevant zeros are established analytically.

\paragraph*{Three residuals.}
For a normalized candidate with reconstructed support \(\mathcal S\), we use
\begin{subequations}
\begin{align}
    r_{\mathrm{int}}
    &\equiv
    \left\|
    H_{\mathcal S\mathcal S}\boldsymbol x-E\boldsymbol x
    \right\|_2,
    \label{eq: search_internal_residual}\\
    r_{\partial}
    &\equiv
    \left\|B_{\mathcal S}\boldsymbol x\right\|_2,
    \label{eq: search_boundary_residual}
\end{align}
\end{subequations}
and independently embed the state in the full resolved sector to evaluate
\begin{equation}
    r_H
    \equiv
    \| (\widehat H-E)|\psi\rangle\|_2.
    \label{eq: search_full_residual}
\end{equation}
For a floating-point candidate, \(r_H\) is evaluated on the untruncated
normalized vector, not on a vector obtained by setting coefficients below
\(\tau_{\rm sup}\) to zero.  The displayed-support threshold therefore cannot
manufacture a smaller full-Hamiltonian residual.  Here \(\|\cdot\|_2\) is the
Hilbert-space norm, equivalently the Euclidean coefficient norm in the
orthonormal product basis.

To distinguish interference caging from a disconnected invariant component, we separately verify $B_{\mathcal S}\neq0$.  The three residuals test, respectively, the internal eigenvalue equation, active-boundary cancellation, and the independently embedded full-Hamiltonian eigenpair.

When \(\mathcal S\) and its complete active boundary contain the full action of
\(\widehat H-E\), the internal and boundary pieces are orthogonal and
\begin{equation}
    r_H^2=r_{\mathrm{int}}^2+r_{\partial}^2.
\end{equation}
A discrepancy beyond numerical rounding diagnoses an incomplete support or
boundary reconstruction, or an inconsistent full-sector embedding.

\paragraph*{Rank convention.}
Eigenspace grouping, rank/nullity, displayed support, and final residual
acceptance use separate numerical conventions.  Their values should be
reported together with the matrix scale, arithmetic precision, solver, and the
singular-value separation supporting each numerical rank decision.  A putative
nullity is not an exact mathematical statement when the smallest retained and
largest discarded singular values are unresolved relative to the declared
threshold.  Analytic or exact-arithmetic constructions can establish exact
vanishing; floating-point residuals certify only to their reported precision.

\paragraph*{What is reported.}
A finite-size certificate should include the candidate-shell labels, matrix
dimensions and nullity, certified caged-subspace dimension, exact support for
analytic or exact-arithmetic vectors or the declared
\(\mathcal S_{\tau_{\rm sup}}\) in floating point, active-boundary
nontriviality, residuals, and subsequent symmetry, sparse-representative,
local-operator, and witness diagnostics.  Such a certificate establishes a
finite-size caged eigenstate; by itself it does not establish a thermodynamic ICQMBS
classification or deformation stability.

\subsection{Graph reductions for candidate compression}
\label{ssec: graph_symmetry_candidate_reduction}

The shell searches already exploit a chiral grading of the kinetic graph and a
diagonal vertex coloring.  Additional weighted automorphisms or equitable
partitions can sometimes expose repeated boundary constraints and reduce the
intermediate candidate problem
\cite{godsil_royle_2001,brouwer_spectra_2012,francis_extensions_2017}.  For a
weighted kinetic matrix \(K\), a weighted automorphism is a permutation
\(P_\sigma\) satisfying
\begin{equation}
    P_\sigma^{\dagger}K P_\sigma=K,
    \label{eq: weighted_graph_automorphism}
\end{equation}
while a partition \(\{\mathcal B_1,\ldots,\mathcal B_m\}\) is equitable when
\begin{equation}
    \sum_{t\in \mathcal B_j}K_{st}=d_{ij},
    \qquad s\in \mathcal B_i,
    \label{eq: equitable_partition_definition}
\end{equation}
with \(d_{ij}\) independent of the chosen \(s\in \mathcal B_i\).  A physical symmetry
induces such a permutation only when it relabels the chosen product basis; a
general unitary symmetry may instead mix basis states.

These reductions are candidate-space aids, never certification criteria.
Orbit equivalence leaves boundary cancellation and the existence of a caged
eigenvector to be verified.  A quotient solution must be lifted to the full
product basis and pass the same boundary and full-Hamiltonian residual checks.
Conversely, an absence result in a quotient applies only to the modes retained
by that quotient unless completeness is proved.  In this work orbit information is
used mainly for interpretation; the complete boundary solve remains
authoritative.  The \(4\times4\) square-QDM example is discussed in
Appendix~\ref{appendix: square_qdm_cage_manifold}.

\section{Spin-1 \texorpdfstring{$XY$}{XY}: exact construction, ensembles, and deformation details}
\label{sec: spin1_xy_analytical_appendix}
\label{appendix: additional_spin1_xy_cages}

This appendix supplies the technical details behind
Sec.~\ref{sec: spin1_xy}.  We first prove that the declared chiral-shell
boundary kernel contains a unique tower member and that its local caging
operators are independent of the outside environment.  We then specialize the thermal
reference and concentration machinery of Appendix~\ref{ssec: appendix_energy_matching_and_canonical_role}
to the positive witnesses derived from the translated $A, Z, Y$ local caging
operators.  In particular, we prove the thermodynamic ICQMBS criteria on a
fluctuation-scale $O(L^{1/2})$ raw window and show that the bounds are uniform on an
open compatible deformation interval.  Narrower-window finite-size diagnostics,
supplementary deformation tests, and additional exact caged eigenstates are
then collected separately; the latter are scope tests only.

\subsection{Boundary-kernel uniqueness of the tower}
\label{ssec: xy_benchmark_appendix}
\label{ssec: xy_boundary_kernel_proof}

\begin{proposition}[Unique interior-shell boundary-kernel state]
\label{prop: xy_unique_boundary_kernel}
For OBC, and for even $L$ under PBC, the kernel of $B_n$ in
Eq.~\eqref{eq: xy_type1_boundary_kernel} is one-dimensional for $0<n<L$.
Up to normalization and an overall phase, it is generated by
Eq.~\eqref{eq: xy_unique_kernel_vector}.
\end{proposition}

\begin{proof}
A boundary configuration obtained from $\mathcal B_n^{\mathrm{XY}}$ contains one adjacent
$\ket{00}$ pair.  It has two parent configurations differing by the local
exchange $\ket{+-}\leftrightarrow\ket{-+}$.  Since their matrix elements in
Eq.~\eqref{eq: xy_local_kinetic_action} are equal, the corresponding boundary
row imposes Eq.~\eqref{eq: xy_adjacent_swap_relation}.  The graph of binary
strings with fixed Hamming weight $n$ is connected under adjacent exchanges,
so this relation fixes all amplitudes relative to one reference configuration.
An adjacent exchange changes $\sum_{r\in\Xi}r$ by an odd integer; the same is
true for the periodic bond only when $L$ is even.  The sign pattern in
Eq.~\eqref{eq: xy_unique_kernel_vector} therefore satisfies every boundary row
and is unique up to scale.
\end{proof}

This proves uniqueness only within the declared chiral-shell candidate set;
it does not classify caged eigenstates outside that search space.

\subsection{Environment-independent local blocks and normalization}
\label{ssec: xy_local_operator_blocks_appendix}

The main text uses the resulting bounded operators directly.  Here we
verify the missing locality step: the same two-site local caging operator works
in every
occupied outside environment.  Because the nearest-neighbor exchange sends the
active binary parents only to $\ket{00}$, it has no matrix element within the
candidate shell,
\begin{equation}
    \widehat P_{\mathcal B_n^{\mathrm{XY}}}\widehat K
    \widehat P_{\mathcal B_n^{\mathrm{XY}}}=0.
    \label{eq: xy_type1_internal_zero}
\end{equation}
Define the unnormalized parent vector
\begin{equation}
    \ket{\mathrm{sym}}=\ket{+-}+\ket{-+}.
    \label{eq: xy_local_parent_vector}
\end{equation}
The operators in Eqs.~\eqref{eq: xy_directed_q_witness} and
\eqref{eq: xy_reduced_iz} then obey
\begin{equation}
\begin{aligned}
    \widehat A_{R_r}^{\dagger}\widehat A_{R_r}
      &=J^2\ket{\mathrm{sym}}\!\bra{\mathrm{sym}},\\
    \widehat Z_{R_r}^{2}
      &=\widehat A_{R_r}^{\dagger}\widehat A_{R_r}
        +2J^2\ket{00}\!\bra{00}.
\end{aligned}
    \label{eq: xy_local_witness_decomposition_appendix}
\end{equation}
At fixed magnetization, every occupied conditional bond state is proportional
to $\ket{++}$, $\ket{--}$, or $\ket{+-}-\ket{-+}$, and
\begin{equation}
\begin{aligned}
    \widehat A_{R_r}\ket{++}
    &=\widehat A_{R_r}\ket{--}=0,\\
    \widehat A_{R_r}(\ket{+-}-\ket{-+})&=0.
\end{aligned}
    \label{eq: xy_complementary_sector_check}
\end{equation}
A tower member has no $\ket{00}$ component, so the reverse process also
vanishes, while every one-site conditional state lies in the kernel of
$\widehat Y_r$.  No outside projector is therefore required.  Dividing
$\widehat A_{R_r}$ and $\widehat Z_{R_r}$ by $\sqrt{2}J$ gives the normalized witnesses in the
main text, each satisfying $0\leq\widehat Q\leq\widehat{\mathbf1}$
on its bounded support.

\subsection{Thermal reference and thermodynamic closure}
\label{ssec: xy_thermal_ensembles_appendix}

This subsection isolates the analytic chain behind the thermodynamic claim.
We first specify the symmetry-resolved reference sector and verify that its
entropy density remains extensive.  We then derive the exact fixed-$M$ local
thermal targets, and finally prove that a still-subextensive raw
microcanonical window converges to those targets and concentrates for every
fixed bounded region.  The narrower numerical windows, joint-dark cleaning,
and finite-$\beta$ checks are intentionally deferred to
Appendix~\ref{ssec: xy_finite_size_thermal_diagnostics_appendix}; they test the
finite-size realization but do not carry the thermodynamic proof.

\subsubsection{Reference symmetry sector}
For the principal family we take even $L$ under PBC, set $h=D=0$, and resolve
the exact unitary symmetries common to the full compatible family: total
magnetization $M$ and translation momentum $k$.  Ordinary inversion is not
resolved at nonzero $\kappa$, because the imaginary second-neighbor exchange
maps $\kappa$ to $-\kappa$ instead of commuting with the Hamiltonian.  The compatible family is
$\widehat H_\kappa=\widehat K_1(J)+\widehat K_3(J_3^{\rm ref})+\widehat K_2(i\kappa)$
with $J_3^{\rm ref}/J=0.1$; the representative point is
$\kappa_\star/J=0.1$.  We keep fixed-$M$, resolved-$(M,k)$, and
microcanonical values distinct throughout.

The endpoint $\kappa=0$ has enhanced symmetry and is used only as a control.
Likewise, $L=6$ is pre-asymptotic for the thermodynamic sequence: the periodic
range-three bond is self-reversing there, so
Eq.~\eqref{eq: xy_range_d_exchange} counts that coupling with a special
small-ring multiplicity.  The thermodynamic finite-size sequence therefore
begins at $L=8$.

Although $\widehat K_2(i\kappa)$ breaks the unitary anticommutation with
$\widehat C_A$, the full family retains an antiunitary spectral reflection.
Complex conjugation $\mathcal K$ leaves the real odd-range terms invariant and
reverses the imaginary even-range term, while $\widehat C_A$ reverses the
odd-range terms and leaves the even-range term invariant.  Hence
\begin{equation}
    (\widehat C_A\mathcal K)\widehat H_\kappa
    (\widehat C_A\mathcal K)^{-1}=-\widehat H_\kappa.
    \label{eq: xy_antiunitary_spectral_reflection_appendix}
\end{equation}
For the selected $M=-2$ sequence, $e^{i\pi M}=1$, so translating
$\widehat C_A$ gives the equivalent $\widehat C_B=\widehat C_A$ operator
inside the fixed-$M$ sector.  Hence $\widehat C_A$ preserves translation
momentum there.  Complex conjugation maps $k$ to $-k$, which is the same sector
for the selected $k=0$ or $\pi$.  Thus $\widehat C_A\mathcal K$ acts inside
the resolved $(M,k)$ reference space and pairs its spectrum about zero.
Although the bare inversion and $\widehat C_A$ anticommutation are broken,
site-centered inversion $\widehat{\mathcal I}$, acting as $r\mapsto-r$ modulo
$L$, gives a commuting antiunitary $\widehat{\mathcal I}\mathcal K$ that
squares to $+1$.  Its product with $\widehat C_A\mathcal K$ gives a surviving unitary spectral
reflection,
\begin{equation}
\begin{aligned}
    \widehat\Gamma&=\widehat{\mathcal I}\widehat C_A,
    &\widehat\Gamma^2&=\widehat{\mathbf1},\\
    \widehat\Gamma\widehat H_\kappa\widehat\Gamma^{-1}
      &=-\widehat H_\kappa.
\end{aligned}
    \label{eq: xy_combined_unitary_chiral_symmetry}
\end{equation}
These operators act within the selected $(M,k)$ sector.  The chiral index
$|\operatorname{Tr}_{M,k}\widehat\Gamma|$ lower-bounds the number of zero
modes.  Character counting gives signed traces $22,-45,171$ at $L=8,10,12$;
the observed zero-block dimensions $22,45,171$ saturate this bound.
Because $\widehat\Gamma$ anticommutes with the Hamiltonian, it pairs the
nonzero-energy halves rather than defining an additional commuting symmetry
block within either half.

\paragraph*{Independent level-statistics check.}
As a diagnostic independent of the local diagonal-ETH criterion, we evaluate
the adjacent-gap ratio at the representative point $\kappa_\star/J=0.1$.
For consecutive nondegenerate energies, each positive gap is the difference
between neighboring energies; $r$ is the smaller divided by the larger of two
consecutive gaps.
After resolving $M=-2$ and the tower momentum, the square-one commuting
antiunitary $\widehat{\mathcal I}\mathcal K$ permits comparison with Gaussian orthogonal ensemble (GOE) bulk statistics away from the
central chiral block.  The operator $\widehat C_A\mathcal K$ is instead the spectral
reflection in Eq.~\eqref{eq: xy_antiunitary_spectral_reflection_appendix}, so
we analyze either nonzero-energy half separately, excluding the central
degenerate zero-energy block of dimensions $22,45,171$ at $L=8,10,12$.
The negative halves contain $37,316,2346$ adjacent-gap ratios and give
$\langle r\rangle=0.5477,0.5324,0.5347$, with bootstrap standard errors
$0.0381,0.0146,0.0052$, respectively.  The reflected positive halves give
the same means by symmetry and are not independent samples.  These values are consistent with
$\langle r\rangle_{\rm GOE}\simeq0.5307$ and well separated from the Poisson
value $2\ln2-1\simeq0.3863$ \cite{oganesyan2007,atas2013}.  This supports a
numerically nonintegrable background, but it is neither a proof of strong ETH
nor an input to the thermodynamic ICQMBS proof below.

\paragraph*{Resolved-sector entropy.}
The spatial symmetry resolution used above does not itself make the thermal
reference exponentially small.  Here $z$ is a formal counting variable:
$\ket{-}$, $\ket0$, and $\ket{+}$ contribute $z^{-1}$, $1$, and $z$,
respectively.  The notation $[z^M]$ extracts the coefficient of $z^M$,
counting product configurations with total magnetization $M$.
At fixed magnetization $M=-2$, the Hilbert-space dimension is
\begin{equation}
    N_{L,-2}
    =[z^{-2}](z^{-1}+1+z)^L
    =3^L\sqrt{\frac{3}{4\pi L}}
      e^{-3/L}\,[1+o(1)],
    \label{eq: xy_fixed_m_entropy_asymptotic}
\end{equation}
by the lattice local central-limit theorem.  For translation momentum $k$, the
character formula gives
\begin{equation}
    \dim\mathcal H_{M,k}
      =\frac{N_{L,M}}{L}+O(3^{L/2}),
    \label{eq: xy_translation_resolved_entropy_asymptotic}
\end{equation}
because every nonidentity translation fixes configurations on at most $L/2$
cycles.  Hence every nonempty momentum subsequence has entropy density
$\log 3$.  The selected tower member contains
$n=(L-2)/2$ staggered pair raises and has one-site translation eigenvalue
$(-1)^n$; it therefore lies in $k=\pi$ for $L=0\pmod 4$ and in $k=0$ for
$L=2\pmod 4$.  Both selected subsequences retain entropy density $\log 3$.
By itself, this full-sector count does not lower-bound the number of levels in
the narrow $\Delta E=(J/2)L^{1/4}$ numerical window.  The thermodynamic proof
below instead combines this count with an $O(L)$ energy-variance bound to show
that an $O(L^{1/2})$ raw window retains a size-independent positive fraction
of the resolved sector, which is sufficient to preserve its entropy density.

For compact notation, define the normalized positive witnesses
\begin{equation}
    \widehat Q^{A}_{R_r}
      =\frac{\widehat A_{R_r}^\dagger\widehat A_{R_r}}{2J^2},
    \qquad
    \widehat Q^{Z}_{R_r}
      =\frac{\widehat Z_{R_r}^2}{2J^2},
    \qquad
    \widehat Q^{Y}_r=\widehat Y_r^2.
    \label{eq: xy_normalized_witnesses_appendix}
\end{equation}
The reported triples $(\tau_A,\tau_Z,\tau_Y)$ are the corresponding
ensemble expectations.

\subsubsection{Exact fixed-\texorpdfstring{$M$}{M} thermal targets}
\label{ssec: xy_fixed_m_reference}

At $D=0$, the selected tower member's energy in a fixed-$M$ sector differs from zero kinetic
energy only by the constant Zeeman shift.  The infinite-temperature state in
that sector is
\begin{equation}
    \rho_{\beta=0,M}
    =\frac{\widehat P_M}{N_{L,M}},
    \qquad
    N_{L,M}
    =[z^M](z^{-1}+1+z)^L,
    \label{eq: xy_fixed_m_infinite_temperature}
\end{equation}
where $N_{L,M}$ is the fixed-$M$ Hilbert-space count, $\widehat P_M$ projects
onto total magnetization $M$, and $[z^M]$ extracts the coefficient of $z^M$.  For a fixed-density sequence $M/L\to q$, the limiting probability of the
local $S^z=0$ state is
\begin{equation}
    p_0(q)=\frac{\sqrt{4-3q^2}-1}{3},
    \qquad |q|<1.
    \label{eq: xy_p0_q}
\end{equation}

\begin{proposition}[Exact fixed-$M$ means and witness values]
\label{prop: xy_thermal_activity}
For any site and nearest-neighbor bond,
\begin{align}
    \operatorname{Tr}(\rho_{\beta=0,M}\widehat Y_r)
    &=-\frac{N_{L-1,M}}{N_{L,M}},
    \label{eq: xy_exact_y_mean}\\
    \operatorname{Tr}(\rho_{\beta=0,M}\widehat Z_{R_r})
    &=0,
    \label{eq: xy_exact_z_mean}\\
    \operatorname{Tr}(\rho_{\beta=0,M}\widehat Q^{Y}_r)
    &=\frac{N_{L-1,M}}{N_{L,M}},
    \label{eq: xy_exact_y_activity}\\
    \operatorname{Tr}(\rho_{\beta=0,M}\widehat Q^{A}_{R_r})
    &=\frac{N_{L-2,M}}{N_{L,M}},
    \label{eq: xy_exact_directed_q_activity}\\
    \operatorname{Tr}(\rho_{\beta=0,M}\widehat Q^{Z}_{R_r})
    &=2\frac{N_{L-2,M}}{N_{L,M}}.
    \label{eq: xy_exact_z_activity}
\end{align}
For $M/L\to q$ with $|q|<1$, these expressions converge to
\begin{align}
    \langle\widehat Y_r\rangle_{\beta=0,M}&\to-p_0(q),\\
    \langle\widehat Q^{Y}_r\rangle_{\beta=0,M}&\to p_0(q),\\
    \langle\widehat Q^{A}_{R_r}\rangle_{\beta=0,M}&\to p_0(q)^2,\\
    \langle\widehat Q^{Z}_{R_r}\rangle_{\beta=0,M}&\to2p_0(q)^2,
\end{align}
with $p_0(q)$ defined in Eq.~\eqref{eq: xy_p0_q}.
\end{proposition}

\begin{proof}
Fixing the selected site to $\ket0$ leaves $L-1$ sites with magnetization $M$,
which gives the ratio $N_{L-1,M}/N_{L,M}$ and hence both the
$\widehat Y_r$ mean and $\widehat Q^Y_r$ witness value.  The state $\rho_{\beta=0, M}$ is diagonal
in the product basis, whereas $\widehat Z_{R_r}$ is off-diagonal, so its mean
vanishes.  The operator $\widehat A^\dagger\widehat A=J^2\ket{\mathrm{sym}}\bra{\mathrm{sym}}$ has two diagonal
contributions, $\ket{+-}$ and $\ket{-+}$, each leaving an $(L-2)$-site
configuration with magnetization $M$.  The additional $\ket{00}\!\bra{00}$
term in $\widehat Z^2$ contributes the same normalized trace, producing the factor of
two.

For the thermodynamic limit, a direct saddle-point evaluation of
$[z^M](z^{-1}+1+z)^L$ at fixed $M/L\to q$ gives the one-site fugacity
probabilities for $s\in\{-1,0,+1\}$,
\begin{equation}
    p_s=\frac{e^{\mu s}}{1+2\cosh\mu},
    \qquad p_+-p_-=q,
\end{equation}
where $\mu$ is the saddle-point Lagrange multiplier fixing the magnetization
density $q$.  For every fixed integer $r$,
$N_{L-r,M}/N_{L,M}\to p_0(q)^r$.  In particular
$p_+p_-=p_0^2$.  Solving the normalization and charge constraints gives
Eq.~\eqref{eq: xy_p0_q} and the limits stated above.  Thus the fixed-$M$
reference values follow directly from coefficient asymptotics and do not rely
on the separate microcanonical--canonical bridge of
Appendix~\ref{ssec: appendix_energy_matching_and_canonical_role}.
\end{proof}

The diagonal route already distinguishes each tower member at first moment,
\begin{equation}
    \langle\widehat Y_r\rangle_{\Psi_n}=0,
    \qquad
    \langle\widehat Y_r\rangle_{\beta=0,M}\to-p_0(q),
    \label{eq: xy_y_direct_mean_witness}
\end{equation}
whereas the Hermitian kinetic operator has zero mean in both states and is
distinguished by its second moment,
\begin{equation}
\begin{aligned}
    \langle\widehat Z_{R_r}\rangle_{\Psi_n}
    &=\langle\widehat Z_{R_r}\rangle_{\beta=0,M}=0,\\
    \langle\widehat Z_{R_r}^2\rangle_{\Psi_n}&=0,\\
    \langle\widehat Z_{R_r}^2\rangle_{\beta=0,M}
    &\to4J^2p_0(q)^2.
\end{aligned}
    \label{eq: xy_z_fluctuation_only_witness}
\end{equation}

Taken alone, these are reference witness values rather than an ICQMBS result;
an energy-matched microcanonical comparison and thermal-background
concentration are still required.  The next subsection supplies both on the
fluctuation-scale thermodynamic proof window.

For the sequence used in the main text, $M=-2$ and hence $q=M/L\to0$, so
Proposition~\ref{prop: xy_thermal_activity} gives the limiting $\beta=0$
witness values $(1/9,2/9,1/3)$ in the order $(A,Z,Y)$.  The next step is to
show that these fixed-$M$ targets are also the local limits of an energy-matched
raw microcanonical window in the selected momentum sector.

\subsubsection{Thermodynamic ICQMBS closure}
\label{ssec: xy_thermodynamic_closure_appendix}

We now close the model-level thermodynamic conditions of
Definition~\ref{def: icqmbs} without extrapolating the narrow numerical
sequence.  Fix any $c>\sqrt{2.1}$, take even $L$ at $M=-2$ and the tower momentum
\begin{equation}
    k_L=
    \begin{cases}
       \pi,&L=0\pmod4,\\
       0,&L=2\pmod4,
    \end{cases}
    \label{eq: xy_selected_momentum_sequence}
\end{equation}
and use the raw window
\begin{equation}
    \mathcal W_L^{\rm th}
      =\left\{E:\ |E|\le c|J|L^{1/2}\right\}.
    \label{eq: xy_wide_thermodynamic_window_appendix}
\end{equation}
The full width is $2c|J|L^{1/2}$ and its energy-density half-width vanishes as
$L^{-1/2}$.  As in the main text, $\rho_L^{\rm mc,th}$ denotes the normalized
projector onto this raw window within the selected $(M,k_L)$ sector.

\begin{proposition}[Fluctuation-scale thermodynamic closure]
\label{prop: xy_wide_window_closure_appendix}
For the compatible family in Eq.~\eqref{eq: xy_reference_family}, the sequence
in Eqs.~\eqref{eq: xy_selected_momentum_sequence} and
\eqref{eq: xy_wide_thermodynamic_window_appendix} satisfies, uniformly for
$0.05<\kappa/J<0.20$,
\begin{align}
    \frac{1}{L}\log N_{\rm win,L}&\longrightarrow\log3,
    \label{eq: xy_wide_window_entropy_limit}\\
    \frac{N_{\rm cage,L}}{N_{\rm win,L}}&\longrightarrow0,
    \label{eq: xy_wide_window_cage_fraction}\\
    \operatorname{Tr}(\rho_L^{\rm mc,th}\widehat Q^A_{R_r})
      &\longrightarrow\frac19,
    \label{eq: xy_wide_window_QA_limit_appendix}\\
    f_{L,\epsilon}[\widehat O_R]
      &\le \frac{C_{R,O,c}+o(1)}{\epsilon^2L}
      \longrightarrow0
    \label{eq: xy_wide_window_concentration_bound}
\end{align}
for every fixed bounded region $R$, bounded Hermitian $\widehat O_R$, and
$\epsilon>0$, where the finite constant $C_{R,O,c}$ depends only on the fixed
region, observable, and fixed window prefactor $c$, not on $L$ or $\kappa$.
The companion witness limits are
$\langle\widehat Q^Z_{R_r}\rangle_{\rm mc,th}\to2/9$ and
$\langle\widehat Q^Y_r\rangle_{\rm mc,th}\to1/3$.  Consequently the selected
tower sequence is an ICQMBS, and the uniformity in $\kappa$ makes it a
deformation-stable ICQMBS family on this open interval.
\end{proposition}

\begin{proof}
The antiunitary reflection in
Eq.~\eqref{eq: xy_antiunitary_spectral_reflection_appendix} preserves the
selected $(M,k_L)$ sector and anticommutes with $\widehat H_\kappa$.  Hence the
resolved infinite-temperature mean energy is exactly zero, equal to the tower
energy.

We first bound its energy variance.  For all sufficiently large even $L$, the
range-$d$ pair terms with $d=1,2,3$ act on distinct unordered site pairs.  From
any $S^z$ product configuration, one pair term produces at most two final
configurations, each with matrix-element magnitude at most $|t_d|$ in the
convention of Eq.~\eqref{eq: xy_range_d_exchange}.  Different unordered pairs
change different site sets and therefore cannot reach the same final product
configuration.  Thus the diagonal row norm obeys
\begin{equation}
   (\widehat H_\kappa^2)_{\sigma\sigma}
   \le 2L\!\left(J^2+(0.1J)^2+\kappa^2\right)
   \le 2.1J^2L
   \label{eq: xy_fixed_m_energy_variance_bound}
\end{equation}
uniformly on the closed interval $0.05\le\kappa/J\le0.20$.

Resolving translation changes this bound only exponentially.  Indeed, for a
bounded operator $\widehat A_X$ supported on a fixed number $s$ of sites, a
product configuration contributing to
$\operatorname{Tr}_M(\widehat T^r\widehat A_X)$ must agree with its $r$-site
translate away from $X$.  Let $C_X$ denote an $L$-independent constant that may
depend on this fixed local operator and its support.  A nonidentity translation
has at most
$\gcd(L,r)\le L/2$ cycles, so
\begin{equation}
   \left|\operatorname{Tr}_M(\widehat T^r\widehat A_X)\right|
   \le C_X 3^{\gcd(L,r)+s}.
   \label{eq: xy_translation_character_defect_bound}
\end{equation}
Expanding $\widehat H_\kappa^2$ gives only $O(L^2)$ products of fixed-support
terms.  Combining Eq.~\eqref{eq: xy_translation_character_defect_bound} with
Eq.~\eqref{eq: xy_translation_resolved_entropy_asymptotic} therefore yields
\begin{equation}
    \operatorname{Var}_{\beta=0,M,k_L}(\widehat H_\kappa)
    \le (2.1+o(1))J^2L,
    \label{eq: xy_resolved_energy_variance_bound}
\end{equation}
where the $o(1)$ is uniform in the declared $\kappa$ interval.  Let
$p_L$ be the fraction of the selected $(M,k_L)$ sector contained in
Eq.~\eqref{eq: xy_wide_thermodynamic_window_appendix}.  Chebyshev's inequality
gives
\begin{equation}
    p_L\ge 1-\frac{2.1+o(1)}{c^2}.
    \label{eq: xy_wide_window_retained_fraction}
\end{equation}
Because $c>\sqrt{2.1}$ is fixed, there is a constant $p_{\min}>0$ such that
$p_L\ge p_{\min}$ for all sufficiently large $L$, uniformly on the declared
$\kappa$ interval.  Thus the window need not asymptotically contain the whole
sector: a size-independent positive fraction already gives
\begin{equation}
    \frac1L\log N_{\rm win,L}
    =\frac1L\log\dim\mathcal H_{M,k_L}
      +\frac1L\log p_L
    \longrightarrow\log3,
    \label{eq: xy_sqrt_window_entropy_argument}
\end{equation}
proving Eq.~\eqref{eq: xy_wide_window_entropy_limit}.  At fixed $M=-2$ the
declared staggered tower contributes exactly one state, so
Eq.~\eqref{eq: xy_wide_window_cage_fraction} follows as well.

Translation averaging is an established route to weak ETH in homogeneous
systems \cite{mori_weak_2016}; here we implement it in the selected fixed-charge
and momentum sector, with exact-energy-block compression and bounds uniform
along the compatible exchange family.  We next establish the local conditional bridge needed because a
fluctuation-scale window need not retain asymptotically unit sector weight.
Within the fixed-$M$ sector, define the translation average
\begin{equation}
    \overline O_L=\frac1L\sum_x
       \widehat T^x\widehat O_R\widehat T^{-x}.
    \label{eq: xy_translation_averaged_local_operator}
\end{equation}
If $\widehat P_{k_L}$ is the momentum projector, then exactly
$\widehat P_{k_L}\widehat O_R\widehat P_{k_L}
 =\widehat P_{k_L}\overline O_L\widehat P_{k_L}$.  The same identity holds
inside every energy block of that momentum sector and therefore removes any
basis ambiguity from exact degeneracies.

For a fixed product pattern $a$ on $s=O(1)$ sites, with total pattern
magnetization $m(a)$, the fixed-$M=-2$ marginal is
\begin{equation}
   \Pr_M(a)=\frac{N_{L-s,-2-m(a)}}{N_{L,-2}}
           =3^{-s}[1+O(L^{-1})],
    \label{eq: xy_fixed_m_local_marginal}
\end{equation}
uniformly over the finite set of local patterns.  Two disjoint fixed regions
therefore have $O(L^{-1})$ covariance.  For a general local operator, only its
magnetization-preserving compression contributes inside the fixed-$M$ sector;
on disjoint supports, off-diagonal local changes cannot be undone by the other
operator, so the same covariance estimate applies.  Only $O(1)$ translated
separations overlap.  Consequently
\begin{equation}
   \operatorname{Var}_{M}(\overline O_L)=O(L^{-1}).
   \label{eq: xy_translation_average_fixed_m_variance}
\end{equation}
Equation~\eqref{eq: xy_translation_character_defect_bound}, applied term by
term to the first and second moments, promotes this to the selected momentum
sector with only an exponentially small correction.  Write normalized
infinite-temperature expectation in that sector as
$\langle\cdot\rangle_0$ and let
$\mu_{M,k}=\langle\overline O_L\rangle_0$.  With
$\widehat P_{\mathcal W}$ the raw-window projector,
$p_L=\langle\widehat P_{\mathcal W}\rangle_0$.  The Hilbert--Schmidt
Cauchy--Schwarz inequality gives
\begin{align}
  \left|
  \langle\overline O_L\rangle_{\rm mc,th}-\mu_{M,k}
  \right|
  &=\frac{\left|
    \left\langle(\widehat P_{\mathcal W}-p_L)
    (\overline O_L-\mu_{M,k})\right\rangle_0
    \right|}{p_L}\notag\\
  &\le
  \sqrt{\frac{1-p_L}{p_L}
  \operatorname{Var}_{0}(\overline O_L)}
  =O(L^{-1/2}),
  \label{eq: xy_sqrt_window_conditional_mean_bridge}
\end{align}
uniformly for the fixed $c$ and declared $\kappa$ interval.  Similarly,
positivity and $p_L\ge p_{\min}$ imply
\begin{equation}
  \left\langle(\overline O_L-\mu_{M,k})^2\right\rangle_{\rm mc,th}
  \le \frac{\operatorname{Var}_{0}(\overline O_L)}{p_L}
  =O(L^{-1}).
  \label{eq: xy_window_conditioned_variance_bridge}
\end{equation}
The variance about the conditioned mean minimizes the mean-square deviation,
so $\operatorname{Var}_{\rm mc,th}(\overline O_L)=O(L^{-1})$ as well.

Taking $\widehat O_R=\widehat Q^A_{R_r}$ in
Eq.~\eqref{eq: xy_sqrt_window_conditional_mean_bridge},
Proposition~\ref{prop: xy_thermal_activity} gives
$\operatorname{Tr}(\rho_{\beta=0,M}\widehat Q^A_{R_r})\to1/9$, while
Eq.~\eqref{eq: xy_translation_character_defect_bound} shows that momentum
resolution changes this fixed-bond expectation only exponentially.  Hence
Eq.~\eqref{eq: xy_wide_window_QA_limit_appendix} follows despite the window
retaining only a fixed positive sector fraction.  The $Z$ and $Y$ limits
follow identically from Eqs.~\eqref{eq: xy_exact_z_activity} and
\eqref{eq: xy_exact_y_activity}.

Finally, for every energy projector $\widehat P_{E,k_L}$ in the raw window,
momentum compression gives
$\widehat P_{E,k_L}\widehat O_R\widehat P_{E,k_L}
 =\widehat P_{E,k_L}\overline O_L\widehat P_{E,k_L}$.  If
$\widehat A=\overline O_L-
\langle\overline O_L\rangle_{\rm mc,th}$, then
\begin{equation}
   \operatorname{Tr}\!\left[(\widehat P_{E,k_L}\widehat A
       \widehat P_{E,k_L})^2\right]
   \le
   \operatorname{Tr}\!\left(\widehat P_{E,k_L}\widehat A^2\right).
   \label{eq: xy_energy_block_compression_bound}
\end{equation}
Summing over blocks and applying Markov's inequality gives
Eq.~\eqref{eq: xy_wide_window_concentration_bound}.  Since the Hermitian
operator space on every fixed finite region is finite dimensional, a basis
estimate extends to every bounded Hermitian $\widehat O_R$.  All constants in
the energy-window step are uniform on the closed $\kappa$ interval, while the
fixed-$M$ local estimates are Hamiltonian independent.  The result is therefore
uniform on the stated open compatible family.
\end{proof}

The theorem does not assert the same thermodynamic limits for the narrower
$\Delta E=(J/2)L^{1/4}$ numerical sequence.  That sequence remains a stricter
finite-size probe of energy locality and is retained in
Appendix~\ref{ssec: xy_finite_size_thermal_diagnostics_appendix} as an
independent stress test of the analytic realization.

\subsection{Finite-size thermal-background diagnostics}
\label{ssec: xy_finite_size_thermal_diagnostics_appendix}

The thermodynamic statement above uses an $O(L^{1/2})$ raw window.
The narrower-window calculations below independently test cleaning,
ensemble matching, local concentration, and finite-energy generality; they
are not premises of Proposition~\ref{prop: xy_wide_window_closure_appendix}.

\subsubsection{Joint-dark cleaning protocol}
\label{ssec: xy_manual_deletion_appendix}

Use the exact-energy-block projector construction of
Appendix~\ref{ssec: appendix_energy_matching_and_canonical_role} with
$\widehat Q_{\rm all}=\sum_r(\widehat Q^A_{R_r}+\widehat Q^Z_{R_r}+\widehat Q^Y_r)$
in the resolved $(M,k)$ sector.  The full window joint-dark projector,
including the selected tower member, is removed only for the cleaned
diagnostic.  Subtracting the selected-member projector is used solely to
report the residual non-target joint-dark rank.  On the sampled $L,\kappa$
grid the full projector has numerical rank one at the reported tolerances.
This basis-independent test inside exact degeneracies concerns the declared
translated $A,Z,Y$ family, not darkness under every possible local operator.

\subsubsection{Narrow-window \texorpdfstring{$\beta=0$}{beta = 0} bridge}
\label{ssec: xy_beta_zero_overlap_appendix}

The principal numerical sequence uses $L=8,10,12,14$,
$\kappa_\star/J=0.1$, and the raw half-width
$\Delta E=(J/2)L^{1/4}$.  The $L=6$ ring and symmetry-enhanced $\kappa=0$
endpoint are separate controls.  Write the general subextensive window as
\begin{equation}
    \mathcal W_L(\gamma,c)
      =\{a:|E_a-E_{\Psi,L}|\le c|J|L^\gamma\},
    \qquad E_{\Psi,L}=0,
    \label{eq: xy_microcanonical_sequence}
\end{equation}
with $0\le\gamma<1$ and fixed $c>0$.  The principal numerical choice is
$(\gamma,c)=(1/4,1/2)$; fixed-width and $\gamma=1/2$ windows are controls.
Table~\ref{tab: xy_window_locality_controls} gives the principal populations,
whose $\log N_{\rm win}/L$ increases from $0.417$ to $0.638$.  The rank-one
joint-dark fraction falls from $1/28$ to $1/7615$; neither finite-size trend
replaces the analytic entropy and sparsity argument.

\begin{figure*}[tp]
    \centering
    \includegraphics[width=0.9\textwidth]{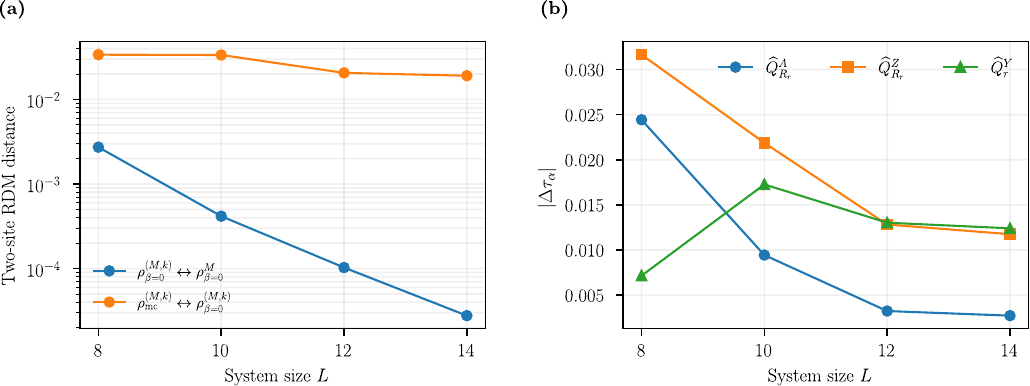}
    \caption{Auxiliary $\beta=0$ local-equivalence comparison at
    $\kappa_\star/J=0.1$ using the principal $\Delta E=(J/2)L^{1/4}$ window.
    \textbf{(a)} Two-site trace distances for the microcanonical-to-resolved-$\beta=0$
    bridge and the resolved-$(M,k)$-to-fixed-$M$ $\beta=0$ bridge, where $M$
    and $k$ denote total magnetization and momentum.  The latter
    falls rapidly below the former.  \textbf{(b)} Absolute differences in the witness values derived from $\widehat A$, $\widehat Z$, and $\widehat Y$
    for the first bridge.  These diagnostics are auxiliary to the
    defining raw microcanonical ETH mismatch.}
    \label{fig: xy_beta_zero_bridges}
\end{figure*}

Both the tower and the fixed-$M$ and resolved-$(M,k)$ $\beta=0$ states have
zero energy density.  For each $\alpha\in\{A,Z,Y\}$, let
$\tau_{\alpha,L}^{{\rm mc},s}=\operatorname{Tr}(\rho_L^{{\rm mc},s}\widehat Q^\alpha)$
with $s\in\{{\rm raw},{\rm clean}\}$ and the operators of
Eq.~\eqref{eq: xy_normalized_witnesses_appendix}.  Compare them with the
corresponding resolved-$\beta=0$ traces:
\begin{equation}
\begin{aligned}
 \varepsilon_{\alpha,L}^{\rm rr}
   &=|\tau_{\alpha,L}^{{\rm mc},{\rm raw}}
       -\tau_{\alpha,L}^{\beta=0,(M,k),{\rm raw}}|,\\
 \varepsilon_{\alpha,L}^{\rm cc}
   &=|\tau_{\alpha,L}^{{\rm mc},{\rm clean}}
       -\tau_{\alpha,L}^{\beta=0,(M,k),{\rm clean}}|,\\
 \varepsilon_L^s&=\max_{\alpha\in\{A,Z,Y\}}\varepsilon_{\alpha,L}^s,
       \qquad s\in\{{\rm rr},{\rm cc}\}.
\end{aligned}
 \label{eq: xy_beta_zero_overlap_appendix}
\end{equation}
The cleaned comparator subtracts precisely the same window joint-dark
projector from the full resolved trace and renormalizes.  This is a
like-with-like diagnostic, not a full-spectrum dark-state census or a
redefinition of the raw thermal reference.  Exact fixed-$M$ counting and the
resolved trace are retained separately in the accompanying records.

Energy matching alone does not prove local equivalence.  On the same
two-site region, distinguish
\begin{equation}
\begin{aligned}
 d_{{\rm tr},L}^{(1)}
   &=d_{\rm tr}(\rho_{\rm mc}^{(M,k)},\rho_{\beta=0}^{(M,k)}),\\
 d_{{\rm tr},L}^{(2)}
   &=d_{\rm tr}(\rho_{\beta=0}^{(M,k)},\rho_{\beta=0,M}),
\end{aligned}
 \label{eq: xy_beta_zero_two_bridges}
\end{equation}
where $d_{\rm tr}(\rho,\sigma)=\tfrac12\|\rho-\sigma\|_1$ and the first
distance uses the raw microcanonical state.

\begin{table}[!htbp]
 \caption{Auxiliary ensemble bridges at $\kappa_\star/J=0.1$ in the principal
 $\Delta E=(J/2)L^{1/4}$ window.  The witness mismatches and two-site
 distances are defined in Eqs.~\eqref{eq: xy_beta_zero_overlap_appendix}
 and \eqref{eq: xy_beta_zero_two_bridges}.}
 \label{tab: xy_beta_zero_bridge_comparison}
 \setlength{\tabcolsep}{7pt}
 \renewcommand{\arraystretch}{1.25}
 \begin{ruledtabular}
 \begin{tabular}{ccccc}
 $L$ & $\varepsilon_L^{\rm rr}$ & $\varepsilon_L^{\rm cc}$
     & $d_{{\rm tr},L}^{(1)}$ & $10^3d_{{\rm tr},L}^{(2)}$ \\
 \hline
 $8$  & $0.0317$ & $0.0267$ & $0.0339$ & $2.73$ \\
 $10$ & $0.0219$ & $0.0209$ & $0.0336$ & $0.417$ \\
 $12$ & $0.0130$ & $0.0128$ & $0.0206$ & $0.103$ \\
 $14$ & $0.0124$ & $0.0124$ & $0.0191$ & $0.0278$
 \end{tabular}
 \end{ruledtabular}
\end{table}

The positive-coupling grid is
$\mathcal G_\kappa/J=\{0.05,0.10,0.15,0.20\}$.
Taking maxima over these anchors gives raw--raw mismatches
$(0.0327,0.0220,0.0132)$ and cleaned--cleaned mismatches
$(0.0278,0.0210,0.0130)$ at $L=8,10,12$.  These sampled envelopes bound
the representative point; open-interval thermodynamic uniformity follows
instead from Proposition~\ref{prop: xy_wide_window_closure_appendix}.

\begin{table}[!htbp]
 \caption{Raw window and locality controls at $\kappa_\star/J=0.1$.
 $N_{\rm p}$ and $w_{\rm p}^{(2,3)}$ use the principal
 $\Delta E=(J/2)L^{1/4}$ window; $w_{\rm f}^{(2)}$ uses the fixed
 $\Delta E=J/2$ window, with populations $(24,101,609,4011)$ in increasing
 $L$.  Superscripts count sites; the three-site calculation stops at $L=12$.}
 \label{tab: xy_window_locality_controls}
 \setlength{\tabcolsep}{7pt}
 \renewcommand{\arraystretch}{1.25}
 \begin{ruledtabular}
 \begin{tabular}{ccccc}
 $L$ & $N_{\rm p}$ & $w_{\rm p}^{(2)}$ & $w_{\rm f}^{(2)}$ & $w_{\rm p}^{(3)}$ \\
 \hline
 $8$  & $28$   & $0.16859$ & $0.17603$ & $0.15042$ \\
 $10$ & $157$  & $0.07633$ & $0.09270$ & $0.07000$ \\
 $12$ & $1083$ & $0.04692$ & $0.06169$ & $0.04077$ \\
 $14$ & $7615$ & $0.01746$ & $0.02373$ & ---
 \end{tabular}
 \end{ruledtabular}
\end{table}

Table~\ref{tab: xy_beta_zero_bridge_comparison} and
Fig.~\ref{fig: xy_beta_zero_bridges} separate the two bridges: the
resolved-to-fixed-$M$ correction becomes much smaller than the
microcanonical-to-resolved mismatch.  The latter decreases overall but
nearly plateaus from $L=12$ to $14$; no limiting offset is inferred.
The independent spectral checks below rule out a missed-window explanation
at the quoted precision, without establishing a thermodynamic limit for
this narrower numerical window.

\subsubsection{Two-site concentration and window systematics}
\label{ssec: xy_block_concentration_appendix}

\begin{figure*}[tp]
    \centering
    \includegraphics[width=0.9\textwidth]{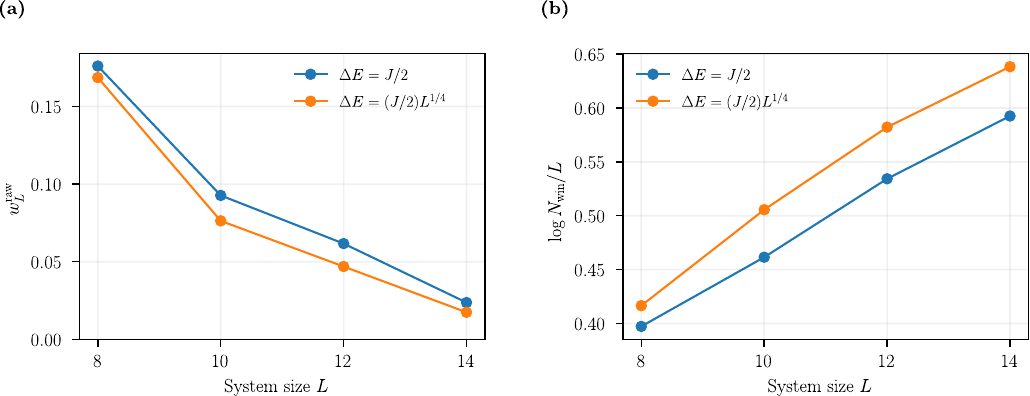}%
    \caption{Window systematics for the representative complete two-site
    concentration test.  \textbf{(a)} Raw worst-case width $w_L^{\rm raw}$ for the
    principal $\Delta E=(J/2)L^{1/4}$ window and the independent fixed-width
    $\Delta E=J/2$ control.  Both sequences decrease monotonically through
    $L=14$.  \textbf{(b)} The corresponding $\log N_{\rm win}/L$ values, showing the
    rapid growth of the raw window population over the available sizes.  No
    power-law exponent is inferred from these four points.}
    \label{fig: xy_concentration_window_systematics}
\end{figure*}

The two-site Hilbert space decomposes by bond magnetization into blocks of
dimensions $1,2,3,2,1$, so the real space of Hermitian operators commuting
with the bond charge has dimension
\begin{equation}
    1^2+2^2+3^2+2^2+1^2=19.
    \label{eq: xy_two_site_operator_basis_dimension}
\end{equation}
Choose a Hilbert--Schmidt-orthonormal basis
$\{\widehat O_a\}_{a=0}^{18}$, with the normalized identity included only as a
zero-width completeness check.  All statements below use Hilbert--Schmidt
normalization on this fixed two-site operator space; because its dimension is fixed,
this is equivalent up to size-independent constants to any other bounded local
operator normalization.  To keep the defining raw window and the finite-rank
cleaning diagnostic distinct, let $s\in\{{\rm raw},{\rm clean}\}$ and set
\begin{equation}
\begin{aligned}
    \widehat P_E^{\rm raw}&=\widehat P_E,\\
    \widehat P_E^{\rm clean}&=\widehat P_E-\widehat P_{{\rm jd},E},\\
    X_{a,E}^{s}
      &=\widehat P_E^{s}
        \left(\widehat O_a-\mu_a^{s}\widehat{\mathbf1}\right)
        \widehat P_E^{s},\\
    \Sigma_{ab}^{s}
      &=\frac{1}{2N_s}\sum_E
        \operatorname{Tr}\!\left(
          X_{a,E}^{s}X_{b,E}^{s}
          +X_{b,E}^{s}X_{a,E}^{s}
        \right),
\end{aligned}
    \label{eq: xy_block_invariant_covariance}
\end{equation}
where $N_s=\sum_E\operatorname{Tr}\widehat P_E^{s}$ and
$\mu_a^{s}=\operatorname{Tr}(\rho_L^{{\rm mc},s}\widehat O_a)$.
The corresponding block-invariant widths are
\begin{equation}
    w_L^{s}(\kappa)
      =\sqrt{\lambda_{\max}\Sigma_L^{s}(\kappa)}.
    \label{eq: xy_local_concentration_width}
\end{equation}
They are invariant under rotations of the numerical eigenbasis inside exactly
degenerate blocks and measure the largest fluctuation among normalized
operators in the full 19-dimensional magnetization-preserving two-site
algebra.  This is also the complete algebra relevant to two-site diagonal
matrix elements inside a fixed-$M$ sector: every local component that changes
the bond magnetization has $\widehat P_M\widehat O_R\widehat P_M=0$ identically.  The raw width
$w_L^{\rm raw}$ is the defining concentration diagnostic used in the main text;
$w_L^{\rm clean}$ is reported only to test sensitivity to the finite-rank
joint-dark subspace.  For either choice of $s$, any
Hilbert--Schmidt-normalized Hermitian operator in this algebra has mean-square
compressed fluctuation bounded by $(w_L^{s})^2$, and hence
$f_L^{s}(\epsilon)\le (w_L^{s})^2/\epsilon^2$ by Markov's inequality.

For cleaned results, Eq.~\eqref{eq: appendix_raw_from_clean_concentration_bound}
and the covariance bound above recover raw concentration when the removed
fraction vanishes; the mean shift obeys
$d_L\le2f_{{\rm jd},L}\|\widehat O_R\|$.
The raw covariance tests concentration directly on the declared two-site
region; neither test alone establishes it on every bounded region.

On the four positive-$\kappa$ anchors, the principal-window raw width ranges
are $[0.16668,0.16859]$, $[0.07168,0.07633]$, and
$[0.04566,0.04697]$ at $L=8,10,12$.  Adding the midpoints
$\kappa/J=0.075,0.125,0.175$ gives, at $L=12$,
\begin{equation}
\begin{aligned}
  0.11272&\le\tau_A^{\rm mc,raw}\le0.11304,\\
  0.21928&\le\tau_Z^{\rm mc,raw}\le0.21958,\\
  0.32750&\le\tau_Y^{\rm mc,raw}\le0.32783,\\
  0.04566&\le w_{12}^{\rm raw}\le0.04706.
\end{aligned}
 \label{eq: xy_kappa_refined_grid_ranges}
\end{equation}
The refined mesh supports weak finite-size variation; it is not a
continuous-parameter bound.  The accompanying records retain the full grid.

Table~\ref{tab: xy_window_locality_controls} and
Fig.~\ref{fig: xy_concentration_window_systematics} compare two homogeneous
window protocols.  Both show monotone two-site narrowing through $L=14$
and rapidly growing populations; no scaling exponent is fitted.

\paragraph*{Spectral validation.}
\label{par: xy_numerical_spectral_validation}
At $L=14$, both windows lie inside the validated returned spectrum: the
principal half-width is $0.967<1.042$, and the worst in-window eigenpair
residuals are $2.17\times10^{-6}J$ and $2.02\times10^{-6}J$ for the
principal and fixed windows, respectively, in the convention of
Eq.~\eqref{eq: xy_reference_family}.  Enlarging the eigenpair budget leaves
the populations unchanged and changes the witness means and matching
distances by at most $2.6\times10^{-11}$ across these and the nested
fixed-width windows.  Varying the examined exact-energy grouping tolerance
leaves the widths unchanged at displayed precision.  The broader $L^{1/2}$
window is not contained at this size and is not used numerically.
The cleaned $L=14$ widths are $0.01741$ and $0.02367$ for the principal
and fixed windows; raw values remain the defining diagnostic.

\subsubsection{Three-site locality check}
\label{ssec: xy_three_site_concentration_appendix}

Repeat the raw covariance test on $R=\{0,1,2\}$ at
$\kappa_\star/J=0.1$.  The local charge blocks have dimensions
$1,3,6,7,6,3,1$, giving the complete real Hermitian operator-space dimension
$1^2+3^2+6^2+7^2+6^2+3^2+1^2=141$.
The principal-window widths in Table~\ref{tab: xy_window_locality_controls}
decrease through $L=12$, with worst in-window residual below
$3.2\times10^{-13}$.  This tests concentration beyond two-site support;
the absolute widths at different support sizes are not a monotonic hierarchy
because their Hilbert--Schmidt normalizations and operator dimensions differ.
These numerical tests cover only their declared regions and narrow windows.
Concentration on every fixed bounded region is supplied separately by
Proposition~\ref{prop: xy_wide_window_closure_appendix}.

\subsubsection{Finite-energy matched-canonical check}
\label{ssec: xy_generic_microcanonical_appendix}
\label{ssec: xy_matched_canonical_appendix}

A uniform $D\ne0$ preserves the tower for compatible real odd-range
exchange, with $e_\psi=D+hM/L$.  The supplementary control uses
$D/J=0.315$, $J_3/J=0.1$, and $h=\kappa=0$.
Its raw window is centered on $E_{\Psi,L}$ with
$\Delta E=(J/2)L^{1/2}$, hence energy-density half-width
$\delta e=J/(2L^{1/2})$.  At $L=8,10,12$, its populations are
$(21,139,918)$ and the largest standard deviations over a fixed spanning
two-site basis are $(0.148,0.0699,0.0267)$.  This fixed-basis envelope,
$\sigma_{\rm basis}^{\max}$, is not the full covariance width.
Changing the window prefactor from $0.75$ to $1.25$ retains positive raw
microcanonical witness values and the same size trend.

The canonical state and energy-matched $\beta_L$ follow
Eqs.~\eqref{eq: appendix_resolved_canonical_state} and
\eqref{eq: appendix_canonical_energy_matching} in the same fully resolved
sector: fixed $M$, the tower momentum, and its compatible inversion parity
for this real Hamiltonian.  Its effective state count is the inverse purity,
$N_{\rm eff}=1/\operatorname{Tr}[(\rho^{\rm ref}_{\beta_L,L})^2]$.
The three witness values for each ensemble are
\begin{center}
\small
\setlength{\tabcolsep}{3.5pt}
\renewcommand{\arraystretch}{1.25}
\begin{ruledtabular}
\begin{tabular}{@{}ccccccc@{}}
$L$ & $\beta_L J$ & $N_{\rm eff}$
    & ensemble & $\tau_A$ & $\tau_Z$ & $\tau_Y$ \\
\hline
\multirow{2}{*}{$8$} & \multirow{2}{*}{$-0.1136$} & \multirow{2}{*}{$48.3$}
    & can & $0.1156$ & $0.2199$ & $0.3280$ \\
    & & & mc & $0.1034$ & $0.1885$ & $0.3050$ \\[0.3em]
\multirow{2}{*}{$10$} & \multirow{2}{*}{$-0.1024$} & \multirow{2}{*}{$332$}
    & can & $0.1223$ & $0.2362$ & $0.3350$ \\
    & & & mc & $0.1175$ & $0.2171$ & $0.3158$ \\[0.3em]
\multirow{2}{*}{$12$} & \multirow{2}{*}{$-0.1146$} & \multirow{2}{*}{$2125$}
    & can & $0.1192$ & $0.2306$ & $0.3341$ \\
    & & & mc & $0.1161$ & $0.2171$ & $0.3206$
\end{tabular}
\end{ruledtabular}
\end{center}
The matched temperatures are negative because the tower lies above the
infinite-temperature mean in this bounded spectrum.  The ensemble differences
decrease, but these three sizes and the fixed-basis envelope provide only a
finite-energy generality check outside the principal $\kappa$ certification.
They establish neither asymptotic ensemble equivalence nor full local
concentration; the raw same-Hamiltonian microcanonical reference remains
the defining ETH test.

\subsection{Supplementary deformation diagnostics and local-operator continuation}
\label{ssec: xy_joint_deformation_appendix}

The main text gives the exact finite-family selection rules for the generalized
tower ansatz.  Here we collect the complementary local continuation formulas
and finite-size first-order Jacobian diagnostics.  An exact compatible family
necessarily supplies tangent directions in the caging kernel, whereas the full
kernel may also contain tangent-only state repairs outside the restricted
tower ansatz.

\subsubsection{Continued local operators}

For the principal imaginary second-neighbor term, the bounded local reductions
are
\begin{equation}
\begin{aligned}
    \widehat A^{(2)}_{r,r+2}(\kappa)
      &=i\kappa\ket{00}_{r,r+2}
        \left(-\bra{+-}+\bra{-+}\right)_{r,r+2},\\
    \widehat Z^{(2)}_{r,r+2}(\kappa)
      &=\widehat A^{(2)}_{r,r+2}(\kappa)
        +\widehat A^{(2)\dagger}_{r,r+2}(\kappa).
\end{aligned}
    \label{eq: xy_kappa_local_operators}
\end{equation}
The scale of both operators vanishes at $\kappa=0$.  The fixed nearest-neighbor
chart provides a common witness comparison across the interval containing
that point.

For a general diagonal deformation,
\begin{equation}
    \widehat V_f=\sum_r f_r(\widehat S_r^z),
    \label{eq: xy_general_diagonal_deformation}
\end{equation}
the support of an interior tower member, $0<n<L$, remains a uniform
potential shell precisely when
\begin{equation}
    f_r(+1)-f_r(-1)=\delta f
    \quad\text{is independent of }r.
    \label{eq: xy_diagonal_deformation_condition}
\end{equation}
The polarized endpoints $n=0,L$ have single-configuration support and
therefore form uniform shells without this condition.
For an interior member satisfying the condition, a shifted local operator is
\begin{equation}
\begin{aligned}
    \widehat Y_r[f]
    ={}&f_r(\widehat S_r^z)-f_r(-1)\widehat{\mathbf1}_r\\
      &-\frac{\delta f}{2}
        \left(\widehat S_r^z+\widehat{\mathbf1}_r\right).
\end{aligned}
    \label{eq: xy_deformed_diagonal_witness}
\end{equation}
It annihilates both $\ket{-}_r$ and $\ket{+}_r$.  For
$f_r(m)=D_rm^2+hm+c_r$ with uniform $h=\delta f/2$, it reduces to
$D_r[(\widehat S_r^z)^2-\widehat{\mathbf1}_r]$.  For $D_r\neq0$, dividing out
this coupling gives the dimensionless operator
$\widehat Y_r=(\widehat S_r^z)^2-\widehat{\mathbf1}_r$ used in the main text.  This
normalized shell operator has a regular $D_r\to0$ continuation even though the
Hamiltonian term itself vanishes there; a purely affine $f_r$ gives a trivial
shifted operator.  This diagonal route is an exact extension but is not
part of the principal $\kappa$ family.

\subsubsection{Blind recovery of the compatible tangent space}

A blind first-order obstruction calculation gives a more direct test of the
analytic phase rule than scanning a guessed path.  Around the representative
$M=-2$, $\kappa/J=0.1$ point, use the six real exchange coordinates
\begin{equation}
    \delta\boldsymbol g=
    (\delta\operatorname{Re}t_1,\delta\operatorname{Im}t_1,
     \delta\operatorname{Re}t_2,\delta\operatorname{Im}t_2,
     \delta\operatorname{Re}t_3,\delta\operatorname{Im}t_3).
    \label{eq: xy_blind_exchange_coordinates}
\end{equation}
Let $\mathcal T_{\rm cage}=\ker\mathfrak O_{\rm cage}$ be the tangent space
returned by the full caged-state repair problem of
Appendix~\ref{ssec: technical_deformation_maps}, and let
$\mathcal T_{\rm fixed}=\ker\mathcal A_\psi$ be its fixed-state slice.  The
analytic tower condition in Eq.~\eqref{eq: xy_complex_exchange_phase_line}
predicts
\begin{equation}
    \mathcal T_{\rm tower}
      =\operatorname{span}\{\delta\operatorname{Re}t_1,
      \delta\operatorname{Im}t_2,
      \delta\operatorname{Re}t_3\}.
    \label{eq: xy_analytic_exchange_tangent_space}
\end{equation}
Write $\theta_{\max}$ for the largest principal angle between the computed
and analytic subspaces.  The blind kernels at two generic even sizes recover
exactly this three-dimensional subspace within numerical precision:
\begin{center}
\setlength{\tabcolsep}{7pt}
\renewcommand{\arraystretch}{1.25}
\begin{ruledtabular}
\begin{tabular}{ccccc}
$L$ & $\dim\mathcal T_{\rm cage}$ & $\dim\mathcal T_{\rm fixed}$
    & \shortstack{$\theta_{\max}$\\$(\mathcal T_{\rm cage},\mathcal T_{\rm tower})$}
    & $\sigma_{\min}^{\perp}$ \\[2pt]
\hline
$8$  & $3$ & $3$ & $1.5\times10^{-6}\,\mathrm{deg}$ & $2.928$ \\
$10$ & $3$ & $3$ & $1.2\times10^{-6}\,\mathrm{deg}$ & $3.266$
\end{tabular}
\end{ruledtabular}
\end{center}
Here $\sigma_{\min}^{\perp}$ is the smallest nonzero singular value of the
obstruction map in the declared coordinate normalization and is quoted only as
a conditioning measure.  The reverse subspace residuals are below
$1.2\times10^{-15}$, while $\mathcal T_{\rm fixed}$ agrees with
$\mathcal T_{\rm tower}$ to machine precision.  Thus allowing arbitrary
first-order amplitude repair produces no additional exchange direction in this
range-three translation-invariant chart: the blind caging kernel, the
fixed-state kernel, and the analytic tower-compatible tangent coincide at both
sizes.  Finite-step evaluations keep compatible directions at roundoff,
whereas a transverse $\delta\operatorname{Im}t_1$ control develops a
residual linear in the step size.

This numerical result is deliberately not identified with the augmented
joint-local tangent space of Appendix~\ref{ssec:
appendix_joint_local_channel_continuation}: the calculation does not evaluate
that bounded-local operator/state co-continuation map.  For the analytic
tower-compatible directions, joint continuation is instead established
exactly by the explicit two-site operators in
Eq.~\eqref{eq: xy_deformed_directed_channel}.  The blind obstruction result is
therefore a predictive finite-size recovery of the exchange-phase selection
rule, not a numerical substitute for the exact local-operator construction.

\subsubsection{Independent odd-range deformation control}

An independent real odd-range scan uses
$J_3/J=0,0.05,0.10,0.15,0.20$, $h=D=\kappa=0$, and $L=8,10,12$.
Each Hamiltonian has its own raw half-width $(J/2)L^{1/2}$ in the
selected $(M,k)$ and inversion-parity sector.  Minimizing each quantity
independently over this full grid gives
\begin{equation}
 (\tau_A,\tau_Z,\tau_Y)\ge(0.0922,0.1906,0.3153),
 \qquad N_{\rm win}\ge21,
\end{equation}
with the witness inequality understood componentwise.  All selected
tower-member residuals vanish at the reported precision.  The full pointwise
table is retained in the accompanying numerical records.
Since the exchange Hamiltonian is off-diagonal at $D=0$,
$\operatorname{Tr}_M H(J_3)=0$: the tower energy and fixed-$M$ $\beta=0$
energy remain matched and the normalized trace targets are unchanged.
This control shows that $J_3^{\rm ref}/J=0.1$ is not an isolated point
along the compatible odd-range direction.  It does not extend the principal
$\kappa$-family concentration theorem to a thermodynamic $J_3$ interval.

\subsection{Additional exact caged eigenstates as scope tests}
The staggered $\pi$-bimagnon tower is the analytically controlled benchmark
used in the main text.  The two constructions below probe the scope of the framework
beyond that benchmark.  The one-magnon-decorated open-chain family is found
within the chiral-shell search but does not cancel bond by bond, whereas the
five-site periodic caged eigenstate lies outside the chiral-shell reduction
because its fixed-magnetization kinetic graph is nonbipartite; the absence
of a physical-chain bipartition is the model-specific origin of the failure of
the real-space $C_A$ construction.  The five-site periodic shell contains a
caged eigenstate with a nonempty internal kinetic graph and nonzero kinetic energy.
These constructions therefore illustrate that the
practical search ansatz is not part of the definition of interference caging:
only the internal eigenvalue and active-boundary equations are fundamental.
The two analytical scope tests are summarized in Table~\ref{table: xy_additional_exact_states}.

\begin{table}[tbp]
    \centering
    \caption{Additional exact spin-1 $XY$ caged eigenstates established in this appendix.
    Here $N_{\pm}\equiv\sum_r(\widehat S_r^z)^2$ counts the nonzero-$S^z$ sites and is constant on each support.}
    \label{table: xy_additional_exact_states}
    \setlength{\tabcolsep}{8pt}
    \renewcommand{\arraystretch}{1.25}
    \begin{tabular}{ll}
        \hline\hline
        State & Exact domain and eigenvalue \\
        \hline
        $\ket{\Phi_{n,A}}$
          & \begin{tabular}[c]{@{}l@{}}OBC, odd $L$, $0\le n\le L-1$;\\
             $(M,N_{\pm})=(-L+2n+1,L-1)$;\\
             $E=h(-L+2n+1)+D(L-1)$.\end{tabular} \\[4pt]
        $\ket{\Phi_0}$, $\ket{\Phi_{-2}}$
          & \begin{tabular}[c]{@{}l@{}}PBC, $L=5$; $(M,N_{\pm})=(-2,4)$;\\
             $E_0=-2h+4D$, $E_{-2}=-2J-2h+4D$.\end{tabular} \\
        \hline\hline
    \end{tabular}
\end{table}

Both constructions are exact for the nearest-neighbor spin-1 $XY$ Hamiltonian
of Eq.~\eqref{eq: xy_hamiltonian} with uniform diagonal terms.  We do not claim
that they survive the $J_3$ or $\kappa$ deformations, nor do we assign them a
thermodynamic classification.  A broader commutant-algebra and
configuration-space-caging classification is available in
Ref.~\cite{mohapatra_unraveling_2025}; the examples here serve only as explicit
scope tests of the present framework.

It is convenient to use normalized local creation operators
\begin{equation}
    \widehat a_r^\dagger=\frac{\widehat S_r^+}{\sqrt{2}},
    \qquad
    \widehat b_r^\dagger=\frac{(\widehat S_r^+)^2}{2},
    \label{eq: xy_normalized_creators}
\end{equation}
so that $\widehat a_r^\dagger\ket{-}_r=\ket{0}_r$ and
$\widehat b_r^\dagger\ket{-}_r=\ket{+}_r$.  This convention removes the
spin-1 ladder-operator factors from the normalization constants below.

\subsubsection{One-magnon-decorated open-chain family}
\label{ssec: spin1_xy_decorated_type1_cages}

For an odd open chain, let $\Omega_A$ denote the real-space sublattice
containing both endpoints,
\begin{equation}
    \Omega_A=\{1,3,\ldots,L\},
    \qquad
    N_A=|\Omega_A|=\frac{L+1}{2}.
\end{equation}
For $r\in\Omega_A$, define the alternating sign on the chain with site
$r$ removed,
\begin{equation}
    \chi_r(s)=
    \begin{cases}
        (-1)^s, & s<r,\\
        (-1)^{s-1}, & s>r,
    \end{cases}
    \qquad
    \widehat J_r^\dagger
    =\sum_{s\ne r}\chi_r(s)\widehat b_s^\dagger,
    \label{eq: xy_deleted_chain_creator}
\end{equation}
and let $\eta_r=(-1)^{(r-1)/2}$.  The exact states are
\begin{equation}
\boxed{
    \ket{\Phi_{n,A}}
    =\frac{1}{\sqrt{N_A\binom{L-1}{n}}}
      \sum_{r\in\Omega_A}
      \eta_r\widehat a_r^\dagger
      \frac{(\widehat J_r^\dagger)^n}{n!}
      \ket{\mathrm{F}_{-}}
}
\label{eq: xy_decorated_tower}
\end{equation}
for
\begin{equation}
    n=0,1,\ldots,L-1.
    \label{eq: xy_decorated_tower_range}
\end{equation}
The upper limit is $L-1$, rather than $L$, because the magnon site cannot also
host a bimagnon.  The divided power in Eq.~\eqref{eq: xy_decorated_tower}
ensures that every $n$-site subset appears once.  Consequently, all nonzero
product-basis amplitudes have the same magnitude, and the displayed
normalization follows from the
$N_A\binom{L-1}{n}$ mutually orthogonal support configurations.

\paragraph*{Uniform-potential shell and energy}

A support configuration contains one local $\ket0$ state, $n$ local
$\ket+$ states, and $L-n-1$ local $\ket-$ states.  Hence
\begin{equation}
    M_n=-L+2n+1,
    \qquad
    \sum_{s=1}^{L}(S_s^z)^2=L-1,
\end{equation}
and the diagonal term is uniform on the support:
\begin{equation}
\begin{aligned}
    \widehat V\ket{\Phi_{n,A}}
      &=E_n''\ket{\Phi_{n,A}},\\
    E_n''&=h(-L+2n+1)+D(L-1).
\end{aligned}
\label{eq: xy_decorated_tower_energy}
\end{equation}
All support configurations lie in one eigenspace of the chiral-grading operator
$\widehat C_A$.  The kinetic term therefore has no matrix elements within the
support shell, and exactness reduces to cancellation of every boundary row.

\paragraph*{Boundary-row cancellation}

Write a support configuration as $\ket{r;\Xi}$, where
$r\in\Omega_A$ is the magnon site and
$\Xi\subset\{1,\ldots,L\}\setminus\{r\}$ is the set of $n$ bimagnon sites.
Its coefficient in Eq.~\eqref{eq: xy_decorated_tower} is proportional to
\begin{equation}
    c(r,\Xi)=\eta_r\prod_{s\in\Xi}\chi_r(s).
    \label{eq: xy_decorated_coefficient}
\end{equation}
The active boundary rows fall into two classes.

First, on a bond not containing $r$, a local
$\ket{+-}$ or $\ket{-+}$ pair is mapped to $J\ket{00}$.  The two parent
configurations differ by exchanging the bimagnon across that bond.  Since the
compressed-chain signs alternate,
\begin{equation}
    \chi_r(s+1)=-\chi_r(s)
    \qquad (r\notin\{s,s+1\}),
\end{equation}
the two parent amplitudes are opposite, and their equal kinetic matrix elements
cancel.  This is the same local bimagnon-exchange relation as in the main
$\pi$-bimagnon tower.

Second, acting on a bond adjacent to the magnon moves the local $\ket0$ state
to the intervening even site.  Each resulting boundary configuration has two
parents, with the original magnon on the left or right neighboring odd site.
Direct substitution of Eq.~\eqref{eq: xy_decorated_coefficient} gives
\begin{equation}
    c(s-1,\Xi_-)+c(s+1,\Xi_+)=0
    \qquad (s\notin\Omega_A),
    \label{eq: xy_decorated_magnon_row}
\end{equation}
where $\Xi_-$ and $\Xi_+$ are the uniquely related bimagnon sets producing the
same output configuration.  The equality follows because the products of
$\chi_{s-1}$ and $\chi_{s+1}$ agree away from the deleted sites, whereas
$\eta_{s+1}=-\eta_{s-1}$.  The representative $n=0$, $L=5$ relation is
\begin{equation}
\begin{aligned}
    \widehat K\big(
       \ket{0----}
       -\ket{--0--}
       +\ket{----0}
    \big)=0.
\end{aligned}
\label{eq: xy_decorated_bimagnon_cancellation}
\end{equation}
Equations~\eqref{eq: xy_decorated_coefficient}--\eqref{eq: xy_decorated_magnon_row}
exhaust the active boundary rows and prove
\begin{equation}
    \widehat K\ket{\Phi_{n,A}}=0.
    \label{eq: xy_decorated_kinetic_zero}
\end{equation}
Together with Eq.~\eqref{eq: xy_decorated_tower_energy}, this establishes the
full eigenvalue equation.

Unlike the main tower $\ket{\Psi_n}$, the state is not annihilated
bond by bond: generally
\begin{equation}
    \widehat k_{s,s+1}\ket{\Phi_{n,A}}\ne0,
    \qquad
    \sum_s\widehat k_{s,s+1}\ket{\Phi_{n,A}}=0.
    \label{eq: xy_decorated_not_frustration_free}
\end{equation}
The second class of rows requires cancellation between neighboring bond terms.
The family nevertheless occupies one grading sector and one uniform-potential
shell, so its internal kinetic block vanishes even though the boundary
cancellation is not bondwise.

\paragraph*{Boundary-condition limitation}

The deleted-chain sign convention in Eq.~\eqref{eq: xy_deleted_chain_creator}
does not define a full periodic tower.  For PBC with $L$ a multiple of four,
the one-magnon endpoint $n=0$ and its spin-reversed endpoint $n=L-1$ are zero
modes of the cycle graph.  For intermediate $n$, however, the same formula
leaves uncanceled rows across the periodic seam.  For example, at $L=4$ and
$n=1$,
\begin{equation}
    \frac{\|\widehat K\ket{\Phi_{1,A}}\|}
         {\|\ket{\Phi_{1,A}}\|}
    =\sqrt{2}\,|J|.
    \label{eq: xy_decorated_pbc_counterexample}
\end{equation}
We therefore claim the family in Eq.~\eqref{eq: xy_decorated_tower} only for
odd $L$ under OBC.  Whether a differently phased periodic generalization
exists is left open.

\subsubsection{Five-site periodic caged eigenstate outside the chiral-shell search}
\label{ssec: spin1_xy_five_site_nonchiral_cage}

For the spin-1 nearest-neighbor exchange, the chiral-shell grading on an open
or even periodic chain is inherited from the $A/B$ bipartition of the physical
chain.  An odd periodic chain has no such physical bipartition.  More strongly,
the kinetic configuration graph is not globally bipartite: in the
$M=-L+1$ sector, the $L$ product configurations with one $\ket0$ and
$L-1$ local $\ket-$ states are connected by nearest-neighbor exchange as the
cycle graph $C_L$, which is odd when $L$ is odd.  The obstruction persists in
the resolved sector relevant to the additional caged eigenstate.  For
$L=5$, $M=-2$, one explicit odd cycle is
\begin{equation}
\begin{aligned}
 \ket{-,0,-,0,0}
 &\longleftrightarrow \ket{0,-,-,0,0}
 \longleftrightarrow \ket{-,-,-,0,+}\\
 &\longleftrightarrow \ket{-,-,-,+,0}
 \longleftrightarrow \ket{-,-,0,0,0}\\
 &\longleftrightarrow \ket{-,0,-,0,0}.
\end{aligned}
\label{eq: xy_odd_pbc_kinetic_five_cycle}
\end{equation}
Every arrow is a nonzero nearest-neighbor exchange matrix element, including
the periodic bond in the second step.  Therefore, this fixed-magnetization kinetic graph
cannot admit the product-basis diagonal $\pm1$ grading required by
Sec.~\ref{ssec: type1_uniform_shell_search}.  The same uniform-potential shell nevertheless contains translation-invariant
caged eigenstates.

It is useful to display two particularly simple members,
because together they separate boundary caging from the auxiliary zero-mode
restriction used in the chiral-shell search.  With all site indices understood
modulo five, define
\begin{align}
    \ket{\Phi_0}
    ={}&\frac{1}{\sqrt{20}}\sum_{r=1}^{5}
      \widehat b_r^\dagger
      \big(
        \widehat a_{r+2}^\dagger
        +\widehat a_{r+1}^\dagger
        -\widehat a_{r-1}^\dagger
        -\widehat a_{r-2}^\dagger
      \big)\ket{\mathrm{F}_{-}},
      \label{eq: xy_nonbipartite_scar_zero}\\
    \ket{\Phi_{-2}}
    ={}&\frac{1}{\sqrt{20}}\sum_{r=1}^{5}
      \widehat b_r^\dagger
      \big(
        \widehat a_{r+1}^\dagger
        -\widehat a_{r+2}^\dagger
        +\widehat a_{r-2}^\dagger
        -\widehat a_{r-1}^\dagger
      \big)\ket{\mathrm{F}_{-}}.
      \label{eq: xy_nonbipartite_scar_nonzero}
\end{align}
Each sum contains the same 20 mutually orthogonal product configurations,
with coefficients of magnitude $1/\sqrt{20}$; the two states differ only in
their sign patterns.  Write $\mathcal S_{\rm nb}$ for this common support
and $\widehat P_{\rm nb}$ for its projector.  In terms of unnormalized spin operators, both prefactors
become $1/(4\sqrt{10})$ because
$(\widehat S_r^+)^2\widehat S_q^+\ket{\mathrm F_-}
=2\sqrt2\,\widehat b_r^\dagger\widehat a_q^\dagger\ket{\mathrm F_-}$.

Every support configuration contains one $\ket+$, one $\ket0$, and three
$\ket-$ states.  Hence both states lie on the same uniform potential shell,
\begin{equation}
    M=-2,
    \qquad
    \sum_{r=1}^{5}(S_r^z)^2=4,
    \qquad
    V_{\rm sh}=-2h+4D.
    \label{eq: xy_nonbipartite_uniform_shell}
\end{equation}
The distinction between the states is entirely kinetic.

\paragraph*{Internal kinetic problem}

\begin{figure}[!tp]
    \centering
    \includegraphics[width=\columnwidth]{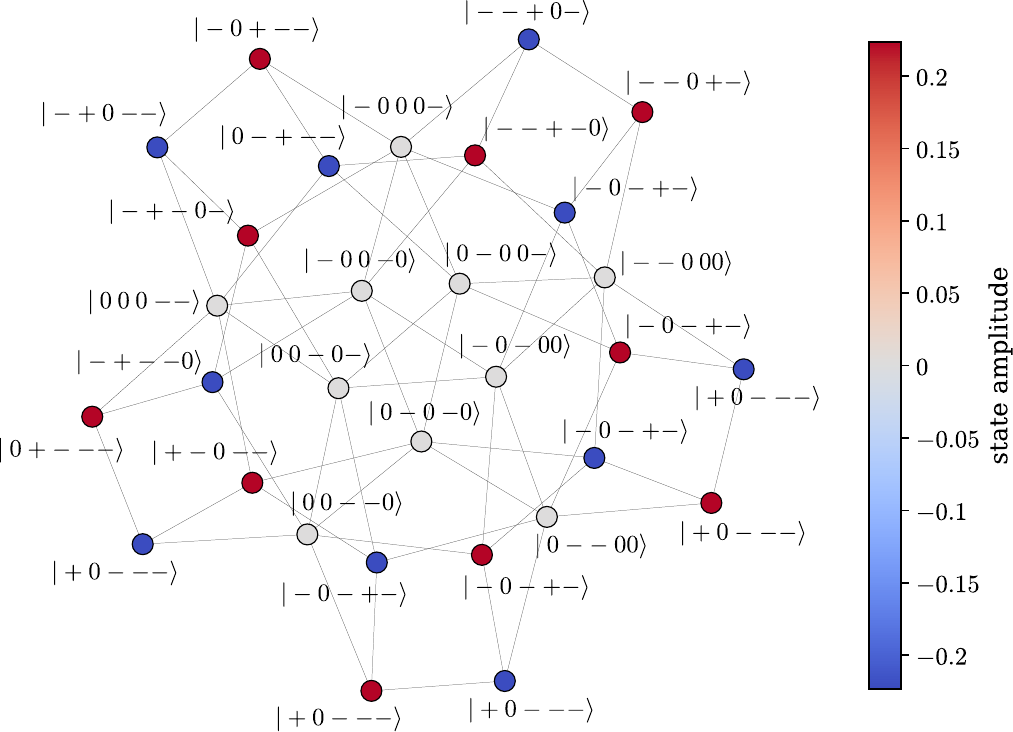}
    \caption{Configuration graph of the $L=5$ periodic chain in the
    fixed-$M=-2$ sector.  The kinetic graph is nonbipartite, as witnessed by the odd cycle
    in Eq.~\eqref{eq: xy_odd_pbc_kinetic_five_cycle}.  The common 20-vertex
    support of Eqs.~\eqref{eq: xy_nonbipartite_scar_zero} and
    \eqref{eq: xy_nonbipartite_scar_nonzero} induces the even cycle $C_{20}$,
    but it is not an independent set: the restricted internal kinetic matrix is
    nonzero.  The amplitudes drawn here are those of $\ket{\Phi_0}$ and
    repeat as $++--$ along a cyclic ordering.  The companion
    $\ket{\Phi_{-2}}$ has the alternating $+-+- $ pattern on the same
    cycle.  The remaining coupled vertices form the active configuration-space
    boundary, on which leakage amplitudes cancel.  Self-loops are omitted.}
    \label{fig: xy_graph_pbc_5_sz-2}
\end{figure}

Label a support configuration by the ordered pair $(r,q)$, with the local
$\ket+$ state at $r$ and the local $\ket0$ state at $q\ne r$.  For five sites,
the common support contains every such ordered pair.  The kinetic term
restricted to this support moves the $\ket0$ state to either neighbor,
including an exchange with the $\ket+$ state when the two are adjacent.
Every support vertex consequently has degree two, and the induced graph is the
connected cycle $C_{20}$ shown in Fig.~\ref{fig: xy_graph_pbc_5_sz-2}.
Crucially,
\begin{equation}
    \widehat P_{\rm nb}\widehat K\widehat P_{\rm nb}\neq0,
    \label{eq: xy_nonbipartite_internal_nonzero}
\end{equation}
so this support is not a chiral independent set.

For the $++--$ pattern of $\ket{\Phi_0}$, each support vertex receives
equal matrix elements from two neighbors with opposite amplitudes, giving
\begin{equation}
    \widehat P_{\rm nb}\widehat K\widehat P_{\rm nb}
    \ket{\Phi_0}=0.
    \label{eq: xy_nonbipartite_internal_zero}
\end{equation}
For the alternating pattern of $\ket{\Phi_{-2}}$, both neighboring
amplitudes have the opposite sign to the amplitude on the target vertex.  With
the convention of Eq.~\eqref{eq: xy_hamiltonian}, each internal edge has matrix
element $J$, and therefore
\begin{equation}
    \widehat P_{\rm nb}\widehat K\widehat P_{\rm nb}
    \ket{\Phi_{-2}}=-2J\ket{\Phi_{-2}}.
    \label{eq: xy_nonbipartite_internal_nonzero_eigenvalue}
\end{equation}
Thus a nonvanishing internal kinetic block is fully compatible with exact
interference caging; the internal problem needs only to close on an eigenvector,
not annihilate it.

\paragraph*{Outer-boundary cancellation}

It remains to impose the actual caging condition.  Boundary states with three
local $\ket0$ configurations receive several equal-magnitude parents from the
support.  For $\ket{\Phi_0}$, representative cancellation rows are
\begin{equation}
\begin{aligned}
    & \widehat k_{2,3}\ket{-\,+\,-\,0\,-}
      -\widehat k_{3,4}\ket{-\,0\,+\,-\,-}\\
    &+\widehat k_{2,3}\ket{-\,-\,+\,0\,-}
      -\widehat k_{3,4}\ket{-\,0\,-\,+\,-}=0,
\end{aligned}
\label{eq: xy_nonbipartite_cancellation_1}
\end{equation}
and
\begin{equation}
    \widehat k_{2,3}
      \big(\ket{-\,+\,-\,-\,0}-\ket{-\,-\,+\,-\,0}\big)=0.
    \label{eq: xy_nonbipartite_cancellation_2}
\end{equation}
The alternating state obeys the same boundary-kernel equation with its own
sign pattern.  Direct evaluation of all boundary rows gives
\begin{equation}
\begin{aligned}
    B_{\rm nb}\ket{\Phi_0}&=0,
    &B_{\rm nb}\ket{\Phi_{-2}}&=0,\\
    B_{\rm nb}&=\widehat P_{\partial\mathcal S_{\rm nb}}
                  \widehat K\widehat P_{\rm nb}.
\end{aligned}
    \label{eq: xy_nonbipartite_boundary_zero}
\end{equation}
Combining the internal and boundary equations yields the exact full-Hamiltonian
relations
\begin{align}
    \widehat H_{\rm XY}\ket{\Phi_0}
      &=(-2h+4D)\ket{\Phi_0},\\
    \widehat H_{\rm XY}\ket{\Phi_{-2}}
      &=(-2J-2h+4D)\ket{\Phi_{-2}}.
    \label{eq: xy_nonbipartite_energies}
\end{align}
Global spin reversal gives the corresponding $M=+2$ partners, with the same
kinetic eigenvalues and with $-2h$ replaced by $+2h$.

The second state is the relevant scope test for the search logic.  It is an
exact interference-caged eigenstate on a nonbipartite fixed-magnetization kinetic graph,
its support has a nonvanishing internal kinetic matrix, and its kinetic energy
is nonzero.  Therefore, neither bipartiteness nor the zero-internal-matrix
(chiral-shell) reduction is part of the definition of interference caging.
Those ingredients only simplify the practical search.  In the general
uniform-shell problem, the fundamental conditions are instead the internal
eigenvalue equation
$\widehat P_{\mathcal S}\widehat K\widehat P_{\mathcal S}\ket\psi
=\varepsilon_K\ket\psi$ together with the active-boundary equation
$\widehat P_{\partial\mathcal S}\widehat K\widehat P_{\mathcal S}\ket\psi=0$.
We do not infer from this isolated finite-size example a thermodynamic family
or a complete perturbative stability classification.

\subsubsection{Independent finite-size residual checks}
\label{ssec: xy_additional_validation}

Direct construction of the full fixed-magnetization Hamiltonian provides an
independent check: for the open-chain family, the relative kinetic residual is
below $10^{-14}$ for every $n=0,\ldots, L-1$ at $L=3,5,7,9$; for the two translation-invariant periodic five-site states, the residuals are
likewise below $10^{-14}$.  These checks validate the closed-form expressions
but are not used in their proofs.

\section{Square QDM: local caging structure, thermal evidence, and deformation details}
\label{appendix: square_qdm_cage_manifold}

This appendix supplies the technical support for the three-stage square-QDM
argument in Sec.~\ref{sec: square_qdm}.  We first resolve the finite-size caged
subspace into compact and collective parts and establish the environment
independence required for bounded physical caging operators.  We then give the
exact circumference-four repetition, the fully resolved finite-temperature
comparison, and the deformation diagnostics.  Secondary nonuniform-potential,
uniform-Peierls, and cage-excised constructions are kept explicitly separate
from the primary checkerboard family.  The fixed-width evidence is not
identified with a true two-dimensional thermodynamic limit.

\subsection{Finite-size caged subspace and local--collective decomposition}

\subsubsection{Chiral-shell caged subspaces}
\label{ssec: qdm_certified_cage_spaces}

For a fixed grading sign $c$ and flippable-plaquette number $n$, the chiral-shell
candidate shell is the space $\mathcal B_{c,n}$ of
Eq.~\eqref{eq: qdm_type1_shell}.  Because its internal kinetic block vanishes,
the complete caged subspace in that shell is
\begin{equation}
    \mathcal E_{(0,n)}
    \equiv
    \ker B_{c,n},
    \qquad
    B_{c,n}
    =\widehat P_{\partial\mathcal B_{c,n}}\widehat K\widehat P_{c,n}.
    \label{eq: qdm_cage_manifold_definition}
\end{equation}
The notation $(0,n)$ records the kinetic and potential eigenvalues.  The
subspace $\mathcal E_{(0,n)}$ is invariant under rotations of the numerical
nullspace basis; individual representatives are introduced only after a sparse basis has been
selected.

For the $4\times4$ torus with periodic boundary conditions in the
$(W_x,W_y)=(0,0)$ sector, the constrained Hilbert-space dimension is $132$.
Table~\ref{tab:qdm_certified_records} lists the shell nullities under the
singular-value convention of Appendix~\ref{ssec: finite_size_search_certification}.

\begin{table*}[tp]
    \caption{Finite-size square-QDM chiral-shell results.
    The sector $(0,0)$ denotes $(W_x,W_y)=(0,0)$, and each signature lists
    (kinetic eigenvalue, flippable-plaquette count).  The fourth column gives the dimension of the linearly independent caged
    subspace in the indicated shell.  The $8\times4$ entries are finite-size
    results and should not be read as a two-dimensional thermodynamic limit.}
    \label{tab:qdm_certified_records}
    \setlength{\tabcolsep}{8pt}
    \setlength{\JustifyingParindent}{0pt}
    \renewcommand{\arraystretch}{1.3}
    \begin{ruledtabular}
    \begin{tabular}{ccccc}
        lattice & boundary/sector & signature & $\dim\mathcal E$ & \parbox[t]{0.40\textwidth}{\justifying established role} \\
        \hline
        $4\times4$ & PBC, $(0,0)$ & $(0,4)$ & $9$ & \parbox[t]{0.40\textwidth}{\justifying $8+1$ local/collective decomposition} \\[3pt]
        $4\times4$ & PBC, $(0,0)$ & $(0,6)$ & $1$ & \parbox[t]{0.40\textwidth}{\justifying additional chiral-shell caged eigenstate} \\[3pt]
        $8\times4$ & PBC, $(0,0)$ & $(0,4)$ & $8$ & \parbox[t]{0.40\textwidth}{\justifying fixed-width padded local representatives} \\[3pt]
        $8\times4$ & PBC, $(0,0)$ & $(0,8)$ & $4$ & \parbox[t]{0.40\textwidth}{\justifying multi-block continuation} \\
    \end{tabular}
    \end{ruledtabular}
\end{table*}

The $(0,4)$ caged subspace is the focus below.  The additional $(0,6)$ state is
an independent chiral-shell caged eigenstate but is not needed for the
local-versus-collective comparison.  These shell dimensions are finite-size
nullities; the compact representative used in the main argument has the
explicit algebraic boundary construction below.  The smallest nonzero boundary
singular values are $0.842$ and $0.874$ in the $(0,4)$ and $(0,6)$ shells,
well separated from zero, and the largest full-Hamiltonian residual among the
representatives reported below is $5.0\times10^{-15}$.

\subsubsection{Sparse representatives and the \texorpdfstring{$8+1$}{8+1} subspace decomposition}
\label{ssec: qdm_ipr_localized_manifold}

Let the columns of $U_{0,4}$ be an orthonormal basis of
$\mathcal E_{(0,4)}$.  We select sparse representatives with the IPR
optimization of Eq.~\eqref{eq: cage_subspace_ipr}, using $U=U_{0,4}$ and
orthogonality constraints after each vector.  This choice displays local
cancellation motifs but does not change the caged subspace or its dimension;
the apparent support and IPR depend on the chosen representative basis.

In a sparse representative basis, vectors $0$--$7$ span a local-motif
subspace $\mathcal E_{\mathrm{loc}}$.  Each representative
can be resolved into bounded real-space cancellation units.  The remaining representative is chosen orthogonal to this span.  The resulting statement is the
basis-independent decomposition
\begin{equation}
\begin{aligned}
    \dim\mathcal E_{(0,4)}&=9,\\
    \dim\mathcal E_{\mathrm{loc}}&=8,\\
    \dim\!\left(\mathcal E_{(0,4)}/\mathcal E_{\mathrm{loc}}\right)&=1.
\end{aligned}
    \label{eq: qdm_appendix_9_8_1}
\end{equation}
Thus the phrase ``ninth state'' is shorthand for a representative of the
one-dimensional quotient after the declared stripe-generated caged subspace has been fixed; it is not an
invariant label attached to an arbitrary eigenbasis.

\subsubsection{Compact representative and local boundary factorization}
\label{ssec: qdm_compact_sector}

Each of the eight local representatives decomposes into two bounded stripe
cancellation blocks with signature $(0,2)$.  The two
active stripes are displaced by one plaquette, while the remaining dimer
configuration acts as a constrained spacer.  More precisely, after the
inactive boundary rows are removed, the boundary equation separates into two
finite local blocks,
\begin{equation}
    B_{c,4}\boldsymbol x
    \simeq
    \left(B^{(1)}_{0,2}\boldsymbol x^{(1)}\right)
    \oplus
    \left(B^{(2)}_{0,2}\boldsymbol x^{(2)}\right)
    =0.
    \label{eq: qdm_two_stripe_boundary_factorization}
\end{equation}
The symbol $\simeq$ denotes the reordering and deletion of rows that are
identically inactive on the chosen support; it does not assert a tensor-product
factorization of the full constrained QDM Hilbert space.  Translations and
lattice rotations generate the eight-dimensional span from the same bounded
motif.

For local equations we retain the enumeration convention introduced in
Sec.~\ref{sssec: qdm_compact_reduced_iz_decomposition}: \(|D_j\rangle\) is the
\(j\)th covering in a fixed basis ordering, and the index has no physical
meaning.

A representative local state is the four-configuration vector introduced in
Eq.~\eqref{eq: qdm_compact_cage_state},
\begin{equation}
    |\psi_{\mathrm{motif}}\rangle
    =\frac12\left(|D_{1}\rangle-|D_{2}\rangle+|D_{3}\rangle-|D_{4}\rangle\right).
    \label{eq: qdm_appendix_local_state}
\end{equation}
All four coverings have the same grading sign and
$N_{\mathrm{fp}}=4$.  The internal kinetic block therefore vanishes and the
only nontrivial condition is cancellation on the eight boundary
configurations reached by one plaquette flip.

\paragraph*{Explicit support--boundary matrix}
\label{par:qdm_explicit_support_boundary_matrix}
\label{ssec: qdm_compact_boundary_matrix}

Order the support columns as
\begin{equation}
\begin{aligned}
    \mathcal S_{\mathrm{loc}}
    &=\bigl(
      |D_{1}\rangle,|D_{2}\rangle,
      |D_{3}\rangle,|D_{4}\rangle
      \bigr),\\
    \partial\mathcal S_{\mathrm{loc}}
    &=\bigl(
      |h_{1}\rangle,|h_{2}\rangle,
      |h_{3}\rangle,|h_{4}\rangle,\\
    &\hspace{2.2em}
      |h_{5}\rangle,|h_{6}\rangle,
      |h_{7}\rangle,|h_{8}\rangle
      \bigr).
\end{aligned}
\label{eq: qdm_compact_ordered_basis}
\end{equation}
In the convention
$\widehat K=-\sum_p\widehat K_p$, the complete support-to-boundary block is
\begin{equation}
    B_{\mathrm{loc}}
    =-
    \begin{pmatrix}
        1&1&0&0\\
        1&1&0&0\\
        0&1&1&0\\
        0&1&1&0\\
        0&0&1&1\\
        0&0&1&1\\
        1&0&0&1\\
        1&0&0&1
    \end{pmatrix},
    \qquad
    \boldsymbol x_{\mathrm{loc}}
    =\frac12
    \begin{pmatrix}
        1\\-1\\1\\-1
    \end{pmatrix}.
    \label{eq: qdm_compact_boundary_matrix}
\end{equation}
Direct multiplication gives
\begin{equation}
    B_{\mathrm{loc}}\boldsymbol x_{\mathrm{loc}}=0.
    \label{eq: qdm_compact_boundary_null}
\end{equation}
Together with the vanishing internal block and the uniform-potential shell,
this proves
\begin{equation}
    \widehat K|\psi_{\mathrm{motif}}\rangle=0,
    \qquad
    \widehat H_{\mathrm{QDM}}|\psi_{\mathrm{motif}}\rangle
    =4\lambda|\psi_{\mathrm{motif}}\rangle.
    \label{eq: qdm_compact_full_certificate}
\end{equation}
Equation~\eqref{eq: qdm_compact_boundary_matrix} is the concise form of the
explicit finite-size support--boundary matrix; entries outside this block
are irrelevant to the eigenvalue equation for this vector.

The paired duplicate rows in Eq.~\eqref{eq: qdm_compact_boundary_matrix}
correspond to distinct boundary configurations generated by the same local
signed cancellation.  They are precisely the structure that allows the exact
configuration-boundary equations to yield a bounded family of real-space
kinetic operators for the locally generated subspace.

\subsubsection{Collective quotient representative as a finite-size counterexample}
\label{ssec: qdm_collective_quotient}

\begin{figure*}[!tp]
    \centering
    \includegraphics[width=\textwidth]{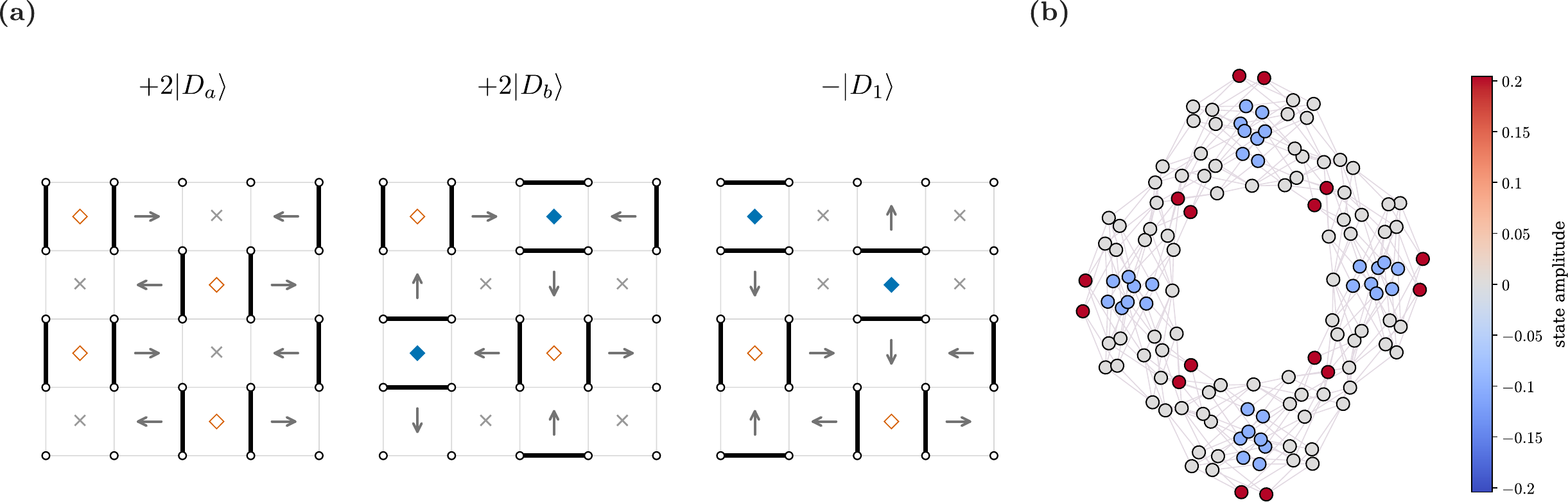}
    \caption{Collective quotient representative as a finite-size counterexample.
    \textbf{(a)} Representative real-space dimer coverings from the collective
    support, labeled by their relative signed coefficients.  \textbf{(b)} The
    corresponding sparse representative of the one-dimensional quotient of
    the target-energy caged subspace by its locally reducible subspace,
    $\mathcal E_{(0,4)}/\mathcal E_{\mathrm{loc}}$, on the configuration graph;
    vertex color encodes the real state amplitude.  Self-loops are omitted.
    This representative remains an exact caged eigenstate on the finite $4\times4$ torus
    even though the known bounded local stripe reduction does not annihilate it.}
    \label{fig:qdm_collective_counterexample}
\end{figure*}

Choose a normalized vector
$|\psi_{\mathrm{col}}\rangle\in\mathcal E_{(0,4)}$ orthogonal to
$\mathcal E_{\mathrm{loc}}$.  Because the quotient in
Eq.~\eqref{eq: qdm_appendix_9_8_1} is one-dimensional, this vector is unique up
to phase and addition of a local-sector vector before the orthogonality convention
is imposed.  For exposition, it is convenient to write
\begin{equation}
    |\psi_{\mathrm{col}}\rangle
    =\sum_{D\in\mathcal S_{\mathrm{col}}} a_D |D\rangle,
    \qquad
    \sum_{D\in\mathcal S_{\mathrm{col}}}|a_D|^2=1,
    \label{eq: qdm_collective_compact_form}
\end{equation}
where $\mathcal S_{\mathrm{col}}$ is the support displayed in
Fig.~\ref{fig:qdm_collective_counterexample}.  Panel (a) shows representative
real-space dimer coverings from this support with their relative signed
coefficients, while panel (b) embeds the same representative in the full
configuration graph and colors vertices by state amplitude.

The finite-size boundary kernel satisfies
\begin{equation}
    B_{c,4}\boldsymbol x_{\mathrm{col}}=0,
    \qquad
    \widehat H_{\mathrm{QDM}}|\psi_{\mathrm{col}}\rangle
    =4\lambda|\psi_{\mathrm{col}}\rangle,
    \label{eq: qdm_collective_appendix_certificate}
\end{equation}
within numerical precision.  Unlike the selected four-covering motif, no
separate algebraic boundary construction is used for this quotient
representative.  Its boundary rows,
however, are linked by cancellations involving several
support configurations.  The same collection of independently annihilating
bounded stripe operators used for $\mathcal E_{\mathrm{loc}}$ does not
annihilate this quotient representative.

This is a positive statement about the present decomposition, not a no-go
theorem.  The calculations establish that the collective quotient lies outside
the span generated by the known local stripe caged eigenstates.  They do not exclude a
different bounded multi-row kinetic operator.  The remaining locality question is the minimum real-space radius and the
minimum number of coupled boundary rows required by another operator class.  The failure of the known single-row/stripe operators
therefore provides a finite-size counterexample to the converse implication
``bounded local reduction fails $\Rightarrow$ no exact caged eigenstate''; it does not
settle whether the collective cancellation admits an
$O(1)$ multi-row local caging operator or instead requires an operator whose
support grows with system size.

\subsubsection{Graph visualization of repeated finite-size motifs}
\label{ssec: qdm_automorphism_orbit}

The Hamiltonian graph provides a useful way to display the product
configurations participating in the finite-size solutions.  In the
$4\times4$ zero-winding graph, the configurations supporting the eight local
representatives display repeated motifs under translations and rotations.
These visual repetitions are neither a necessary nor a sufficient caging
criterion.  Exactness follows only from the
signed boundary equation.

\subsection{Bounded local caging operators and the collective counterexample}

\subsubsection{Environment independence and local transport}
\label{ssec: qdm_complementary_sector_certificate}

\begin{figure}[!htbp]
    \centering
    \includegraphics[width=\columnwidth]{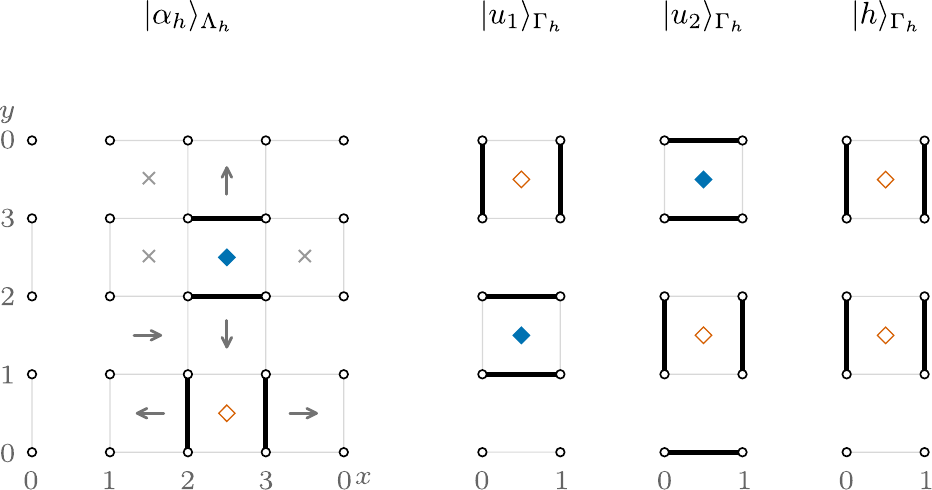}
    \caption{Environment independence for the square-QDM
    compact-motif caged eigenstate.  For the representative boundary row
    $h$, the common outside state separates from the two
    configuration-support parents and their boundary image; the outside
    projector is removed only after establishing this environment
    independence.}
    \label{fig:qdm_appendix_environment_independence}
\end{figure}

The support--boundary equations ensure cancellation on the configuration
graph.  A physical local caging operator requires the stronger statement that
the same operator acts in every conditionally allowed outside environment.
For one representative boundary row, the conditioned graph transition is
\begin{equation}
    \widehat A_{h}
    =\widehat P_{\alpha_{h};\Lambda_{h}}
    \otimes
    \left(
      |h\rangle\langle u_1|
      +|h\rangle\langle u_2|
    \right)_{\Gamma_{h}},
    \label{eq:qdm_directed_local_witness_example}
\end{equation}
as shown in Fig.~\ref{fig:qdm_appendix_environment_independence}. For the compact stripe row, the constrained-basis enumeration shows that each
occupied environment either has zero conditional weight of the caged eigenstate or its conditional
state is annihilated by the same two-parent operator.  The outside projector in
Eq.~\eqref{eq:qdm_directed_local_witness_example} can therefore be replaced by
the identity on the complement of the local cluster.

The resulting operators act on the eight-link region
$R=R_x^{(\eta)}$ in Eq.~\eqref{eq: qdm_eight_link_region}.
Writing a link as its unordered endpoint pair, its links are explicitly
\begin{equation}
\begin{aligned}
 R={}&\bigl\{\{(x,y),(x+1,y)\}:y=0,1,2,3\bigr\}\\
 &\cup\bigl\{\{(x+a,\eta+2j),(x+a,\eta+2j+1)\}:\\
 &\hspace{7em}a,j=0,1\bigr\},
 \label{eq: qdm_appendix_eight_link_coordinates}
\end{aligned}
\end{equation}
where $y$ is taken modulo four and $\eta$ fixes the chosen placement.
The two adjacent vertex columns describe the geometric envelope, not the
larger twelve-link full-column operator region.

For the active local chart in
Fig.~\ref{fig:qdm_appendix_environment_independence}, write $v$ and $h$
for the two vertical and two horizontal dimers on a selected plaquette.
The two parents are $|u_1\rangle=|vh\rangle$ and
$|u_2\rangle=|hv\rangle$, and their common child is
$|b\rangle=|vv\rangle$ (the boundary configuration labeled $h$ in the
figure).  Exchanging the plaquettes only exchanges the parent labels.
After removing the outside projector, the dimensionless normalization used
in the thermal data is
\begin{equation}
\begin{aligned}
 \widehat A_R&=\frac{|b\rangle(\langle u_1|+\langle u_2|)}{\sqrt2},\\
 \widehat Z_R&=\widehat A_R+\widehat A_R^\dagger,
 \qquad \|\widehat A_R\|=\|\widehat Z_R\|=1.
 \label{eq: qdm_appendix_normalized_local_chart}
\end{aligned}
\end{equation}
Both operators vanish on all other allowed local patterns.  In every
occupied outside environment, the active conditional component is
proportional to $|u_1\rangle-|u_2\rangle$, or the chart has zero weight;
the child has zero caged-state amplitude.  These facts give darkness under
both $A_R$ and $A_R^\dagger$.  Transporting this fixed chart, rather than
rescaling with the strip length, defines the kinetic operators at every size.

For the localized compact-motif caged eigenstate, the corrected shifted shell is
\begin{equation}
    \widehat Y_R^{\mathrm{shell}}
    =\frac12\left[
      \widehat F_{p_{(x,0)}}+\widehat F_{p_{(x,2)}}
      -2\widehat{\mathbf1}
    \right].
    \label{eq: qdm_appendix_transported_Y}
\end{equation}
Every occupied localized support configuration has both selected plaquettes
flippable, and the shell residual vanishes through $12\times4$.  The same
fixed local operator is not dark after momentum projection: its residual is
$0.75$ and the expectation of its positive square is $0.5625$ at these lengths.
It is therefore excluded from the primary symmetry-resolved thermal and
joint-dark claims.  The bounded fixed-width construction for that route uses
$\widehat A_R,\widehat Z_R$; the shifted shell is retained for the localized,
translation-breaking potential family.

This construction applies to the compact locally generated subspace.  It does
not imply that the collective quotient admits the same bounded local
operator.

\subsubsection{Finite-size annihilator radius}
\label{ssec: qdm_local_witness_continuation}
\label{ssec: qdm_collective_locality_appendix}

Within the plaquette-linear operator class, the two mechanisms separate
already on the $4\times4$ torus.  A radius-$r$ cluster about a recorded
plaquette center contains the plaquettes whose maximum shortest periodic
coordinate distance from that center is at most $r$ (Chebyshev radius).
The centers may differ between radii and representatives; the retained
search table records the center used for each quoted residual.  For the compact representative, a radius-one cluster has a
null combination of two active plaquette operators with residual
$3.3\times10^{-16}$.  For the collective representative, radius zero and one
have trivial kernel; the best radius-one residual is $0.675$.  A null vector
appears at radius two only when all 16 plaquette operators are active, with
residual $1.4\times10^{-15}$.  Thus, on the $4\times4$ torus, the first collective null vector in this
operator class spans the whole system.  This finite-size radius result
does not determine how the minimum support scales with $L_x$ and therefore is
not by itself a no-go theorem for a bounded multi-row reduction along a larger
sequence.

\subsection{Fixed-width continuation and thermal evidence}

\subsubsection{Repeated periodic-product construction}
\label{ssec: qdm_fixed_width_appendix}
\label{ssec: qdm_continuation_diagnostics}

The $8\times4$ shell dimensions are listed in
Table~\ref{tab:qdm_certified_records}.  The same compact construction repeats
along the circumference-four sequence, whose finite-size
constraint matrices are
\begin{center}
\small
\setlength{\tabcolsep}{4.5pt}
\renewcommand{\arraystretch}{1.3}
\begin{tabular}{@{}ccccc@{}}
\hline\hline
system & \shortstack{support\\size} & \shortstack{kinetic\\parameters}
    & \shortstack{compatible\\dimension}
    & \shortstack{boundary\\conditioning\\scale} \\[2pt]
\hline
$4\times4$  & $4$   & $16$ & $14$ & $2$ \\
$8\times4$  & $16$  & $32$ & $28$ & $2$ \\
$12\times4$ & $64$  & $48$ & $42$ & $2$ \\
$16\times4$ & $256$ & $64$ & $56$ & $2$ \\
\hline\hline
\end{tabular}
\end{center}
Along this circumference-four sequence, the boundary kernel is one-dimensional.
Two equal-coupling conditions
repeat in each four-column cell, while the full local
flippability-potential basis remains compatible.  The support sizes
$4,16,64,256$ make explicit that the global state is not a four-covering
compact localized object at large $L_x$: compactness belongs to the repeated
local motif and to its bounded physical caging operator.

This finite-size pattern follows from an exact local repetition argument.
Within one four-column motif, only plaquette columns $x=0,2$ are flippable on
the product support; columns $x=1,3$ are inactive for every support
configuration.  Hence no plaquette term crosses an inter-motif seam.  On each
active column the two support parents have opposite exact amplitudes and every
kinetic child receives their paired contributions.  For the checkerboard
pattern $\chi_{p_{(x,y)}}=(-1)^{x+y}$ these parents acquire the same phase because
$\chi_{p_{(x,y+2)}}=\chi_{p_{(x,y)}}$, so their integer coefficients cancel identically.
Every support configuration has four flippable plaquettes per motif.  Therefore
for every positive integer $N$ and arbitrary real $\varphi$,
\begin{equation}
    \widehat H_{\varphi,\lambda}\ket{\psi_{4N}}
    =4N\lambda\ket{\psi_{4N}},
    \qquad
    e_\psi=\lambda/4.
    \label{eq: qdm_appendix_exact_periodic_product}
\end{equation}

The same local repetition proves continuation of the bounded kinetic caging
operators at arbitrary strip length.  Choose a translated eight-link region
$R_x^{(\eta)}$ of Eq.~\eqref{eq: qdm_eight_link_region} inside one repeated
cell and decompose the periodic-product state by its
outside dimer environment $\alpha$,
\begin{equation}
    |\psi_{4N}\rangle
    =\sum_{\alpha} c_{\alpha}
      |\alpha\rangle_{\bar R}\otimes|\phi_R(\alpha)\rangle.
    \label{eq: qdm_appendix_conditional_repetition}
\end{equation}
By the environment-independence result of
Sec.~\ref{ssec: qdm_complementary_sector_certificate}, every nonzero
conditional state $|\phi_R(\alpha)\rangle$ is either inactive for the directed
transition or contains the same paired support parents with opposite
coefficients.  Thus $\widehat A_R|\phi_R(\alpha)\rangle=0$ for every occupied
environment separately.  The corresponding boundary child $|h\rangle$ lies
outside the periodic-product support, so the reverse transition also has no
support amplitude and $\widehat A_R^\dagger|\psi_{4N}\rangle=0$.  Since the
checkerboard-compatible parent flips carry the same phase,
$\chi_{p_{(x,y+2)}}=\chi_{p_{(x,y)}}$, this conditional identity is unchanged
for arbitrary real $\varphi$.  Consequently
\begin{equation}
    \widehat A_R|\psi_{4N}\rangle
    =\widehat Z_R|\psi_{4N}\rangle=0,
    \qquad N\in\mathbb N,
    \label{eq: qdm_appendix_allN_local_annihilation}
\end{equation}
for the same bounded eight-link operator at every system size.  The residual checks
at $L_x=4,8,12$ therefore test this analytic identity numerically; they are not
a finite-size substitute for it.

\begin{figure*}[tp]
    \centering
    \begin{minipage}[t]{0.45\textwidth}
        \raggedright
        \small\textbf{(a)}\par\smallskip
        \centering
        \includegraphics[width=\linewidth]{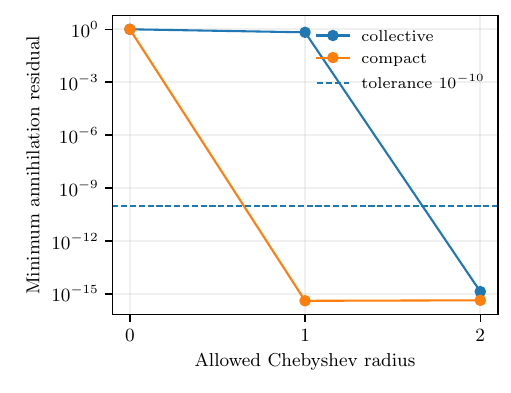}
    \end{minipage}
    \hspace{6pt}
    \begin{minipage}[t]{0.45\textwidth}
        \raggedright
        \small\textbf{(b)}\par\smallskip
        \centering
        \includegraphics[width=\linewidth]{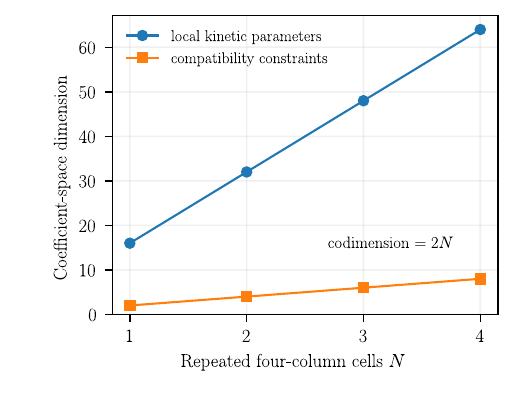}
    \end{minipage}
    \caption{Finite-size locality and fixed-width compatibility for the
    square-QDM compact-motif caged eigenstate.  \textbf{(a)} In the
    plaquette-linear annihilator search on the $4\times4$ torus, the compact
    representative has a radius-one null vector, whereas the collective
    representative first acquires a null vector on the radius-two region
    containing all plaquettes.  This is a finite-size statement within the
    declared plaquette-linear class.  \textbf{(b)} Along the
    circumference-four strip sequence, the number of local kinetic parameters
    grows linearly while compatibility imposes two independent constraints per
    repeated four-column cell.  Equivalently, the compatible kinetic dimension
    remains a fixed fraction of the local parameter count over this
    sequence; no two-dimensional thermodynamic scaling is inferred.}
    \label{fig:qdm_appendix_locality_scaling_certificates}
\end{figure*}

Its kinetic codimension along the circumference-four sequence nevertheless
grows with the number of repeated cells:
\begin{equation}
    \text{size extensibility}
    \not\Rightarrow
    \text{finite-codimension protection}.
    \label{eq: extensibility_not_finite_codimension}
\end{equation}
Figure~\ref{fig:qdm_appendix_environment_independence} isolates the
local environment-independence argument used to remove the common outside
projector.  The finite-size radius test and fixed-width compatibility
results are shown in Fig.~\ref{fig:qdm_appendix_locality_scaling_certificates}(a)
and (b), respectively.

\subsubsection{Resolved finite-size thermal comparison}
\label{ssec: qdm_microcanonical_activity_appendix}

At $\lambda_\star/J=1$, the unprojected strip $\beta=0$ witness values
approach $(\tau_A,\tau_Z)=(0.04488,0.08977)$, but transfer counting through
$L_x=256$ gives
\begin{equation}
 (e_{\beta=0}-e_\psi)/J\simeq5.98\times10^{-3}.
 \label{eq: qdm_appendix_beta0_energy_failure}
\end{equation}
This counting baseline is not the energy-matched thermal reference.

For $\varphi>0$, resolve the checkerboard-preserving translations
$T_{\rm diag}=T_xT_y$ and $T_y^2$ in
$(k_{\rm diag},T_y^2)=(\pi/2,+1)$, then the transverse reflection $S_y$.
The translation-sector dimensions $(15,1125,114483)$ at $L_x=4,8,12$
reduce to $(1,875,39667)$ on selecting the target parities
$s_y=(-1,+1,-1)$.  The one-dimensional $4\times4$ sector contains no
thermal background and is omitted below; $\varphi=0$ is a separate
symmetry-enhanced endpoint.

\begin{table}[!htbp]
 \caption{Resolved square-QDM thermal comparison at $\lambda_\star/J=1$,
 $\varphi_\star=0.05$, and raw half-width $J/4$ about $E_\psi=L_xJ$.
 $\mathcal H_{\rm ref}$ is the selected symmetry sector and
 $r_{\rm jd}$ the translated-$A,Z$ joint-dark rank.  Canonical means are
 exact at $8\times4$ and stochastic at $12\times4$.
 $\Delta$ is defined in Eq.~\eqref{eq: qdm_local_ensemble_mismatch}.}
 \label{tab: qdm_resolved_thermal_comparison}
 \setlength{\tabcolsep}{8pt}
 \renewcommand{\arraystretch}{1.15}
 \begin{ruledtabular}
 \begin{tabular}{lcc}
 Quantity & $8\times4$ & $12\times4$ \\
 \hline
 $s_y$ & $+1$ & $-1$ \\
 $\dim\mathcal H_{\rm ref}$ & $875$ & $39667$ \\
 $N_{\rm win}$ & $47$ & $1959$ \\
 $r_{\rm jd}$ & $1$ & $3$ \\
 $\beta J$ & $0.079144$ & $\simeq0.046123$ \\
 $\tau_A^{\rm mc,raw}$ & $0.049315$ & $0.037115$ \\
 $\tau_Z^{\rm mc,raw}$ & $0.101265$ & $0.073664$ \\
 $\tau_A^{\rm can}$ & $0.057212$ & $0.038553$ \\
 $\tau_Z^{\rm can}$ & $0.111345$ & $0.075915$ \\
 $w^{\rm raw}$ & $0.070856$ & $0.024825$ \\
 $w^{\rm clean}$ & $0.056869$ & $0.024483$ \\
 $\Delta$ & $0.010080$ & $0.002251$
 \end{tabular}
 \end{ruledtabular}
\end{table}

Table~\ref{tab: qdm_resolved_thermal_comparison} uses the complete raw
windows, with energy-density half-width $J/(16L_x)$.  The $L_x=8$
spectrum is exact.  At $L_x=12$, subspaces of dimensions $4096$ and $5120$
give the same window population and raw observables within
$9.5\times10^{-15}$.  The latter spectrum covers
$[11.342636J,12.657313J]$, beyond both edges of
$[11.75J,12.25J]$; the maximum in-window eigenpair residual is
$6.18\times10^{-10}J$ and the Frobenius orthogonality residual is
$3.20\times10^{-13}$.

The $L_x=12$ canonical comparator uses $16$ random-state samples
(stochastic trace estimation) and an $81$-point inverse-temperature grid.
Its ratio-jackknife standard errors, conditional on matched $\beta$, are
$5.00\times10^{-5}$ and $5.02\times10^{-5}$ for $A,Z$; sample refinement
is checked separately.  Changing from $41$ to $81$ grid points changes
either mean by less than $1.9\times10^{-7}$.

Neighboring half-widths $J/10,J/5,J/2$ give positive raw microcanonical witness values.
Including the primary $J/4$ window, the mismatch ranges are
$[0.005195,0.010080]$ and $[0.002066,0.002372]$ at $L_x=8,12$.
The $L_x=8$ scan $\varphi=0.025,0.05,0.075,0.1$ gives primary-window
$\Delta\in[0.010080,0.010821]$ and
$w^{\rm raw}\in[0.070855,0.072375]$; no $L_x=12$ phase scan is available.
Across the window-width controls, the raw width ranges are
$[0.052291,0.094604]$ and $[0.022846,0.026022]$, respectively.

The raw reference retains the joint-dark fractions $1/47$ and $3/1959$.
Their removal changes a tested witness mean only by
$\tau_\alpha^{\rm clean}=\tau_\alpha^{\rm raw}/(1-r_{\rm jd}/N_{\rm win})$,
up to numerical darkness tolerance.  The cleaned covariance widths in
Table~\ref{tab: qdm_resolved_thermal_comparison} are independent controls.

For the eight-link region $R$, group its $27$ allowed occupation patterns
by boundary flux: the number of occupied links in $R$ incident on each
boundary vertex.  Local changes preserve the unchanged outside environment
precisely within these blocks.  Of the $18$ blocks, eleven have dimension one, six
dimension two, and one dimension four, so the complete real Hermitian
space $\mathcal A_R=\bigoplus_f\operatorname{Herm}(d_f)$ has dimension
$11+6\cdot2^2+4^2=51$.
With local inner product $(O,O')_R=\sum_f\operatorname{Tr}_f(OO')$, let
$\Phi_L(O)=P_L\iota_R(O)P_L$, where $P_L$ selects the spatial sector and
$\iota_R$ embeds the constrained local operator.  The covariance uses the
quotient of this \emph{local} Hilbert--Schmidt metric,
\begin{equation}
 \|[O]\|_{R,L}=\min_{K\in\ker\Phi_L}\|O+K\|_{\mathrm{HS},R}.
 \label{eq: qdm_local_hs_quotient_metric}
\end{equation}
Retain an orthonormal coefficient basis of $(\ker\Phi_L)^\perp$.
The projected many-body Gram matrix identifies null directions only;
its nonzero eigenvalues do not whiten or rescale the operators, so the
local normalization stays fixed with size.  Both lengths have projected
rank $25$, or $24$ covariance directions after centering the identity.
Applying Eqs.~\eqref{eq: xy_block_invariant_covariance}--\eqref{eq: xy_local_concentration_width}
to this quotient defines $w^{\rm raw}=\sqrt{\lambda_{\max}\Sigma^{\rm raw}}$
with exact-energy-block compression.  Completeness concerns this eight-link
region only, not the twelve-link full column or every bounded region.

These two lengths give positive raw microcanonical witness values and a
narrowing background, without a size fit.  A thermodynamic ICQMBS classification still requires
the raw-window conditions of Definition~\ref{def: icqmbs}; a canonical
continuation additionally requires bounded-local equivalence to that raw
sequence.  Uniform thermal continuation on an open checkerboard interval
and a growing-circumference two-dimensional limit remain unestablished.

\subsection{Deformation and joint local-operator continuation}

\subsubsection{Cage obstruction and fixed-state joint equations}
\label{ssec: qdm_deformation_maps}

On the $4\times4$ torus, the kinetic deformation basis contains 16 real flip
amplitudes and 16 real Peierls-phase directions.  We compare the full cage
obstruction, which allows the state amplitudes to readjust, with the
fixed-state map
\begin{equation}
    \mathcal A_{\psi}(\delta\widehat K)
    =
    (\widehat{\mathbf1}-\widehat P_{\psi})
    \delta\widehat K|\psi\rangle,
    \qquad \widehat P_{\psi}=|\psi\rangle\!\langle\psi|.
    \label{eq: state_resolved_deformation_map}
\end{equation}
Let $m$ denote the number of real deformation coordinates in the selected
sector.  The state and joint local-operator dimensions are summarized in
the accompanying table.

\begin{center}
\small
\setlength{\tabcolsep}{4.5pt}
\renewcommand{\arraystretch}{1.25}
\begin{tabular}{@{}lcccc@{}}
\hline\hline
\shortstack[l]{deformation\\sector} & $m$ & \shortstack{full\\preserving}
    & \shortstack{fixed\\state} & \shortstack{joint fixed\\state/operator} \\[2pt]
\hline
\multicolumn{5}{@{}l}{\emph{Compact representative}} \\[2pt]
amplitudes & $16$ & $14$ & $14$ & $14$ \\
phases & $16$ & $16$ & $14$ & $14$ \\
potentials & $16$ & $16$ & $16$ & $16$ \\
amplitudes+phases & $32$ & $30$ & $28$ & $28$ \\[3pt]
\hline
\multicolumn{5}{@{}l}{\emph{Collective representative}} \\[2pt]
amplitudes & $16$ & $5$ & $1$ & $1$ \\
phases & $16$ & $15$ & $0$ & $0$ \\
potentials & $16$ & $9$ & $9$ & $9$ \\
amplitudes+phases & $32$ & $20$ & $1$ & $1$ \\
\hline\hline
\end{tabular}
\end{center}

The discrepancy between the full and fixed-state columns is the finite-size
tangent-lift effect of Sec.~\ref{ssec: deformation_general_criteria}.  These
ranks concern a selected sparse representative.  A basis-independent claim for
the complete nine-dimensional caged subspace would instead use
\begin{equation}
    \mathcal A_{\mathcal E_{(0,4)}}(\delta\widehat K)
    =(\widehat{\mathbf1}-\widehat P_{\mathcal E_{(0,4)}})
      \delta\widehat K \widehat P_{\mathcal E_{(0,4)}}.
    \label{eq: subspace_resolved_deformation_map}
\end{equation}

\subsubsection{Checkerboard path and gauge qualification}
\label{ssec: qdm_boundary_holonomy}

The finite checkerboard kinetic path is obtained from the fixed-state
compatible phase space after projecting away the image of local link
rephasings.  Explicitly,
$\widehat G(\boldsymbol\theta)=\exp(i\sum_\ell\theta_\ell\widehat n_\ell)$ induces a linear map
from link phases to plaquette-flip phases.  On the $4\times4$ torus, the
selected compatible unit vector is proportional to
\begin{equation}
    \chi_{p_{(x,y)}}=(-1)^{x+y}.
\end{equation}
With Euclidean normalization $\chi/4$, let $g_{\rm cb}$ denote the path
coordinate, so that the corresponding couplings are
$\exp(ig_{\rm cb}\chi_p/4)$ on plaquette $p$.  Over
$-0.35\le g_{\rm cb}\le0.35$, the compact-motif residual
remains below $3\times10^{-15}$ and the fixed $A,Z$ residuals vanish.  The gauge obstruction of the checkerboard \emph{direction} can in fact be
seen analytically within this declared local class.  Let
$\mathsf M_{\rm lg}$ map link phases $\boldsymbol\theta$ to the induced
plaquette-flip phases.  A physical link belongs to two neighboring plaquettes
with the same incidence sign in the corresponding column of
$\mathsf M_{\rm lg}$, whereas their checkerboard signs are opposite.
Consequently
\begin{equation}
    \boldsymbol\chi^{\mathsf T}\mathsf M_{\rm lg}=0,
    \qquad
    \boldsymbol\chi^{\mathsf T}\boldsymbol\chi=N_p>0,
    \label{eq: qdm_checkerboard_link_gauge_obstruction}
\end{equation}
for the even periodic lattices considered here, where $N_p$ is the number of
plaquettes.  Hence
$\boldsymbol\chi$ lies in the left kernel of the local link-to-plaquette map
and cannot lie in its linear image.  At $L_x=4,8,12$, the finite-size quotient has unit relative distance and
rank $N_p-1$, consistent with this analytic obstruction.  This is the precise scoped meaning of the non-gauge checkerboard
direction used in the paper; it is not a theorem against arbitrary
configuration-dependent diagonal rephasings, and finite phase patterns are
understood modulo the usual $2\pi$ identifications.

A size-compatible phase family keeps the local phase amplitude fixed rather
than renormalizing the full phase vector at each size.  The size-compatible
family therefore uses
\begin{equation}
    \phi_{p_{(x,y)}}=\varphi(-1)^{x+y},
    \label{eq: qdm_appendix_checkerboard_family}
\end{equation}
so that on the $4\times4$ normalization above, $\varphi=g_{\rm cb}/4$.

The torus obstruction can be represented explicitly as a boundary holonomy.
Let $\theta^h_{x,y}$ and $\theta^v_{x,y}$ be the phases on links from $(x,y)$
to $(x+1,y)$ and $(x,y+1)$, respectively, with coordinates understood modulo
the lattice lengths.  Since $\widehat U_p=|v\rangle\langle h|$, conjugation
by $\widehat G(\boldsymbol\theta)$ changes the plaquette phase by
\begin{equation}
\begin{aligned}
    \delta\phi_{p_{(x,y)}}
      ={}&\theta^v_{x,y}+\theta^v_{x+1,y}\\
         &-\theta^h_{x,y}-\theta^h_{x,y+1}.
\end{aligned}
    \label{eq: qdm_link_phase_incidence_explicit}
\end{equation}
Choose the periodic link assignment, using representatives $0\le x<L_x$,
\begin{equation}
    \theta^v_{x,y}=\varphi x(-1)^{x+y},
    \qquad \theta^h_{x,y}=0.
    \label{eq: qdm_checkerboard_seam_rephasing}
\end{equation}
For $\phi'_p=\varphi\chi_p+\delta\phi_p$, all bulk plaquette phases cancel:
\begin{equation}
    \phi'_{p_{(x,y)}}=
    \begin{cases}
      0,&0\le x<L_x-1,\\
      L_x\varphi\chi_{p_{(x,y)}},&x=L_x-1.
    \end{cases}
    \label{eq: qdm_checkerboard_seam_phases}
\end{equation}
The diagonal flippability term is unchanged, so
\begin{equation}
\begin{aligned}
    \widehat G\widehat H_{\varphi,\lambda}\widehat G^\dagger
      &=\widehat H_{0,\lambda}+\widehat V_{\rm seam},\\
    \|\widehat V_{\rm seam}\|&\le 2|J|L_y.
\end{aligned}
    \label{eq: qdm_checkerboard_boundary_norm_bound}
\end{equation}
Only $L_y$ flip terms differ, each by operator norm at most $2|J|$.
At $L_y=4$, this bounds the difference of the full winding-sector free-energy
densities by $2|J|/L_x$ at fixed temperature.  The remaining invariant phase
product is $\exp(i\sum_p\chi_p\phi_p)=\exp(i4L_x\varphi)$; thus a
nontrivial finite-torus holonomy is consistent with the linear obstruction in
Eq.~\eqref{eq: qdm_checkerboard_link_gauge_obstruction}.
This phase family tests continuation under a boundary twist, not a distinct
bulk phase coupling.  The rephasing is exact at each size, although its
derivative with respect to $\varphi$ is not uniformly bounded as $L_x$ grows.
It preserves the dimer constraint and winding, but generally conjugates the
translation and reflection projectors.  The boundary representation therefore
leaves open whether the microcanonical and canonical local means
in the original fixed sectors, including the witness values, vary with $\varphi$.

The all-size state and fixed-$A,Z$ continuation follow from
Eqs.~\eqref{eq: qdm_appendix_exact_periodic_product} and
\eqref{eq: qdm_appendix_allN_local_annihilation}; the finite-size residuals
check these identities.  Table~\ref{tab: qdm_resolved_thermal_comparison}
collects the same-family raw thermal results, including the target-energy
joint-dark subspace at $L_x=12$.

\subsubsection{Secondary deformation controls}

The nonuniform-potential path uses one translation-breaking real pattern of
the 16 flippability coefficients.  Over $-0.3\le g\le0.3$, the localized caged eigenstate
and its kinetic operators remain dark.  For the pair
$p_1=p_{(x,0)}$ and $p_2=p_{(x,2)}$, the corrected diagonal construction
must be deformation adapted,
\begin{equation}
    \widehat Y_R(g)
    =v_{p_1}(g)\widehat F_{p_1}+v_{p_2}(g)\widehat F_{p_2}
     -[v_{p_1}(g)+v_{p_2}(g)]\widehat{\mathbf1}.
    \label{eq: qdm_appendix_adapted_Y}
\end{equation}
The kinetic witnesses have positive microcanonical values in the $4\times4$
windows.  Larger-strip transport and concentration for this secondary path
remain open.  Because
translation is broken, this path uses the localized caged eigenstate and winding sectors,
not the symmetry-resolved compact representative of the primary kinetic route.

\paragraph*{Uniform Peierls control and gauge qualification.}
The uniform phase path remains exact on the compact representative over the
stated interval.  A rank-15 map from 32 link phases to the 16 plaquette-phase
directions places the uniform vector inside the link-gauge image, with distance
$2.3\times10^{-15}$.  It is therefore retained only as a finite-size
thermal-background control.  The checkerboard direction instead has the
finite-torus holonomy described in
Sec.~\ref{ssec: qdm_boundary_holonomy}, and supplies the compatible boundary-twist
reference family.

\subsubsection{Boundary conditioning}
\label{ssec: qdm_boundary_singular_gap}

For a preserving path parameterized by $g$, define the smallest positive
boundary singular value
\begin{equation}
    \sigma_B^{+}(g)
    =\min_{\sigma_j(B_{\mathcal S}(g))>0}
      \sigma_j(B_{\mathcal S}(g)).
    \label{eq: qdm_boundary_singular_gap}
\end{equation}
This is a conditioning scale after compatibility has been established.  It is
not an energy gap and does not imply exact continuation, bounded operator
support, a positive microcanonical witness value, or topological protection.  Exact preservation is
decided by the caging and local-operator equations; the singular spectrum only
measures their numerical conditioning.

\FloatBarrier

\section{Conditional entanglement bounds for caged eigenstates}
\label{appendix: caging_entanglement_consequences}

Interference caging and a bounded local caging operator do not by themselves imply low entanglement.  This appendix records only sufficient bounds that follow when the configuration support has additional real-space structure.

\subsection{Support and real-space bounds}
\label{ssec: appendix_caging_entanglement}
\label{sssec: appendix_active_region_entanglement}

Fix a bipartition of the ambient microscopic tensor factors into $A$ and $B$; in a constrained or gauge system, the assignment of degrees of freedom crossing the geometric cut must be declared explicitly \cite{donnelly2012,casini2014}.  For
\begin{equation}
    \ket\psi=\sum_{s\in\mathcal S}x_s
      \ket{a(s)}_A\otimes\ket{b(s)}_B,
    \qquad x_s\neq0,
    \label{eq: appendix_entanglement_support_expansion}
\end{equation}
write $\pi_A(\mathcal S)=\{a(s):s\in\mathcal S\}$ and similarly for $B$.

\begin{proposition}[Configuration-support bound on Schmidt rank]
\label{prop: caging_support_rank_bound}
The Schmidt rank obeys
\begin{equation}
    \chi_{A|B}
    \leq\min\{|\mathcal S|,|\pi_A(\mathcal S)|,|\pi_B(\mathcal S)|\},
    \label{eq: appendix_support_schmidt_rank_bound}
\end{equation}
and hence every R\'enyi entropy of order $\alpha>0$ satisfies
\begin{equation}
    S_A^{(\alpha)}(\psi)
    \leq\log\chi_{A|B}
    \leq\log\min\{|\mathcal S|,|\pi_A(\mathcal S)|,|\pi_B(\mathcal S)|\}.
    \label{eq: appendix_support_entropy_bound}
\end{equation}
\end{proposition}
\begin{proof}
Use the projected configurations as row and column labels of
$\Psi_{ab}=\sum_{s\in\mathcal S}x_s\delta_{a,a(s)}\delta_{b,b(s)}$.
Its matrix rank is the Schmidt rank and is bounded by the number of nonzero rows, columns, or entries.  A rank-$\chi$ reduced density matrix has entropy at most $\log\chi$.
\end{proof}

Thus subexponential growth of the projected support across every macroscopic cut gives a sub-volume-law bound; in particular,
\begin{equation}
    \log\min\{|\mathcal S_L|,|\pi_A(\mathcal S_L)|,|\pi_B(\mathcal S_L)|\}
    \le C_{\rm area}|\partial A_L|+O(1)
    \label{eq: appendix_area_support_condition}
\end{equation}
with a size-independent constant $C_{\rm area}$ implies an area-law upper
bound; $\partial A_L$ is the real-space boundary of $A_L$.  These support counts are basis-dependent sufficient diagnostics, not entanglement invariants, and the boundary equation $B_{\mathcal S}\boldsymbol x=0$ alone constrains none of them.

Write $S_A$ for the von Neumann entanglement entropy of the reduced state
on $A$.  Three common real-space refinements follow immediately from rank bounds.  If all support configurations differ from one reference product state only inside $R_L$, with local dimension at most $d_*$, then
\begin{equation}
    S_A(\psi_L)
    \le \min\{|R_L\cap A_L|,|R_L\cap B_L|\}\log d_*.
    \label{eq: appendix_active_region_entropy_bound}
\end{equation}
If the coefficient matrix decomposes into $N_\gamma$ matching sectors with $\operatorname{rank}\Psi^{(\gamma)}\le r_\gamma$, then
\begin{equation}
    S_A(\psi)\le\log\!\left(\sum_{\gamma=1}^{N_\gamma}r_\gamma\right).
    \label{eq: appendix_boundary_label_entropy_bound}
\end{equation}
Finally, if at most $n_{\partial A}$ bounded entangled blocks of Schmidt rank $D$ cross the cut, then $S_A\le n_{\partial A}\log D$; for a coherent superposition of $M$ such patterns,
\begin{equation}
    S_A\le\log M+n_{\partial A}^{\max}\log D.
    \label{eq: appendix_block_superposition_entropy}
\end{equation}
These statements require the corresponding real-space organization independently; a small configuration-graph radius or fixed strip width does not establish it.

\subsection{Present models}
\label{sssec: appendix_entanglement_model_status}

For the spin-1 $XY$ tower, let $L_A=|A|$ and $L_B=|B|$.  Write
$\ket{D_{A,q}}$ for the normalized equal-weight superposition of configurations
with $q$ bimagnons on $A$, carrying the tower's staggered phases, and similarly
for $B$.  Grouping configurations by $q$ gives the exact Schmidt decomposition
\begin{equation}
    \ket{\Psi_n}
    =\sum_{q=q_{\min}}^{q_{\max}}
      \sqrt{\frac{\binom{L_A}{q}\binom{L_B}{n-q}}{\binom{L}{n}}}
      \ket{D_{A,q}}\otimes\ket{D_{B,n-q}},
    \label{eq: appendix_xy_schmidt_decomposition}
\end{equation}
with $q_{\min}=\max(0,n-L_B)$ and $q_{\max}=\min(n,L_A)$.  Hence
\begin{equation}
\begin{aligned}
    \chi_{A|B}&=q_{\max}-q_{\min}+1
    \le\min\{L_A,L_B,n,L-n\}+1,\\
    S_A&\le\log\chi_{A|B}.
\end{aligned}
    \label{eq: appendix_xy_entanglement_bound}
\end{equation}
So every tower sequence has at most logarithmic entanglement across the cut, including finite-density members.  This follows from the tower's coherent number structure, not from interference caging itself, consistent with Ref.~\cite{schecter_weak_2019}.

For the four-configuration $4\times4$ QDM representative, Proposition~\ref{prop: caging_support_rank_bound} gives $S_A\le\log4$ for any bipartition of the declared ambient link tensor factors.  The repeated periodic-product sequence would inherit a bounded-block area-law estimate only if an actual tensor-product or bounded-bond-dimension block representation were established; the support--boundary factorization alone does not do so, and no entanglement-scaling claim is made for the collective quotient.

Entanglement is therefore an auxiliary structural diagnostic here.  The witness and its microcanonical value give the direct test of the thermodynamic local ETH mismatch.

\section{Relations to other QMBS and constrained-dynamics formalisms}
\label{appendix: relations_other_formalisms}

This appendix records the precise logical relations between the local-caging
framework and neighboring viewpoints: many-body/Fock-space caging,
Shiraishi--Mori projector embedding, commutant algebras,
quasiparticle/ladder constructions, and Hilbert-space fragmentation
\cite{ben2025many,mohapatra_unraveling_2025,serbyn_quantum_2021,moudgalya2022quantum,chandran_quantum_2023}.
Recent many-body/Fock-space caging studies emphasize interference localization
on sparse subgraphs of a many-body state graph; in the spin-1 setting,
Ref.~\cite{mohapatra_unraveling_2025} also identifies simultaneous eigenstate
structure of noncommuting local operators through a commutant algebra.
Local annihilators, positive parent constraints, and QMBS-preserving
Hamiltonian families are therefore established ingredients
\cite{shiraishi_systematic_2017,moudgalya_exhaustive_2023,mohapatra_unraveling_2025}.
The result-specific additions here are the constructive removal of the
outside conditioning from an actual active-boundary cancellation, the raw
thermal comparison for the resulting bounded observable, and the joint
continuation of that chosen operator with a readjusting state.
Table~\ref{tab: formalism_comparison} summarizes these comparisons.

\begin{table*}[tp]
\caption{Logical relation between local caging and neighboring QMBS/constrained-dynamics formalisms.  None of the comparison structures follows solely from configuration-space confinement.}
\label{tab: formalism_comparison}
\small
\setlength{\tabcolsep}{8pt}
\setlength{\JustifyingParindent}{0pt}
\renewcommand{\arraystretch}{1.25}
\begin{tabular}{@{}lll@{}}
\hline\hline
\parbox[t]{0.16\textwidth}{\justifying Formalism} &
\parbox[t]{0.35\textwidth}{\justifying Defining structure / extra condition} &
\parbox[t]{0.42\textwidth}{\justifying Relation to local caging} \\[3pt]
\hline
\parbox[t]{0.16\textwidth}{\justifying Many-body/Fock-space caging} &
\parbox[t]{0.35\textwidth}{\justifying Interference-localized support on a many-body state graph.  The ICQMBS bridge additionally requires a bounded physical-space annihilator with nonzero same-Hamiltonian local ensemble contrast.} &
\parbox[t]{0.42\textwidth}{\justifying Configuration-graph caging is the common starting point. The present framework constructs a witness in physical space and tests the thermodynamic background.} \\
\noalign{\vskip 0.5em}
\parbox[t]{0.16\textwidth}{\justifying Shiraishi--Mori} &
\parbox[t]{0.35\textwidth}{\justifying Local projectors embedding a target, together with the Hamiltonian decomposition in Eq.~\eqref{eq: sm_general_form}.} &
\parbox[t]{0.42\textwidth}{\justifying $\widehat Q_a$ supplies a positive parent constraint, but need not identify the minimal target or put the original Hamiltonian in Shiraishi--Mori form.} \\
\noalign{\vskip 0.5em}
\parbox[t]{0.16\textwidth}{\justifying Commutant algebra} &
\parbox[t]{0.35\textwidth}{\justifying Bond algebra $\mathcal A_H$ and commutant $\mathcal A_H'$; equivalence requires equality of the relevant generated algebras or commutants.} &
\parbox[t]{0.42\textwidth}{\justifying Local constraints imply $\operatorname{End}(\mathcal T)\subseteq\mathcal A_Q'$, not $\mathcal A_Q=\mathcal A_H$.} \\
\noalign{\vskip 0.5em}
\parbox[t]{0.16\textwidth}{\justifying Quasiparticle/ RSGA} &
\parbox[t]{0.35\textwidth}{\justifying A physical creator $\widehat{\mathcal R}^{\dagger}$ and tower; equal target-space spacing for an RSGA.} &
\parbox[t]{0.42\textwidth}{\justifying A creator organizes what is repeatedly added; caging identifies which leakage amplitudes cancel.  Neither implies the other.} \\
\noalign{\vskip 0.5em}
\parbox[t]{0.16\textwidth}{\justifying Fragmentation} &
\parbox[t]{0.35\textwidth}{\justifying Invariant dynamical blocks $\mathcal H_\mu$ defined independently from the local dynamics.} &
\parbox[t]{0.42\textwidth}{\justifying Fragmentation is subspace-wide invariance.  Caging may occur inside a large fragment; a disconnected frozen component has no active boundary.} \\[3pt]
\hline\hline
\end{tabular}
\end{table*}

The common input for the operator-algebra comparisons below is such a bounded
local caging operator and its positive constraint,
\begin{equation}
    \widehat Q_a=\widehat L_a^\dagger\widehat L_a\ge0,
    \qquad
    \widehat L_a\widehat P_{\mathcal T}=0,
    \label{eq: target_iz_annihilation}
\end{equation}
for a target subspace $\mathcal T$ after the relevant exact global symmetries and superselection labels have been resolved.  The support--boundary derivation of such operators is given in Sec.~\ref{sec: interference_caging} and Appendix~\ref{appendix: local_iz_technical_details}; it is not repeated here.
For the comparison below, $\widehat\Pi_a$ denotes a bounded local projector,
$\mathcal A_H$ the Hamiltonian bond algebra, and $\mathcal A_Q$ the algebra
generated by the positive constraints.  The bond algebra is the unital
algebra generated by the declared local Hamiltonian terms and their adjoints.
A prime denotes the commutant, the operators commuting with every generator.
The notation $\operatorname{alg}^{*}$ means the generated unital algebra closed
under adjoints, while $\operatorname{End}(\mathcal T)$ contains all linear
operators on $\mathcal T$, extended by zero outside it.  We use RSGA for a restricted spectrum-generating algebra and
$\mathcal H_\mu$ for an invariant dynamical block.

\subsection{Projector embedding and constraint algebra}
\label{appendix: iz_projector_embedding}
\label{appendix: iz_commutant_relation}

Every positive constraint has the spectral-support projector
\begin{equation}
    \widehat\Pi_a^{\rm spec}
    =\widehat{\mathbf1}-\widehat P_{\ker\widehat Q_a}.
    \label{eq: q_spectral_support_projector}
\end{equation}
In an unconstrained tensor-product Hilbert space, strict locality of $\widehat Q_a$ gives a corresponding local projector.  After restriction to a constrained physical Hilbert space, spectral projection need not preserve a microscopic tensor-factor representation; the Shiraishi--Mori bridge therefore requires a bounded microscopic projector $\widehat\Pi_a$ satisfying
\begin{equation}
    \ker\widehat\Pi_a=\ker\widehat Q_a
    \quad\text{on the physical Hilbert space}.
    \label{eq: q_local_projector_condition}
\end{equation}
Hidden and local projector-embedding constructions provide related routes to
such QMBS subspaces \cite{omiya2023_PRA,omiya2023_PRB}.
Positivity already gives a frustration-free parent with
\begin{equation}
    \mathcal K_Q\equiv\bigcap_a\ker\widehat Q_a
    =\ker\!\left(\sum_a\widehat Q_a\right)
    \supseteq\mathcal T,
    \label{eq: iz_parent_kernel}
\end{equation}
and exact identification of the intended target additionally requires $\mathcal K_Q=\mathcal T$.  The Shiraishi--Mori construction further requires \cite{shiraishi_systematic_2017}
\begin{equation}
    \widehat H_{\rm SM}
    =\widehat H_0+\sum_a\widehat\Pi_a\widehat h_a\widehat\Pi_a,
    \qquad
    [\widehat H_0,\widehat\Pi_a]=0.
    \label{eq: sm_general_form}
\end{equation}
Thus local caging can supply projector data when Eq.~\eqref{eq: q_local_projector_condition} holds, but it does not by itself establish kernel completeness or the Hamiltonian decomposition above.  Conversely, a Shiraishi--Mori projector is boundary-row-derived only when its dark kernel can be traced to a bounded combination of the actual configuration-boundary processes in the declared microscopic basis.

The positive constraints generate
\begin{equation}
    \mathcal A_Q
    =\operatorname{alg}^{*}(\widehat{\mathbf1},\{\widehat Q_a\}),
    \qquad
    \operatorname{End}(\mathcal T)\subseteq\mathcal A_Q',
    \label{eq: target_matrix_algebra_in_commutant}
\end{equation}
because every $\widehat Q_a$ vanishes on $\mathcal T$ from both sides.  A useful converse statement is the following.

\begin{proposition}[Common scalar sector and shifted annihilators]
\label{prop: scalar_sector_annihilator_equivalence}
Let $\mathcal A=\operatorname{alg}^{*}(\widehat{\mathbf1},\{\widehat g_\alpha\})$, where each Hermitian generator $\widehat g_\alpha$ has bounded physical support $R_\alpha$.  The following are equivalent:
\begin{enumerate}[label=(\roman*)]
    \item $\operatorname{End}(\mathcal T)\subseteq\mathcal A'$;
    \item $\widehat g_\alpha$ preserves $\mathcal T$ and acts there as a scalar,
    $\widehat g_\alpha\widehat P_{\mathcal T}=\widehat P_{\mathcal T}\widehat g_\alpha=\lambda_\alpha\widehat P_{\mathcal T}$;
    \item the shifted local operators $\widehat g_\alpha-\lambda_\alpha\widehat{\mathbf1}_{R_\alpha}$ annihilate $\mathcal T$.
\end{enumerate}
\end{proposition}
\begin{proof}
Condition (i) includes commutation with $\widehat P_{\mathcal T}$, so each generator preserves $\mathcal T$; commutation with the full matrix algebra on $\mathcal T$ forces scalar action, giving (ii), and subtraction gives (iii).  Conversely, Hermiticity and two-sided scalar action imply commutation with every operator in $\operatorname{End}(\mathcal T)$.
\end{proof}

This proposition constructs local annihilators from scalar action but does not identify the Hamiltonian bond algebra.  If $\mathcal A_H=\operatorname{alg}^{*}(\widehat{\mathbf1},\{\widehat h_\ell\})$ is generated by declared local Hamiltonian terms, full algebra-level identification requires
\begin{equation}
    \mathcal A_Q=\mathcal A_H
    \quad\Longleftrightarrow\quad
    \mathcal A_Q'=\mathcal A_H'
    \label{eq: iz_bond_algebra_equality}
\end{equation}
for finite-dimensional unital $*$-algebras \cite{Moudgalya2022,moudgalya_exhaustive_2023,moudgalya_numerical_2023}.  Kernel completeness does not imply this equality.

\subsection{Ladders and fragmentation}
\label{appendix: iz_rsga_relation}
\label{appendix: iz_fragmentation_relation}

A quasiparticle construction organizes states as
\begin{equation}
    \ket{\psi_n}\propto(\widehat{\mathcal R}^{\dagger})^n\ket{\psi_0},
\end{equation}
while an RSGA additionally requires a fixed level spacing $\omega$ on the
target tower \cite{mark_unified_2020,moudgalya2020eta,chandran_quantum_2023},
\begin{equation}
    [\widehat H,\widehat{\mathcal R}^{\dagger}]\widehat P_{\mathcal T}
    =\omega\widehat{\mathcal R}^{\dagger}\widehat P_{\mathcal T},
    \label{eq: appendix_rsga_conditions}
\end{equation}
together with closure of $\widehat{\mathcal R}^{\dagger}$ along the tower.  This yields equal spacing, but neither a quasiparticle interpretation nor equal spacing is required for interference caging.  Conversely, a ladder state is caged only if its microscopic support has a nonempty active boundary whose leakage amplitudes cancel.  If the common dark kernel $\mathcal K_Q$ exceeds the tower, preservation of that larger kernel is the additional condition
\begin{equation}
    [\widehat Q_a,\widehat{\mathcal R}^{\dagger}]
    \widehat P_{\mathcal K_Q}=0
    \quad\forall a.
    \label{eq: iz_ladder_compatibility}
\end{equation}

Fragmentation instead concerns invariant blocks of the local dynamics.  For independently specified projectors $\widehat P_\mu$,
\begin{equation}
    \mathcal H_{\rm sec}=\bigoplus_\mu\mathcal H_\mu,
    \qquad
    [\widehat P_\mu,\widehat h_\ell]=0\quad\forall\ell,
    \qquad
    \widehat P_\mu\in\mathcal A_H'.
    \label{eq: appendix_fragment_decomposition}
\end{equation}
Product-basis fragmentation corresponds to disconnected configuration-graph components and hence $B_{\mathcal C_\mu}=0$; more general quantum fragmentation can require an entangled basis or an algebraic commutant resolution \cite{Moudgalya2022,chen2024}.  The distinction from caging is captured by
\begin{equation}
\begin{aligned}
    (\widehat{\mathbf1}-\widehat P_\mu)\widehat H\widehat P_\mu&=0,\\
    (\widehat{\mathbf1}-\widehat P_{\mathcal S})\widehat H\widehat P_{\mathcal S}\ket\psi&=0,\\
    (\widehat{\mathbf1}-\widehat P_{\mathcal S})\widehat H\widehat P_{\mathcal S}&\neq0.
\end{aligned}
    \label{eq: fragment_vs_cage_projector}
\end{equation}
The first is subspace-wide invariance; the second is state-selective cancellation on an active boundary.  If an independently defined fragment is the correct reference space, the ETH comparison must be performed inside that fragment, and its energy window must still have positive entropy density and raw bounded-local concentration.  Fragmentation alone therefore supplies neither an ICQMBS classification nor a thermal background.

\subsection{Present models and claim boundary}

For the spin-1 $XY$ tower, each nearest-neighbor kinetic bond annihilates every tower member.  The local support projector
\begin{equation}
    \widehat\Pi_{r,r+1}
    =\widehat{\mathbf1}-\widehat P_{\ker\widehat k_{r,r+1}}
    \label{eq: xy_bond_support_projector}
\end{equation}
is a two-site projector, and the original Hamiltonian admits the corresponding Shiraishi--Mori decomposition because $\widehat k_{r,r+1}=\widehat\Pi_{r,r+1}\widehat k_{r,r+1}\widehat\Pi_{r,r+1}$ and the diagonal two-site term commutes with $\widehat\Pi_{r,r+1}$.  The same family has a staggered-bimagnon creator and ladder description \cite{schecter_weak_2019,chandran_quantum_2023,mark_unified_2020}.  These are complementary: the creator organizes the tower, whereas the local caging operators identify the leakage processes that cancel.

Neither the known tower nor this local parent relation by itself supplies
the raw-window conditions proved in
Proposition~\ref{prop: xy_wide_window_closure_appendix}: a nonzero limiting
raw microcanonical witness value, positive window entropy density, and
concentration on every fixed bounded region, all in the selected $(M,k_L)$ sector and uniformly
along the compatible exchange interval.  The translation-averaging route to
weak ETH is established \cite{mori_weak_2016}; the model-specific work here
is its conditional-window bridge, exact-degeneracy control, and uniform
application to the same deformed Hamiltonians.  This permits a thermodynamic
local diagnosis without extrapolating the narrow numerical windows or
assuming chaotic dynamics.

The continuation problem likewise has a narrower output than an exhaustive
commutant classification.  QMBS-preserving bond algebras organize local
Hamiltonians for a fixed target \cite{moudgalya_exhaustive_2023}; the cage
Jacobian instead allows support amplitudes and energy to readjust and asks
whether a chosen bounded boundary rule can readjust with them.  Its cokernel
detects first-order failure within the declared support and operator class.
This distinguishes loss of a particular local explanation from loss of the
eigenpair, without classifying all possible QMBS-preserving interactions.

For the square QDM, the compact representative has a support--boundary certificate, bounded local kinetic operators, and an exact circumference-four periodic-product continuation under a compatible torus boundary twist.  The constrained-model addition is an explicit eight-link rule valid in every occupied outside environment, accompanied by finite-size raw-window and canonical comparisons in the target spatial sector.  The present results do not identify a bounded operator family whose common kernel is exactly the full eight-dimensional locally generated $4\times4$ caged subspace, and bounded multi-row reducibility remains open for the one-dimensional collective quotient.  Accordingly, no minimal Shiraishi--Mori target, exhaustive commutant-algebra equality, quasiparticle creator, fragmentation claim, or thermodynamic QDM ICQMBS classification is inferred from those results.

Finally, none of these algebraic relations is itself an ETH statement.  The ICQMBS classification additionally requires the raw-window thermodynamic conditions of Definition~\ref{def: icqmbs}, evaluated in the same independently specified reference sector or fragment.  The structures in Table~\ref{tab: formalism_comparison} are therefore complementary model-dependent properties, not interchangeable names for the same mechanism.

\FloatBarrier

\makeatletter
\immediate\write\@auxout{\string\citation{apsrev42Control}}
\makeatother
\bibliographystyle{apsrev4-2}
\bibliography{references}

\end{document}